# Hierarchical Crystal Structure Prediction of Zeolitic Imidazolate Frameworks Using DFT and Machine-Learned Interatomic Potentials

Yizhi Xu,[a,b] Jordan Dorrell,[c,d] Katarina Lisac,[b] Ivana Brekalo,[b] James P. Darby,[e] Andrew J. Morris[c]* and Mihails Arhangelskis[a]*

[a]Faculty of Chemistry, University of Warsaw, Warsaw 02-093, Poland.

[b]Division of Physical Chemistry, Ruđer Bošković Institute, Zagreb 10000, Croatia.

[c]School of Metallurgy and Materials, University of Birmingham, Birmingham B15 2TT, UK.

[d]School of Chemistry and Chemical Engineering, University of Southampton, Southampton, SO17 1BJ, UK.

[e]Department of Engineering, University of Cambridge; Trumpington Street, Cambridge CB2 1PZ, UK.

Corresponding authors:

Andrew J. Morris: a.j.morris@bham.ac.uk

Mihails Arhangelskis: m.arhangelskis@uw.edu.pl

**Abstract:** Crystal structure prediction (CSP) is emerging as a powerful method for the computational design of metal-organic frameworks (MOFs). In this article we employ CSP to perform high-throughput exploration of the crystal energy landscape of zinc imidazolate (Zn**Im**$_2$). As the most polymorphic member of the zeolitic imidazolate framework (ZIF) family, Zn**Im**$_2$ has at least 24 reported structural and topological forms, and new polymorphs still being regularly discovered. With the aid of custom-trained machine-learned interatomic potentials (MLIPs) we have performed a high-throughput sampling of over 3 million randomly-generated crystal packing arrangements and identified 9609 energy minima characterized by 1484 network topologies, including 855 topologies that have not been reported before. All but one experimentally-reported structures of Zn**Im**$_2$, falling within the search boundaries, were ultimately matched with the predicted structures, demonstrating the power of the CSP method in sampling experimentally-relevant ZIF structures. Finally, through a combination of topological analysis, density and porosity considerations, we have identified a set of structures representing promising targets for future experimental screening. as well as demonstrated how structures of mechanochemically-synthesized MOFs could be identified via matching experimental powder diffraction patterns with simulated patterns from the predicted structures.

## Introduction

Metal-organic frameworks (MOFs) are highly-versatile materials with applications in gas storage[1] and separation,[2] catalysis,[3,4] water purification,[5] removal of harmful agents from air,[6] energy storage,[7] light harvesting,[7] fuels[8,9] and more.[10,11] Such functional diversity is directly related to the modular nature of MOFs, which are constructed from metal nodes interconnected by organic linker molecules, giving rise to a vast number of node and linker combinations, resulting in materials with diverse short-range interaction geometries, long-range crystal packing and associated functional properties.[12]

The structural and functional variability of MOFs, however, is not limited to node-and-linker variations: a further dimension of structural diversity comes from polymorphism, where the same building blocks give rise to multiple crystallographic arrangements. A prominent class of MOFs renowned for polymorphic and topological diversity are zeolitic imidazolate frameworks (ZIFs),[13] which are geometrically and topologically related to zeolites, thanks to the tetrahedral geometry of the metal nodes and angled coordination geometry of the imidazolate linkers. An archetypal example of ZIF polymorphic diversity is zinc imidazolate (Zn**Im**$_2$) which, to date, has been represented by at least 24 crystallographically-distinct forms in 19 topologies, isolated *via* solution crystallization, template-assisted synthesis,[14] solvothermal methods,[15] high-pressure-and-temperature experiments[16] and mechanochemical screening.[17] The regular discovery of new polymorphs of Zn**Im**$_2$ suggests that many more such forms can be discovered in the future. Yet, without knowing the crystal structures of the not-yet-discovered polymorphs of Zn**Im**$_2$ it is difficult to systematically target materials with specific functional characteristics, including surface area and pore volume.

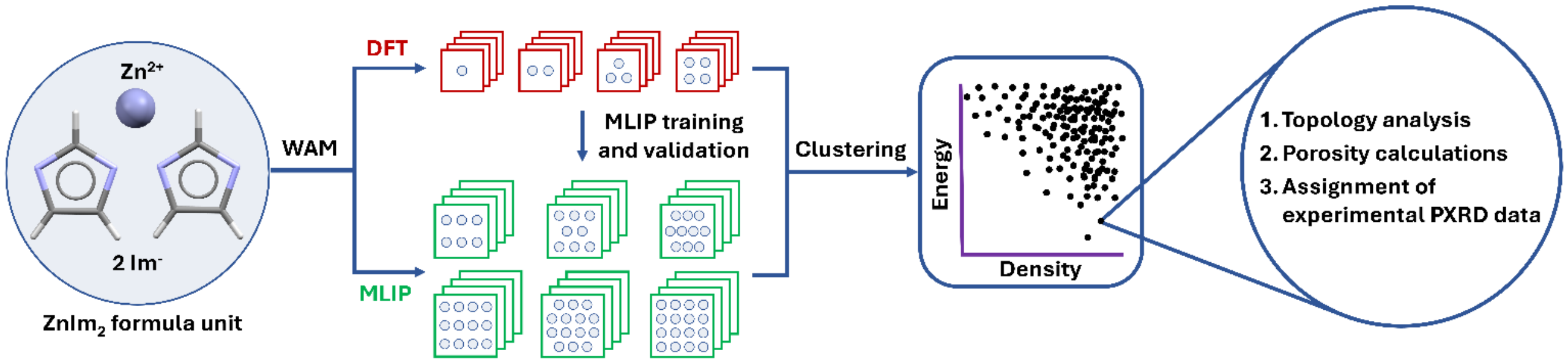

**Figure 1**. Diagram depicting the key steps of crystal structure prediction for ZnIm2. We begin with the crystal structure generation using AIRSS+WAM method. The generated structures containing 1 - 4 formula units of ZnIm2 per primitive crystallographic unit cell, are then subjected to periodic DFT optimization. The MLIPs are trained against a subset of DFT data, and then validated against the remainder of DFT data, not used during training. The resulting MLIPs are then used to optimize and energy-rank all the trial structures, containing up to 16 formula units per cell. Duplicate structures are removed during the clustering step, resulting in a crystal-energy landscape of unique structures in a synthetically-relevant energy range of 45 kJ mol-1 from the global minimum. The predicted structures are then characterized in terms of network topology and porosity, as well as matched against experimental powder diffraction data.

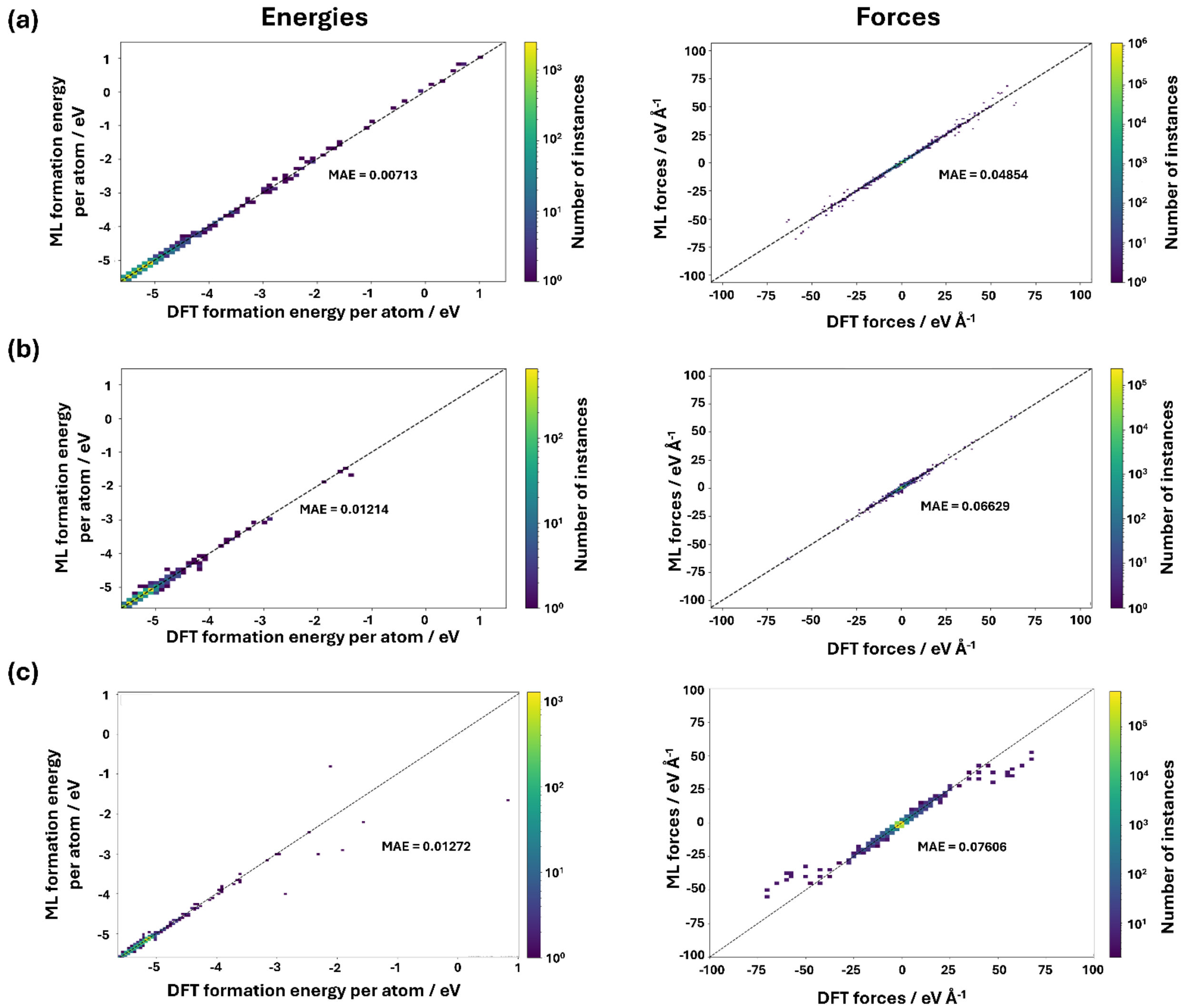

**Figure 2.** Correlation between the ML and DFT energies and forces in a) training set, b) validation set and c) testing set. The strong correlation between DFT and ML forces is a prerequisite for accurate and reliable structure optimization by the ML potential, while agreement between DFT and ML energies is a prerequisite for accurate energy ranking of the optimized structures.

The discovery of new ZIF forms with the desired functional characteristics can be accelerated through the use of crystal structure prediction (CSP), a method which has been widely used for the discovery of new crystal forms of organic molecular materials,[18] including pharmaceutical solids, porous organic materials,[19] inorganic solid electrolytes[20] and high pressure mineral phases.[21] Yet, unlike for purely organic and inorganic materials, where CSP has become an established method for materials design, the development of CSP methods capable of addressing the hybrid node-and-linker composition of MOFs mainly relied on topology-based[22–26] structure generation, limiting the generated structures only to derivatives of known topologies. To address this issue in 2020[27] we developed a new CSP approach for structure generation of MOFs based on the *ab initio* random structure searching (AIRSS) method,[28] supplemented by the Wyckoff alignment of molecules (WAM) procedure,[27] which utilizes the point group symmetry of linkers when generating putative structures. The structures were then optimized by periodic density-functional theory (DFT) calculations, resulting in an energy ranking of the generated structures, and thus a prediction of the most thermodynamically stable crystal forms. Emphasizing its utility, this approach soon after allowed for the first CSP-driven discovery of functional hypergolic MOFs.[29]

Our choice in using periodic DFT for the energy ranking was motivated by its excellent accuracy in reproducing experimentally-measured MOF polymorph energies,[30–33] yet the high computational cost is a major limitation in terms of the system sizes amenable to CSP. This limitation is particularly relevant in the study of the highly polymorphic Zn**Im**$_2$ materials, where reported system sizes range up to 40 formula units per primitive crystallographic unit cell.[34] Since our key focus is on the wide adaptation of CSP-based MOF design, as a complementary approach to experimental structure screening, accurate, yet computationally more efficient alternatives to DFT-based energy ranking are indispensable. Such an alternative has been presented in the form of machine-learned interatomic potentials (MLIPs), which have recently gained traction in computational materials discovery.[35–42]

Here we present a hierarchical high-throughput CSP study of Zn**Im**$_2$ (Figure 1), involving training of a custom-made MLIP (Figure 2). The chemical equivalence of all ZIF structures and low computational cost of MLIP-based geometry optimizations allowed us to target structures with up to 16 formula units of Zn**Im**$_2$ per primitive crystallographic unit cell. This is a significant advancement compared to our previous DFT-based CSP studies that have been limited to 1-4 MOF formula units per cell.[27,29] Inclusion of structures comprising larger unit cells and higher atomic content increased the chances of locating experimentally-relevant structures and expanded the topological diversity of the predicted structures.

The validity of our WAM-MLIP approach was tested by comparing the MLIP predictions with the results of select periodic DFT geometry optimizations for large structures, and by comparing the geometry parameters of the MLIP-generated structures to statistical experimental values.

We verify the robustness of the presented CSP approach by reproducing multiple experimentally-observed polymorphs of Zn**Im**$_2$, and use the exploration of the topology and porosity characteristics of the other predicted structures to propose likely targets for future experimental synthesis. Finally, we present the assignment of experimental powder X-ray diffraction (PXRD) data for the mechanochemically-synthesized polymorphs of Zn**Im**$_2$ against the predicted structures. The presented protocol, based on the variable cell powder-based similarity index (VC-GPWDF) method,[43] highlights the utility of CSP in analyzing the outcomes of mechanochemical reactions, where the polycrystalline nature of their products makes the experimental structure determination particularly challenging.

## Results and discussion

### Training and validation of the ML potential

Previously, the AIRSS method has been used to predict structures of a wide variety of materials, including solid electrolytes,[44,45] materials under high pressure,[46–49] extra-terrestrial minerals,[50] hybrid perovskites,[51] organic molecular crystals,[52,53] non-metal organic frameworks based on porous organic salts[19] and MOFs.[27,29,54] This diversity signifies the versatility of AIRSS, which is based on placing the structural building blocks at random positions within the trial unit cell with randomly defined unit cell parameters, followed by relaxation of the geometry of such trial structures. The structure generation step is then repeated until the search is converged. The key strength of AIRSS lies in the ability to apply structural constraints suitable for a particular system, *e.g.* defining geometries of the structural building blocks in the form of isolated atoms, atomic clusters or extended molecules.

In the context of MOFs, the natural building blocks for an AIRSS search are metal nodes and organic molecular linkers. In addition, given that MOFs are known for their high crystallographic symmetry, we apply symmetry constraints via the Wyckoff Alignment of Molecules (WAM)[27] method, that allow symmetric building blocks to occupy special Wyckoff positions, enabling structures to be realized in unit cells with fewer formula units.

The key feature of our AIRSS+WAM methodology is that we are not making any assumptions about metal coordination number, coordination geometry or framework topology. The only input information is the atomic composition and geometry of individual nodes and linkers, as well as overall number of these fragments to be placed in the trial unit cell, subject to minimum separation (MINSEP) constraints. The connectivity between individual building blocks and, ultimately, framework topology, is established during the subsequent structure optimization steps.

This is in stark contrast to a family of structure-building methods,[22,23,25,26] where the nodes and linkers are initially placed at the positions defined by the desired network topology, and this topological connectivity is then preserved during the structure optimization step. Building structures from isolated nodes and linkers allows us to sample structures from a wider range of topologies, as well as discover new topologies, not yet found in databases such as Reticular Chemistry Structure Resource (RCSR)[55] or Topological Types Database (TTD).[56] We also see promise in addressing polymorphism within the same network topology: with the recent discovery of two new forms of zinc imidazolate with **crb** topology,[17] this material now has five crystallographically-distinct **crb** polymorphs, emphasizing the importance of considering this type of polymorphism in computational screening of MOF structures.

The major challenge in the development of MOF CSP has been directly related to their covalent node-and-linker, hybrid organic-inorganic character. MOFs, being covalent 3D-polymeric structures, cannot be broken into fragments held only by non-covalent interactions. Because of this, MOF CSP cannot utilize force-field potentials in a way they are used for molecular crystals. Our initial strides in CSP for MOFs were, therefore, made using periodic DFT for energy ranking of putative structures, due to its excellent accuracy in reproducing experimentally-measured MOF polymorph energies,[30–33] and despite the high computational cost.

The CSP calculation process for Zn**Im**$_2$ initially followed a similar approach to our earlier CSP studies for zinc triazolate and tetrazolate,[27] as well as copper(II)-based hypergolic ZIFs.[29] The initial structure search spanned the space of 1-4 formula units per primitive crystallographic cell, with these structures geometry-optimized via plane-wave periodic DFT calculations in CASTEP19, using the LDA functional. However it quickly became apparent that the search limited to 1-4 formula units per cell would not be sufficient to cover the relevant structural space, where many of the previously reported polymorphs of Zn**Im**$_2$ have been found. Indeed the esperimentally-reported polymorphs span a much larger structural space, including 8 formula units (**gis**, CSD HIFVUO;[57] **crb**, CSD VEJYEP;[58] **moc**, CSD KUMXEW[59]) and 16 formula units (**coi**, CSD IMIDZB07;[60] **zni**, CSD IMIDZB02;[60] **crb** GITTEJ;[61] **cag**, CSD VEJYUF;[58] **dft**, CSD VEJYOZ;[58] **mer**, CSD VEJZIU; [58] **crb**, CSD VEJYIT[58]) as well as several examples of extra-large structures spanning 20-40 formula units per primitive cell (**nog**, CSD HIFWAV;[62] **zec**, CSD

HICGEG;[62] **hlw**, CSD ZAVBUX;[63] **can**, CSD PAJRUQ;[64] **afi**, CSD IMIDZB13;[64] **10mr**, CSD GOQSIQ[34]).

Expanding the standard search method to a higher number of formula units would be met with two major obstacles: first, with increased number of formula units, the number of atoms in the unit cells and the size of the unit cells themselves get larger, resulting in a higher cost of DFT optimization for each structure; second, a larger number of formula units leads to more structural degrees of freedom, making it necessary to optimize more structures in order to obtain good coverage of the potential energy surface (PES). Overall, the computational cost of exploring the structural landscape of Zn**Im**$_2$ with DFT-based energy ranking would quickly become prohibitively expensive, motivating us to seek a different strategy.

Broadly, there exist two approaches for the use of MLIPs in computational materials design: one can train a custom-made potential, based on DFT data gathered for a particular atomic or molecular composition, using one of the available codes for the construction of interatomic potentials based on artificial neural networks (ANN), for example ænet[65] or SchNetPack.[66] Alternatively, one can utilize one of the generalized foundation models, such as MACE,[67] CHGnet[68] or NequIP,[69] that are trained on large databases of *ab initio* calculations, such as Materials Project,[70] OMOL[71] and SPICE.[72]

The generalized foundation models, owing to their large training datasets (it is now typical to train models on millions of structures, containing most of the elements of the Periodic Table) offer better generality over a range of systems they can target, but their accuracy needs to be tested for a particular problem at hand. Additionally, it is often possible to use fine-tuning procedures to improve the performance of these models when describing specific characteristics of a given material class.

On the other hand, custom-made potentials offer a more accurate description for materials closely-related to the structures contained in the underlying training set, with no additional fine-tuning needed. As a downside, introduction of different types of chemical bonding can cause difficulties, and addition of new elements not present in the original training set is generally impossible.

However, in our particular case, the similarity of all ZIF structures in terms of chemical connectivity (each structure is based on tetrahedral Zn nodes connected by imidazolate linkers via Zn-N bonds), and the need for a very accurate model (due to the very small energetic differences between different ZIF polymorphs) encouraged us to use custom-made potentials. This was made easier by the fact that extensive DFT data from the optimizations of 1-4 formula units was available as a basis to train a custom-made MLIP. We hypothesized that a MLIP trained on DFT data sampled from structures containing 1-4 Zn**Im**$_2$ formula units would be capable of predicting structures containing a much larger number of formula units. The rationale for this comes from the fact that all Zn**Im**$_2$ ZIF structures are chemically-equivalent, regardless of their size: zinc metal nodes are expected to form four Zn-N coordination bonds, with consistent bonding lengths and angles, with each imidazolate linker utilizing both nitrogen atoms to connect with the metal nodes. Since the connectivity of the Zn**Im**$_2$ building blocks is the same regardless of their total number, a custom-made MLIP should have no scaling problems within this system.

To test this hypothesis, we will check the validity of our MLIP methodology by performing selected periodic DFT geometry optimizations for larger structures containing different numbers of formula units in their crystallographic cells, and comparing our MLIP predictions with the results of DFT calculations. We will also check the statistics of the geometric parameters of the MLIP-optimized structures to see if they conform to the ranges expected based on a survey of experimental ZIF structures. To ensure that building a custom-made potential instead of using a generalized foundation model was valid, we will evaluate the performance of our MLIP against the MACE model (SI Section S2).

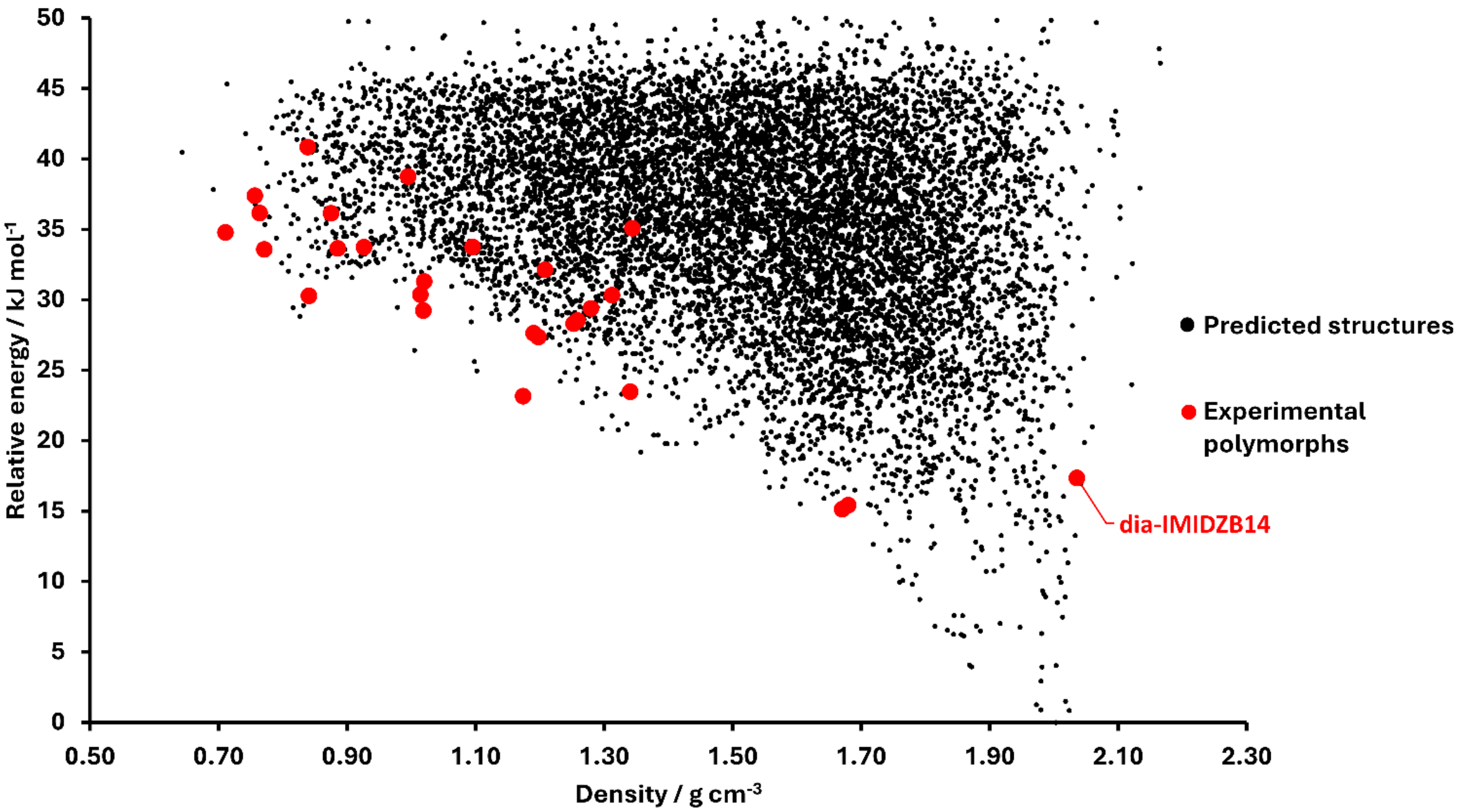

**Figure 3.** Crystal energy landscape of Zn**Im**$_2$, where structures were geometry-optimized and energy-ranked with MLIPs, shows the calculated relative energies of predicted crystal structures against their density. The energies and densities of experimental structures of Zn**Im**$_2$ from CSD are shown in red. Experimental structures are clustered along the lower end of the energy-density envelope, with the exception of the highest density structure **dia**-IMIDZB14, which has been only experimentally synthesized under high pressure. This structure is specifically highlighted with a CSD REFCODE.

We have selected the deep neural network atomistic simulation code SchNetPack[66] with the built-in polarizable interaction neural network (PaiNN) architecture[73] for constructing the MLIPs. The PaiNN neural network allowed us to use both energy and force data from the DFT calculations to train the MLIP models. In the end, we have constructed two separate MLIPs: one potential trained exclusively on DFT forces, which we used to optimize the trial structures, and another one trained on energies, used for energy ranking of the optimized structures (see SI Sections S1 and S2 for details). To validate the accuracies of energy and forces MLIPs, comparisons between ML predicted energies and forces with DFT values were obtained (Figure 2) from structures in training, validation and test sets. This resulted in low mean absolute errors (MAE) of 7.13 meV/atom, 12.14 meV/atom and 12.72 meV/atom from the energy MLIP for the training, validation and test set, respectively. In the case of force MLIP, the MAEs were 48.54 meV/Å, 66.29 meV/Å and 76.06 meV/Å for the training, validation and test set, respectively. The MAEs obtained herein are similar to those reported in earlier MLIP studies.[40,74] Based on these encouraging results from validating the accuracy of our MLIPs, all putative ZIF structures were geometry optimized. The optimized structures were then ranked using the energy MLIP, with the final set of structures ranging up to 45 kJ $mol^{-1}$ above the global energy minimum retained for detailed analysis. Such an energy window was selected based on the prior results of DFT calculations and experimental calorimetric measurements of ZIF polymorph stability.[30–33,75–80]

**General trends within the predicted energy landscape**

The first, immediately apparent feature of the energy landscape (Figure 3) is the trend where lower density structures tend to be higher in energy. This is consistent with our previous calculations on MOFs,[30,31,81] as well as CSP studies of porous molecular crystals.[52,82,83] Based on void fraction analysis from the software PLATON,[84] 8247 structures out of 9609 from the CSP energy landscape are considered porous with non-zero void fractions. However, we must keep in mind that the porous nature of MOF structures poses certain challenges for energy ranking in CSP. While in close-packed materials, predicted structures with higher densities tend to have the lowest energies due to a larger number of short-range interatomic contacts, the situation is more complex with MOFs and other porous materials. The challenge is associated with the possibility of guest inclusion within the voids of the porous structures: while in a conventional CSP calculation, such voids are assumed to be empty, under the conditions of experimental synthesis, the structural voids can be readily occupied by solvent molecules, or other small molecule guests present in the reaction mixture. The inclusion of guests within the voids leads to additional stabilization of the structure via host-guest interactions, effectively making porous structures more stable than they appear under the energy calculations based on structures with empty voids. The effect of guest inclusion has been recognized as a challenge in previous CSP studies of porous molecular crystals,[85–88] as well as during the DFT-based energy ranking of MOF polymorphs obtained in the mechanochemical screening via liquid-assisted grinding (LAG).[17] Given the known propensity of Zn**Im**$_2$ and other ZIF systems to form porous structures, renowned for their sorption capacity,[89,90] it will be imperative to consider the effect of host-guest stabilization on the calculated energy landscape of Zn**Im**$_2$ in this study.

The energy-density trend discussed above, correlates well with the characteristics of the experimentally-reported forms of Zn**Im**$_2$. When placing the MLIP-optimized structures of experimentally-reported polymorphs of Zn**Im**$_2$ on the energy landscape of CSP structures (Figure 3), it is evident that these structures are concentrated at the lower diagonal part of the energy-density plot, signifying that there exists a relationship between relative stability, density and porosity of ZIF structures found experimentally. This suggests that the lower-diagonal region of the energy landscape is where the candidates for the future discovery of new polymorphs of Zn**Im**$_2$ are most likely to be found.

**Topology distributions and structures matching experimentally-reported forms of ZnIm$_2$**

Having investigated the general stability trends throughout the calculated energy landscape, we turned our attention to the topological analysis of the predicted structures and the geometrical matching between the predicted structures and experimentally-determined ZIF polymorphs (Figure 4). The topological analysis revealed a remarkable diversity with 1484 distinct topological nets, of which 855 were found to be new topologies, not contained in the ToposPro[91] TTD database. We performed a statistical analysis of the geometrical parameters of the predicted structures, using the CCDC suite of programs.[92] Our analysis shows that these structures contain node and linker connectivity appropriate for ZIFs, with the Zn-N bond length and N-Zn-N bond angle distributions (SI Figures S4 and S5) very similar to the experimental structures of Zn**Im**$_2$ found in the Cambridge Structural Database (CSD),[93] with the $\tau$ geometry index indicative of tetrahedral coordination geometry (SI Figure S6). Agreement of the geometrical statistics of our predicted structures with experiment provides confidence to the validity of our MLIP approach.

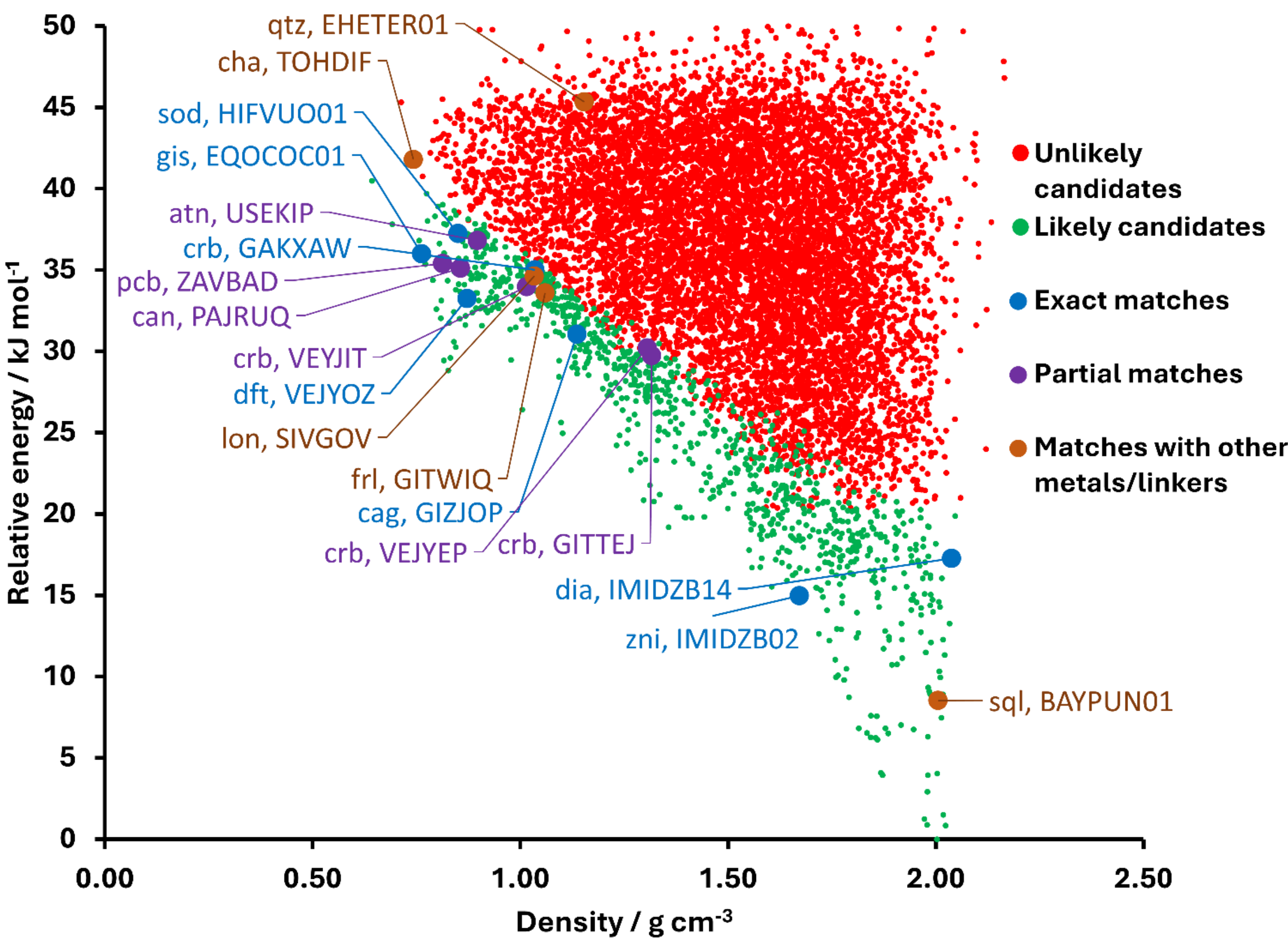

**Figure 4.** Crystal energy landscape of Zn**Im**$_2$, where structures were geometry-optimized and energy-ranked with MLIPs. The structures colored by their synthetic feasibility, based on the relationship between the energy and void fraction calculated for the experimentally-observed polymorphs. The structures colored in green are most likely to be synthesizable, while those colored in red are deemed synthetically less likely, given their higher relative energy and lower calculated porosity. In particular, it is evident that the hypothetical **qtz** structure is unlikely to exist based on the combined energy-porosity criterion for synthetic feasibility.

We next focused on the exploration of individual topologies and their distribution within the overall CSP energy landscape, (SI Figures S9-S23) while highlighting the predicted structures matching the experimentally-observed polymorphs, and suggesting structures that appear as likely candidates for future synthesis of new polymorphs of Zn**Im**$_2$. In order to make such comparisons more robust, the experimental structures from the CSD were optimized with the same MLIP as used for crystal structure prediction. The optimized structures were then compared with the structures from the CSP crystal energy landscape (SI Table S2).

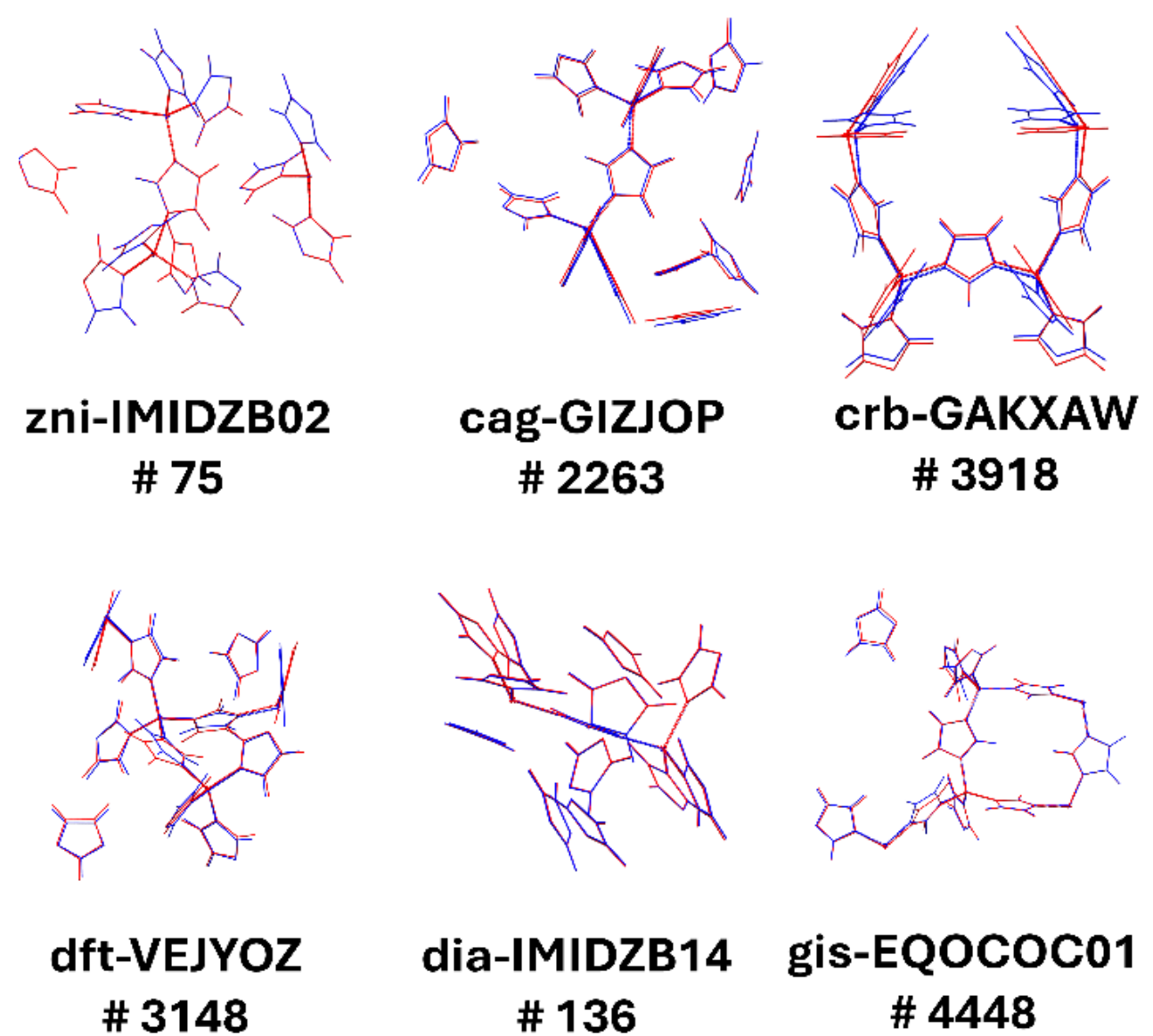

**Figure 5.** Overlays of the predicted (red) and experimentally-reported (blue) structures from CSD, where the experimental structures were optimized with the same MLIP as used for CSP. The topology and CSD REFCODE are written underneath each overlay picture. The CSP-generated structures are shown in blue, and the experimental MLIP-optimized structures are shown in red. The numbers underneath the structures refer to the entries in the supporting Chemiscope file.

The first major observation was the abundance of 2D structures throughout the CSP landscape, with 1294/9626 structures belonging to the **sql** topology, including the global minimum structure and all other structures within 4.1 kJ mol$^{-1}$ above it (SI Figure S22). To the best of our knowledge, no **sql** polymorph has been isolated for Zn**Im**$_2$ so far, but **sql** structures of unsubstituted imidazolate ZIFs have been reported for other transition metal centers, notably for Ni(**Im**)$_2$ (CSD ALIDUU)[94] and Hg(**Im**)$_2$ (CSD BAYPUN01).[95] In the latter case both experimental simulations and periodic DFT calculations have shown that the **sql**-Hg(**Im**)$_2$ form is more stable than its 3D polymorph with **dia** topology.[95] Interestingly, the Zn**Im**$_2$ structure isomorphous to the reported **sql**-Hg(**Im**)$_2$ form is found in our CSP energy landscape with the energy of 8.50 kJ mol$^{-1}$ above the global minimum, and below any of the experimentally obtained Zn**Im**$_2$ structures. This implies that it should in principle be possible to prepare a 2D polymorph of Zn**Im**$_2$.

Going up in energy, at 14.98 kJ mol$^{-1}$ the structure of **zni** topology Zn**Im**$_2$, matching the experimental structure (CSD IMIDZB02) was found. The **zni** form is currently regarded as one of the two densest and most thermodynamically stable reported polymorphs of Zn**Im**$_2$, along with the **coi** form.[75,76,78] The low energy of the **zni** form is evidenced both by experimental dissolution calorimetry measurements[75] and periodic DFT calculations.[17] Notably, the CSP energy landscape contained 15 different crystal structures with the **zni** topology, but it was the lowest energy structure among them that matched the experimentally-reported form (SI Figure S23).

Further inspection of the energy landscape revealed several more matches (Figure 5) to the experimentally-observed polymorphs of Zn**Im**$_2$, including the high pressure doubly-interpenetrated **dia** polymorph (17.26 kJ mol$^{-1}$ above the global minimum, matching structure CSD IMIDZB14),[16] **cag** (31.04 kJ mol$^{-1}$, matching CSD GIZJOP);[61] **dft** (33.25 kJ mol$^{-1}$, matching CSD

VEJYOZ);[58] **gis** (36.00 kJ mol$^{-1}$, matching CSD EQOCOC01)[58] and **sod** (37.24 kJ mol$^{-1}$, matching CSD HIFVUO01).[57] In the latter case, it should be noted, that the pure **sod** polymorph of Zn**Im**$_2$ composition has not actually been obtained so far, however, an **sod** material of the composition Zn(**Im**)$_{1.7}$(**mIm**)$_{0.3}$ (where **mIm** = 2-methylimidazolate) has been synthesized through solvent-assisted linker exchange (SALE) procedure starting from Zn(**mIm**)$_2$ (ZIF-8), achieving 85% replacement of the **mIm**$^-$ linker with **Im**$^-$.[57] In the light of that result, the presence of **sod** structure in the crystal energy landscape of Zn**Im**$_2$ is fully justified.

Further matches were found among the structures represented by the **crb** topology. This topology is rather unique in a sense that five crystallographically-distinct polymorphs of Zn**Im**$_2$ have been reported so far,[17,58,61] making it the highest number of ZIF polymorphs sharing the same topology, to the best of our knowledge. The CSP landscape included a match to one of the forms of **crb**-Zn**Im**$_2$ just recently reported as **crbT** in our earlier publication (CSD GAKXAW).[17] This structure, experimentally obtained by liquid-assisted grinding of zinc oxide with imidazole in the presence of toluene liquid additive is found at 35.00 kJ mol$^{-1}$ relative energy. For other existing polymorphs with **crb** topology, exact matches could not be found among the predicted structures, however partial matches were located for three out of four remaining **crb** structures: 29.73 kJ mol$^{-1}$, matching CSD GITTEJ;[61] 30.20 kJ mol$^{-1}$, matching CSD VEJYEP[58] and 33.97 kJ mol$^{-1}$, matching CSD VEJYIT.[58] In addition, a partial match was identified for the **pcb**/**aco** topology (CSD ZAVBAD) at 35.39 kJ mol$^{-1}$, and for the **can** topology (CSD PAJRUQ), with the relative energy of 35.10 kJ mol$^{-1}$. In all of these cases, some of the imidazolate linkers were oriented differently in the CSP-generated structures, compared to their experimental counterparts, as seen from the overlays in Figure 6, with the details given in the SI Table S3. Imidazolate linker rotation around the Zn-Zn axis can bring the structure to a new energy minimum without breaking the covalent bonds and changing the network topology, therefore such partial matches, whilst less rewarding than complete matches discussed above, are nonetheless instructive from the point of view of structural and topological diversity of zinc imidazolate crystal energy landscape.

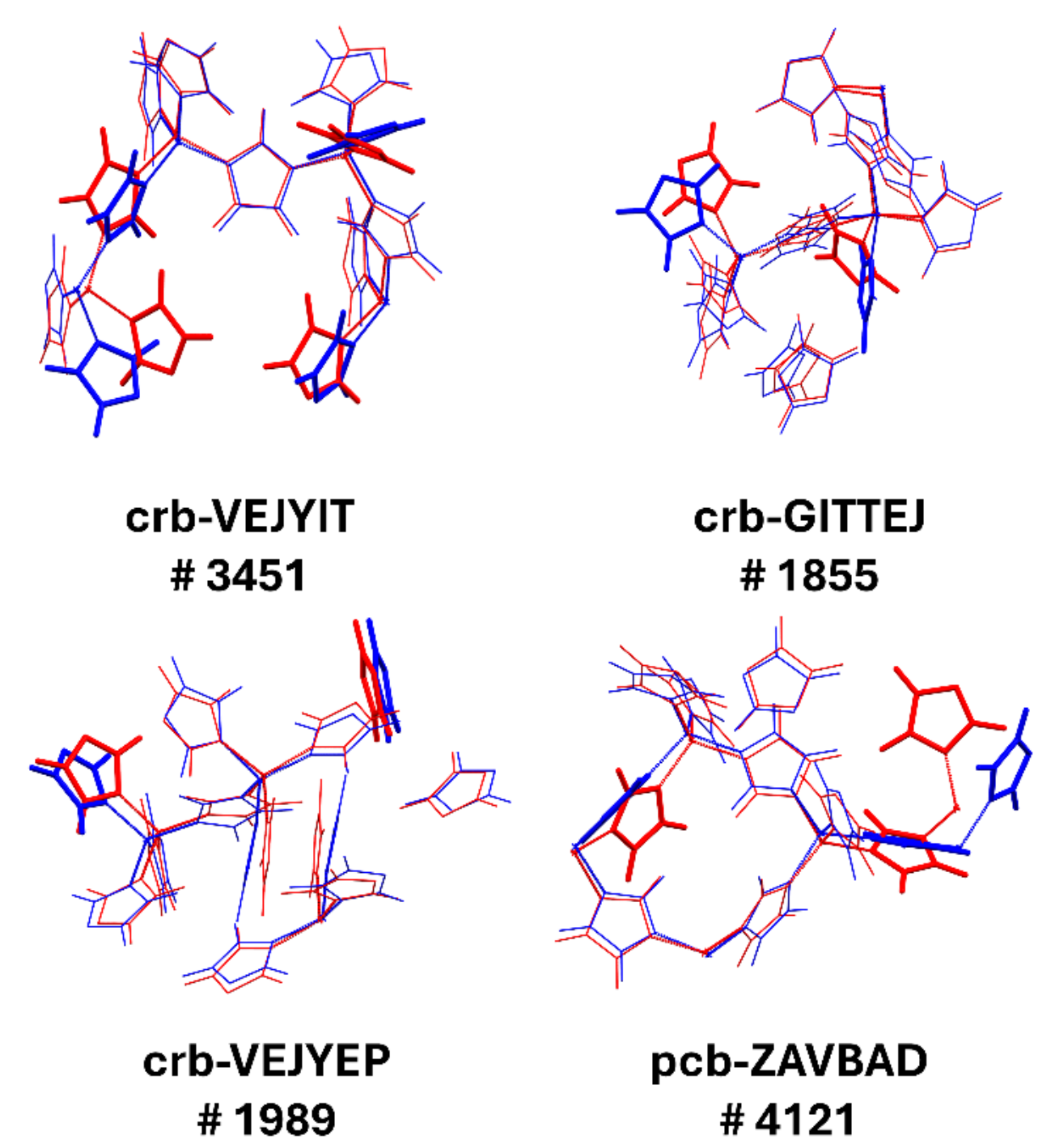

**Figure 6.** Partial overlay between the experimental (blue) and predicted (red) structures of **crb** and **pcb** topologies. The imidazolate linkers drawn with thicker bonds are those whose orientations do not agree between the predicted and experimental structure. The numbers underneath the structures refer to the entries in the supporting Chemiscope file.

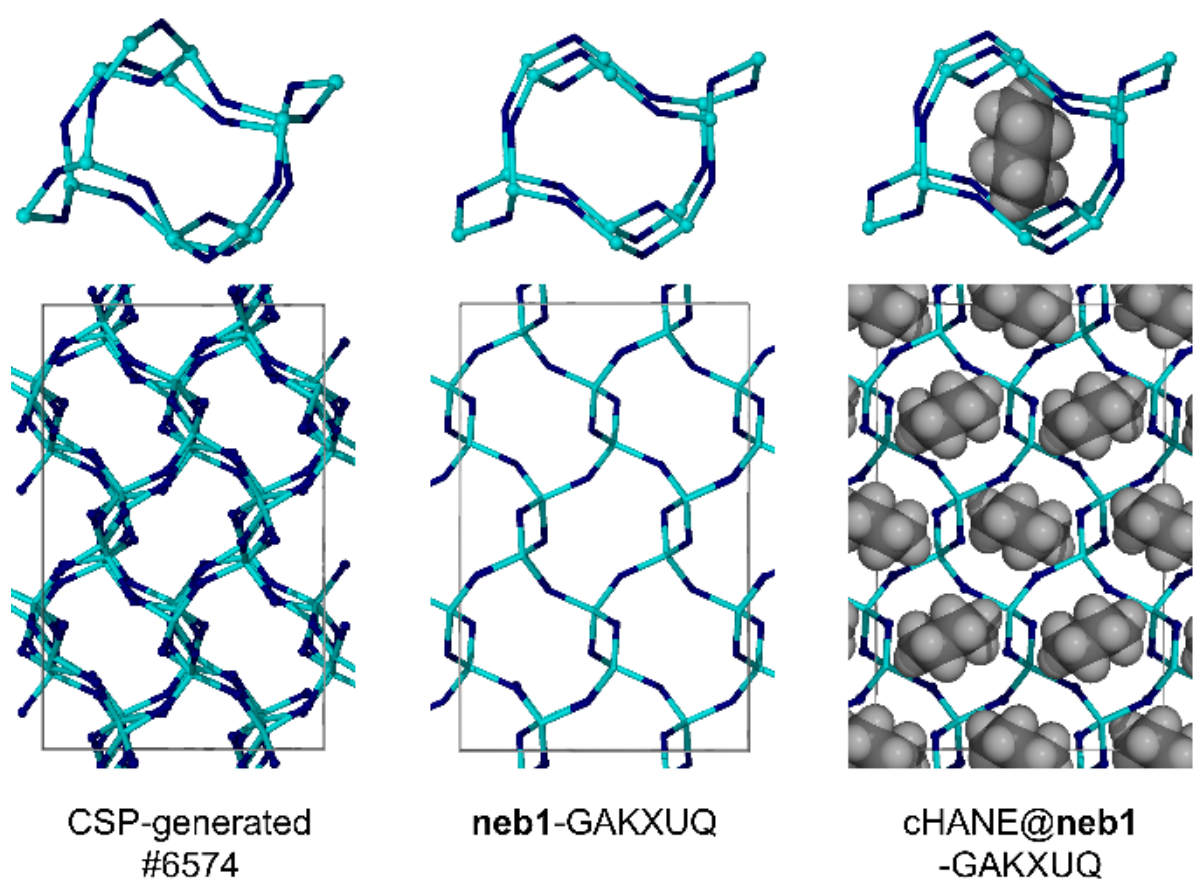

**Figure 7.** Comparison of the predicted and experimental **neb** polymorphs of Zn**Im**$_2$, highlighting the structural similarity when viewing down the c-axis.

A particularly interesting case is that of the **neb** topology. Experimentally, the neb topology was found in two distinct forms, **neb1** (CSD KUDJOK[96] - with morpholine – and GAKXUQ[17] - with cyclohexane, cHANE) and **neb2** (CSD KEVLEE[78] - with pyridine), primarily dictated by the type of included guest. Our CSP search found three different structures with **neb** topology, one of which, while not an exact match, appears to be structurally related to the **neb1** experimental polymorph. Namely, the structure (Chemiscope file from the supporting information entry 6574) is in the same *Fdd2* space group as cHANE@**neb1**-Zn**Im**$_2$ (GAKXUQ), and has the following unit cell parameters: a = 16.87 Å, b = 26.72 Å, c = 28.48 Å, while the cHANE@**neb1**-Zn**Im**$_2$ (GAKXUQ) unit cell parameters are a = 17.74 Å, b = 27.46 Å, c = 9.11 Å. It therefore appears that the predicted structure has similar a and b unit cell axes, but triple the c axis of the cHANE **neb1** polymorph. A graphical inspection of the predicted structure shows that its unit cell can be divided into three roughly repeating layers along the c axis, where each layer is a slightly distorted **neb1** unit cell, with rotations of imidazolate ligands causing differences between the layers. A comparison of the node and linker representations of the predicted **neb** structure with the empty and cyclohexane occupied **neb1**-GAKXUQ structure (Figure 7) shows that the **neb1** cage is preserved in both structures, but is conformationally distorted in the CSP generated structure. We hypothesize that the source of the distortion is the lack of guest modelled in the CSP generated structure. It is very likely that the ordering of **neb** cages in the **neb1**-GAKXUQ structure arises from the incorporated guest. If the guest is not there, like in our CSP calculations, the linkers and nodes have much more freedom to move and distort, resulting in a structural mismatch, despite the fairly accurate crystal structure prediction.

Having identified the matching experimental polymorphs of Zn**Im**$_2$ among the predicted structures, we need to investigate the structures that have not been located and discuss the reasons for their absence within the predicted crystal energy landscape. The main reason for missing some of the existing polymorphs was limiting the search space to 16 formula units per primitive cell, as several experimentally-determined structures contain more formula units. Specifically, the polymorphs with **zec** (CSD HICKEG) and **nog** (CSD HIFWAV) topologies contain 20 formula units per primitive cell, while structures with **gme** (CSD DOTCIC), **hlw** (CSD ZAVBUX) and **afi** (CSD IMIDZB13) topologies contain 24 formula units. Finally, the **10mr** framework (CSD GOQSIQ) represents the most complex structure as a 10-nodal net with 40 formula units per primitive cell.

While the limit on the size of the structural search space explains the majority of the missing structures, there were two experimental polymorphs of Zn**Im**$_2$ containing 16 formula units per primitive cell, which have not been located in our CSP search. These were **mer** (CSD DOTBOH) and **coi** (CSD IMIDZB07). The **mer** structure, while representing a uninodal net, with just one Zn atom in the crystallographic asymmetric unit, has all its imidazolate linkers disordered with respect to rotation around Zn-Zn axis. Since the predicted structures are necessarily ordered, we may suggest the inability to match the disorder of the experimental structure as the reason for our inability to reproduce the structure of the **mer**-Zn**Im**$_2$ framework.

The experimental form with **coi** topology (CSD IMIDZB07), also missing from our CSP landscape, represents a 4-nodal network, meaning the corresponding crystal structure must contain at least 4 symmetry-independent Zn nodes, resulting in a large asymmetric unit, which is harder to generate during the AIRSS+WAM structure generation. Given the importance of the **coi** form as the lowest energy structure among the experimentally-synthesized polymorphs of Zn**Im**$_2$ so far,[78] we decided to perform an additional structural search in order to better understand the challenges associated with the discovery of low symmetry MOF structures by CSP.

The dedicated search for the **coi** polymorph included generation of additional 100,000 structures containing 16 formula units of Zn**Im**$_2$ in the space group $I4_1$, the settings consistent with the experimental structure **coi**-IMIDZB07. For comparison, our original CSP search contained 12102 structures in these crystallographic settings, therefore the additional search corresponded to an 8-fold increase in the number of trial structures. Gratifyingly, this additional search resulted in the location of a **coi** structure as the overall energy minimum among the newly sampled structural space (Figure 8). This result implies that missing the **coi** structures in the initial CSP search was not caused by the limitations of AIRSS and WAM methods, but rather by the restricted number of generated structures. Increasing the number of trial structures can certainly increase our chances of locating low symmetry structures, yet the benefits of searching more structures have to be balanced with the higher computational cost of the calculation.

To summarize, the presented CSP search located all but two experimental crystal forms of Zn**Im**$_2$ within the imposed 16 formula unit limit, with the **coi** polymorph subsequently recovered in a more targeted search. Given that our previous CSP studies of MOFs, utilizing periodic DFT energy ranking were limited to 4 formula units per primitive cell, the introduction of MLIP provided for a great expansion of the search space. Indeed, if we had performed this search at the DFT level and kept the search complexity limit to 4 formula units, we would not have found any of the experimental polymorphs of Zn**Im**$_2$, highlighting the importance of MLIPs for geometry optimizations and energy rankings for high-throughput CSP of MOF materials.

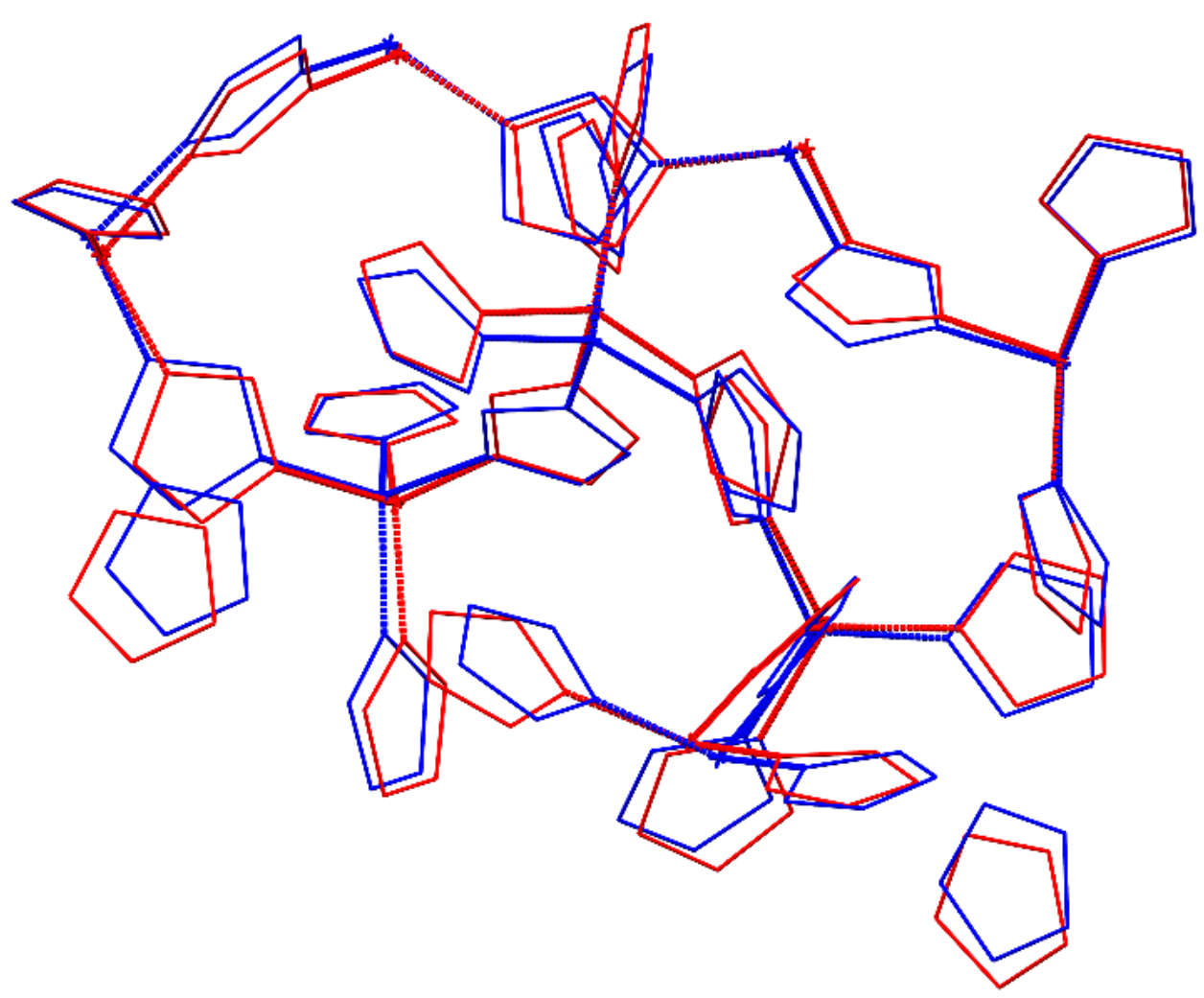

**Figure 8.** Overlay of the experimental **coi**-Zn**Im**$_2$ (shown in blue, CSD IMIDZB07) with the structure generated during the additional search in *I41* symmetry (shown in red).

**Predicted structures that are likely to be found in the future**

The primary purpose of performing CSP calculations is to find likely candidates for future synthesis. With over 9000 structures found in the CSP landscape, we need a way to narrow down the search space for future synthetic efforts.

The first thing to notice is the presence of multiple **sql** structures near the bottom of the energy landscape, including the global minimum. While no **sql** polymorphs have been isolated for Zn**Im**$_2$ so far, the existence of 2D **sql** structures for Ni**Im**$_2$ and Hg(**Im**)$_2$ suggests that an **sql**-Zn**Im**$_2$ could be isolated, perhaps through seeding experiments.

The next general observation, arising from the location of the experimentally-matching structures, is that synthetically-viable structures appear at the bottom end of the energy-density envelope. This means that high energy structures can be experimentally-feasible, as long as they have low density and, correspondingly, high void volume and surface area, that lead to energy stabilization through host-guest interactions with structural templates or solvent guest molecules. Higher energy non-porous structures, however, cannot benefit from such host-guest stabilization, and are therefore less likely to be produced during experimental synthesis.

One promising candidate for future synthesis may be the predicted structure with *lon* topology, with an energy of 34.62 kJ mol$^{-1}$ above the global minimum. This structure is isomorphous to CSD SIVGOV, an experimental framework containing 2-methyltetrazolate linker. The hypothetical structure of **lon**-Zn**Im**$_2$ has a similar energy to multiple experimentally-observed polymorphs (e.

g. **dft**, **can** and **gis**), has a low calculated density of 1.03 g $cm^{-3}$ density and a high calculated void fraction of 49%, making it highly accessible to guest inclusion.

Another structure deemed promising based on similar arguments is one with *cha* topology. With an even lower density of 0.86 g $cm^{-3}$ and 58% calculated void fraction, this structure is isomorphous to CSD TOHDIF, a ZIF based on mixed 2-methylimidazolate and 5-methylimidazolate linkers.

The true value of CSP, however, is not in identifying individual structures isomorphous with CSD entries with different metal nodes or organic linkers, but rather providing a range of targets that are structurally-distinct from anything that has been experimentally obtained before, yet feasible from a synthetic standpoint. In order to narrow down the range of predicted structures and rank the remaining structures in the order of synthetic feasibility, we devised an empirical equation combining the relative lattice energy and calculated void volume:

$$E' = E_{rel} - k_V \times f_{void}$$

where $E_{rel}$ is the calculated energy of the structure relative to the global minimum (in kJ $mol^{-1}$) and $f_{void}$ is a calculated void fraction (range from 0 to 1). The $k_V$ parameter was fitted by evaluating the $E_{rel}$ and $f_{void}$ values of the existing experimental polymorphs of Zn**Im**$_2$, resulting in the best fit value of $k_V$ = 34.05 kJ $mol^{-1}$, with the mean value for E' = 16.35 $\pm$ 3.66 kJ $mol^{-1}$ (see SI Section S7 for details). The significance of the descriptor E' is that structures with high lattice energy $E_{rel}$ can be stabilized by solvent inclusion if they contain solvent-accessible void volume, the higher the void fraction, the greater the stabilization offered by guest inclusion. For non-porous structures E' is equal to $E_{rel}$, while for more porous structures the difference between E' and $E_{rel}$ becomes progressively higher. Empirically we considered structures within one standard deviation from the mean E' value for the experimentally observed polymorphs of Zn**Im**$_2$ to be considered as viable candidates for future synthesis. This resulted in 8-fold reduction of the number of structures under consideration from 9626 to 982, narrowing down the set of structures worthy of consideration for experimental screening.

In the future such analysis may be expended by explicitly modelling the host-guest interactions via molecular-dynamics (MD) and Grand Canonical Monte Carlo (GC-MC) simulations,[97,98] however the force field potentials used in these simulations will first have to be tested for compatibility with the MLIP model used to describe the guest-free MOF structures. Alternatively, a generalized model capable of simultaneously modelling the structures of guest-free and guest-filled MOFs, as well as the bulk liquids, needs to be developed. These calculations are beyond the scope of the present study, while the empirical model allows us to predict the guest-induced stabilization of Zn**Im**$_2$ structures and narrow down the range of structures deemed feasible from an experimental synthesis standpoint.

The significance of the synthesizability criterion can best be highlighted by looking at the predicted structure with **qtz** topology, that was found to be isomorphous to the experimentally-reported **qtz** polymorph of Zn(**EtIm**)$_2$ (CSD EHETER, **EtIm** = 2-ethylimidazolate).[99] While **qtz**-Zn(**EtIm**)$_2$ is known to be a stable dense structure,[30] our predicted **qtz**-Zn**Im**$_2$ analogue is found very deep in the region of non-synthesizable structures on the energy landscape (Figure 4), It has a high relative energy, yet has only 6% solvent-accessible void volume, suggesting that guest-stabilization is likely to be unsuccessful, and that **qtz**-Zn**Im**$_2$ is unlikely to be synthesized in the future.

The likely synthesizable structures were further analyzed for porosity, with 517 structures having non-zero calculated surface area, and 291 structures having non-zero network-accessible surface area. Among these, 21 structures exceeded network accessible surface area of 2000 $m^2$ $g^{-1}$, with the maximum surface area found in a predicted structure with **dei** topology, at 2538.62 $m^2$ $g^{-1}$. This structure had a three-dimensional pore network with a limiting pore diameter of 7.35 Å and maximum pore diameter of 12.39 Å. Calculated porosity characteristics for all predicted structures can be found in the Chemiscope file, attached as the supporting information.

Finally, periodic DFT optimizations using the PBE functional[100] with D3[101] dispersion correction were performed for the 982 likely synthesizable structures, in order to further verify the accuracy of our MLIPs.

In general, we have observed a strong correlation between relative ML and DFT energies, as shown in SI Figure S2, where the energy of the experimental matching **zni** structure (CSD IMIDZB02) was used as a reference. The slope of the linear fit slightly deviated from one, suggesting that our MLIP trained on data obtained by using the LDA functional, gives a somewhat "compressed" energy scale, compared to the dispersion-corrected PBE functional. However, the standard error for the MLIP energies compared to the PBE+D3 DFT energies was found to be only 3.09 kJ $mol^{-1}$, which is quite small, considering the overall energy landscape spans a window of 45 kJ $mol^{-1}$.

Furthermore, this new set of DFT-optimized larger structures, allowed us to verify that the accuracy of MLIP predictions does not depend on the number of formula units contained in the crystallographic unit cell. Since our potentials were trained exclusively on structures containing 1-4 formula units, we tested that the accuracy of the predictions remains high for structures with up to 16 formula units. By separately comparing the MLIP and PBE+D3 DFT energies for structures with 1-4, 6, 8, 12 and 16 formula units (there were only three structures with 14 formula units among the likely-synthesizable forms of Zn**Im**$_2$, making statistical analysis meaningless for this group), we found that MLIPs perform similarly for all of these groups, both in terms of linear regression parameters and error distributions (SI Figures S4 - S6). This highly rewarding result, validates our training strategy, with the potentials trained on 1-4 formula units performing comparably across the entire range of 1-16 formula units per cell.

Analysis of the new DFT-based energy ranking supports the observation made with MLIP ranking, that there are hypothetical structures of Zn**Im**$_2$ more stable than the lowest-energy reported **zni** and **coi** forms. Moreover, the global minimum based on DFT ranking has **sql** topology, further supporting the possibility of discovering **sql**-Zn**Im**$_2$ experimentally.

**Identification of unknown experimental structures through the assignment of powder diffraction patterns**

An important challenge in the synthesis of new materials is structure determination of products of high-throughput syntheses. The synthesis does not always result in diffraction quality single crystals, instead producing polycrystalline materials. This is often the case, during solvothermal syntheses, but especially when using mechanochemistry. In that case, structure solution from powder X-ray diffraction (PXRD) data must be performed, which can be challenging. Recently, advances in electron diffraction methods provided a potential alternative, however this technique is still not widely available, and porous materials are often challenging subjects for ED, as they are extremely susceptible to electron beam damage.[7] Instead, we propose a combination of CSP and a PXRD based structure matching protocol as an alternative to *ab initio* structure solution. We here demonstrate the assignment of experimental PXRD patterns of mechanochemically-synthesized MOFs against the thousands of predicted structures in the CSP landscape, as a way to determine the structures of new materials. The assignment is based on the GPWDF algorithm,[43] implemented in Critic2.[102,103] We first tested the protocol on a selection of experimental PXRD patterns with known crystal structures from our recent publication[17] on the mechanochemical solid form screening of Zn**Im**$_2$, in order to test the sensitivity and precision of this assignment method. Then, we performed an assignment for a pattern with an unknown crystal structure.

In the initial test we included PXRD patterns of four Zn**Im**$_2$ materials that were both identified by CSP and found in our mechanochemical screening, namely **zni** (CSD IMIDZB02), **crbT** (CSD GAKXAW) and two different solvates of the **cag** topology material (CSD VEJYUF01, prepared by milling with DMF and chloroform liquid additives). First, PXRD patterns for all CSP predicted structures were simulated using the CSD Python API.[104] The simulated patterns were then compared to the selected experimental patterns, and the structures were ranked in ascending similarity order based on the variable-cell similarity index (DIFF), from the function of COMPAREVC[43] in Critic2. A variable-cell similarity index of 0 indicates a perfect match between the experimental powder pattern and the predicted structure, while a score of 1 means a full dissimilarity. The detailed results of these PXRD assignments, with the DIFF rankings of the matching predicted structures are shown in SI Table S5. Among these assignments, the predicted structure with **zni** topology showed the lowest DIFF score of 0.02 getting ranked as 6$^{th}$ best match overall. The CSP structure matching the recently-reported **crbT** form synthesized via liquid-assisted grinding with toluene, was ranked as the 199$^{th}$ best match to the corresponding experimental pattern. Finally, the predicted structure for **cag-**Zn**Im**$_2$ ranked as 473$^{rd}$ and 763$^{rd}$, respectively, when compared against the patterns for two different preparations for the **cag** form, milled with dimethylformamide and chloroform, respectively.

The results demonstrate considerable variations in the ranking of correct structural matches against the experimental data, and gives us an indication of what to expect, when using this method for PXRD patterns whose true experimental structure is not yet known and needs to be determined. The lowest overall ranking for the matching of the **zni** structure is attributed to the lack of porosity in this structure, resulting in a very good match of the calculated diffraction peak intensities.

Conversely, for the **cag** and **crbT** forms, the experimental patterns were collected on materials with guest molecules occupying the structural voids, while the predicted structures are modelled with empty voids. This leads to a discrepancy between the simulated and experimental diffraction intensities and leads to higher DIFF scores. Given that most synthesized MOFs will have their pores occupied by guest molecules, this will be an important consideration for the future method development for the assignment of experimental PXRD patterns against CSP results for MOFs and porous materials in general.

Additionally, not only do guests inside MOF pores contribute electron density and thus change the intensity profile of the PXRD patterns, they can also have a direct impact on the MOF framework itself. This is particularly true in the case of flexible MOFs, such as ZIFs. The two tested **cag** solvates are an excellent example, as we see that the CSP structure matching well with

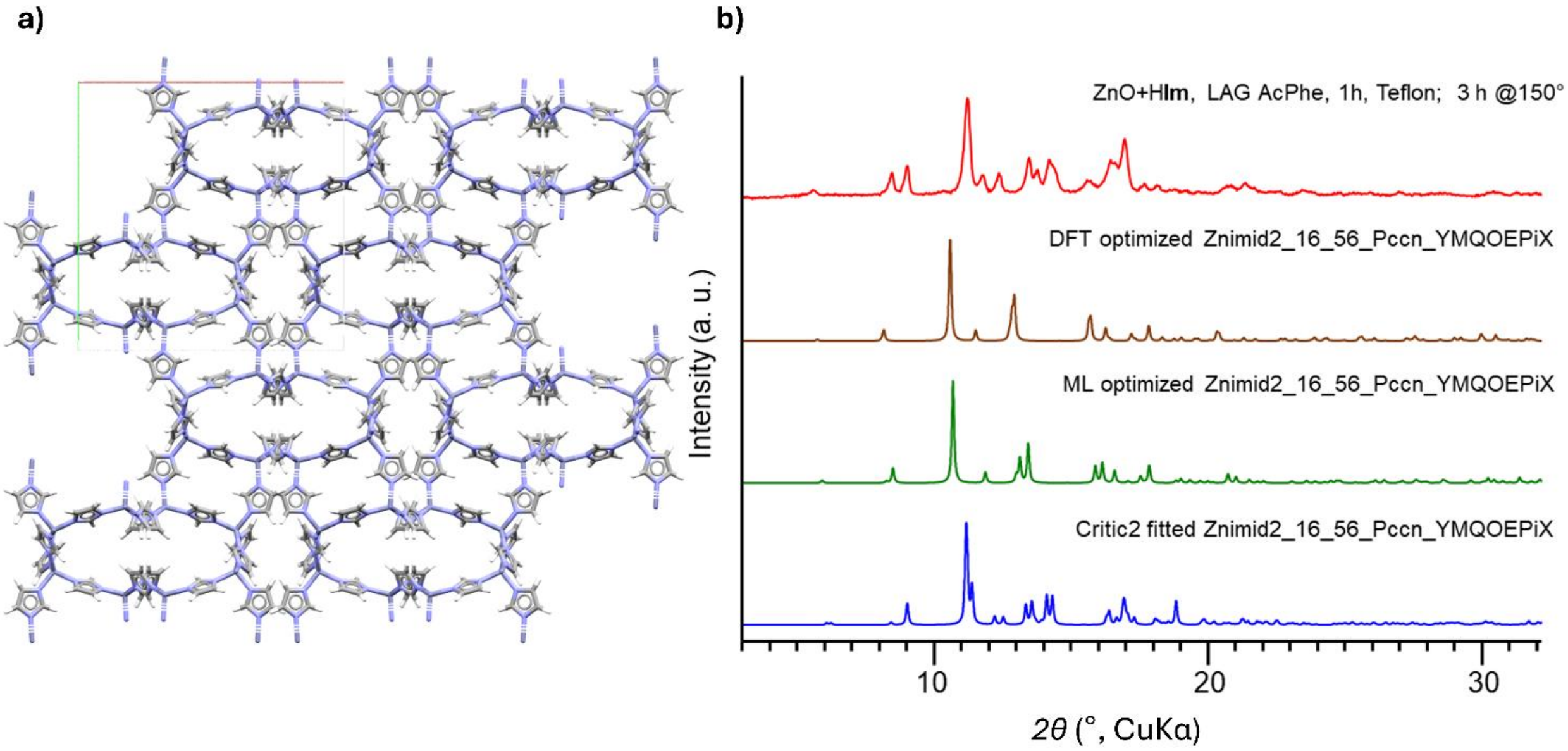

**Figure 9.** a) ML optimized crystal structure with a Chemiscope file data label 2470 (name Znimid2_16_56_Pccn_YMQOEPiX), viewed along c axis; b) PXRD patterns from top to bottom: 1) (red) experimentally collected unknown phase from our previous work,[17] obtained by heating **crbA**-Zn**Im**$_2$ at 150° for 3 hours; 2) (brown) DFT optimized predicted matching structure; 3) (green) ML optimized matching structure; 4) (blue) Predicted matching structure after cell relaxation in Critic2.

0.5DMF@**cag**-Zn**Im**$_2$ (CSD: VEJYUF01) has very different DIFF scores when compared to the DMF and $CHCl_3$ solvates of cag-Zn**Im**$_2$. Namely, the DMF solvate provides a much better match. This is unsurprising when we take into consideration that the VEJYUF01 structure is exactly a DMF solvate of **cag**-Zn**Im**$_2$. Even without actually modelling the DMF guest in the CSP calculation, the effect of the guest on the conformation of the framework is visible in the quality of the match with the experimental structure.

With these observations in mind, we continued the exploration of our CSP energy landscape, aiming to gain more structural insights for experimentally unknown structures. We selected several PXRD patterns with unknown structures from our internal experimental findings, one of which yielded a CSP match, shown in Figure 9. The DIFF score of the structure was 0.078 and it was found as the 33$^{rd}$ lowest ranked structure among the whole list of CSP entries. Experimentally, the unknown structure was obtained by heating the acetophenone solvate of **crbA**-Zn**Im**$_2$ (CSD: GAKXOX) for 3 hours at 150 degrees, resulting in a guest-free porous material of unknown structure. Besides the PXRD similarity, several other factors pointed in favor of this assignment: first, the predicted matching structure had the **crb** topology, same as the parent phase from which the new material originated via thermal transformation; and second, the matching CSP structure fell into the set of structures deemed synthesizable based on the energy-porosity criteria described earlier (Figure 4, SI Section S7). However, Rietveld refinement[105] against the experimental PXRD data using the CSP matched structure (SI Figure S28) was not in full agreement with this assignment. Not all the experimental PXRD peaks could be matched against the predicted structure, suggesting that a lower symmetry transformation may be needed to obtain the structural model fully matching the experimental data.[106]

The candidate structures described above are certainly not the only possible options for the future synthesis of new polymorphs of Zn**Im**$_2$. The great opportunities presented by CSP are brought by the diversity of structures found in these calculations. However, there also lies a challenge: CSP has been known to produce more structures than could be experimentally isolated,[107] this being a general phenomenon, applicable to molecular crystals, inorganic materials and, most certainly, MOFs. The reasons for this are both experimental (inability to sample all possible synthetic conditions) and computational: predicted energy minima may be separated by very high energy barrier, making them kinetically-inaccessible, or, alternatively, the barriers may be too low, making some of the predicted polymorphs inherently kinetically unstable at any temperature above 0 K.[108] In line with these thoughts we are releasing the entire CSP dataset, hoping that our results will prove useful to the experimental MOF community, helping in the interpretation of existing results and acting as a guideline for the design of future experiments.

## Methods

### Crystal Structure generation

Structure generation was performed using the *ab initio* random structure search (AIRSS)[28] algorithm, with randomly-chosen symmetry constraints. The Wyckoff alignment of molecules (WAM)[27] method was used to exploit the point group symmetry of the individual building blocks and allow them to occupy appropriate special Wyckoff sites.[27]

Structures were generated separately for 1, 2, 3, 4, 6, 8, 10, 12, 14 and 16 formula units per primitive crystallographic unit cell, with each formula unit including one Zn atom and two imidazolate molecular fragments, placed at random positions within the trial unit cell. Further details of the structure generation are given in the SI section S1.1.

### Generation of the DFT training dataset

The randomly-generated structures containing 1-4 formula units per primitive cell were geometry-optimized using periodic density-functional theory (DFT) calculations within the code CASTEP19.[109] Calculations were performed using LDA functional, with the plane-wave basis set truncated at 400 eV cutoff. The ultrasoft pseudopotentials were used from the internal QC5 library of CASTEP. The first electronic Brillouin zone was sampled with a $2\pi x 0.07$ Å$^{-1}$ Monkhorst-Pack k-point grid.[110] Structures were optimized with respect to unit cell parameters and atomic positions, subject to the symmetry constraints imposed by WAM space group assignment. The following convergence criteria were used: maximum energy change $2 \times 10^{-5}$ eV atom$^{-1}$; maximum atomic force 0.05 eV Å$^{-1}$; maximum atom displacement $10^{-3}$ Å; maximum residual stress 0.1 GPa.

### Training of machine-learned potentials

Periodic DFT optimization of Zn**Im**$_2$ structures, containing 1 - 4 formula units, provided us with 6000 distinct DFT trajectories, from which the crystal structures, representing individual optimization steps, were extracted. The data was then split into training, validation and test sets, making sure that each of the optimization trajectories was only used in one of those sets. This resulted in the training set, containing 10972 structures, validation set with 2739 structures and test set with 5600 structures.

The training was performed in SchNetPack[66] using the polarizable interaction neural network (PaiNN) architecture[73] with a distance cutoff for pairwise interatomic interactions set to 5 Å.

Separate MLIPs were constructed for energies and forces, the reasons for training separate potentials rather than a single one are discussed in SI Section S2.

**Geometry optimization and energy ranking**

The MLIPs were used to optimize and rank the energies of structures containing 6-16 formula units, which would be too computationally expensive to study with periodic DFT. The structures generated by AIRSS+WAM were optimized using the MLIP trained on atomic forces, initially with a force tolerance of 0.05 eV $Å^{-1}$. Subsequently, the structures were energy-ranked by performing single point calculations using the energy MLIP. Structures found to be within 45 kJ $mol^{-1}$ from the global energy minimum were then clustered with the aid of simulated powder diffraction pattern (PXRD) comparison method, implemented in the code Critic2. Structures containing 1 to 4 formula units, which were originally used to train the ML potentials, were subjected to the same procedure, in order to have a complete structural landscape from 1 to 16 formula units per primitive cell.

The final set of structures was re-optimized using the force MLIP with a tighter force tolerance of 0.005 eV $Å^{-1}$, followed by single point calculation using the energy MLIP. At that point a presence of low-density 1D and 2D structures was noticed, with large separations between chains or layers. These structures were subjected to geometry optimization under increased pressure, as explained in SI Section S1.4.

The final combined set of structures was again clustered using PXRD similarity algorithm, followed by geometry comparison via the COMPACK algorithm,[111] accessed through the CCDC Python API.[112] This resulted in 9626 unique structures.

Finally, structures were assessed in terms of chemical connectivity. Structures containing 3-, 5- or 6-coordinate zinc atoms were removed, leaving only frameworks with 4-coordinate nodes. In addition, structures containing Zn-C and N-N bonds were also removed from the final CSP set. The number of such erroneous structures, however, was small, with only 17 structures removed, leaving us with a final set of 9609 structures with proper ZIF connectivity, with 4-coordinate metal centers and doubly-coordinated imidazolate linkers.

Finally, the void volumes and packing coefficients for the remaining structures were calculated using PLATON.[84]

**Post-processing of the predicted structures**

Network topologies were determined for the final set of the predicted Zn**Im**$_2$ structures using ToposPro.[91] Coordination networks were established using the default settings of the AutoCN module. The resulting nets were then simplified, and analyzed using the default settings of the ADS module. The topological descriptors were compared against the built-in TTD database, as well as https://topcryst.com/ online server.

The structures were merged into a database using CSD Editor software. This database was then analyzed in ConQuest,[92] in order to extract bond length and bond angle distributions (SI Figures S4 and S5). The bond length and angle distributions from the predicted structures of Zn**Im**$_2$ were compared with the experimental structures found in Cambridge Structural Database (CSD).[93]

**Periodic DFT optimization of the structures deemed likely experimental candidates**

The 982 predicted structures deemed to be synthesizable based on the energy-porosity criterion (SI Section S7) were geometry-optimized with periodic DFT. These calculations used the PBE functional,[100] combined with Grimme D3 dispersion correction.[101] The plane-wave basis set truncated at 800 eV cutoff, and CASTEP default ultrasoft pseudopotentials were used. The electronic k-point grid was sampled with a 2πx0.06 $Å^{-1}$ Monkhorst-Pack k-point grid.[110] Convergence criteria were set as follows: maximum energy change $2 \times 10^{-5}$ eV $atom^{-1}$; maximum atomic force 0.05 eV $Å^{-1}$; maximum atom displacement $10^{-3}$ Å; maximum residual stress 0.05 GPa.

**Mechanochemical synthesis**

All results of mechanochemical syntheses presented herein, utilize the data reported in our previous publication.[17] Mechanochemical ball milling reactions were performed by mixing zinc oxide (75.0 mg, 0.92 mmol), imidazole (125.5 mg, 1.84 mmol) and 100 µl of a liquid additive (methanol, toluene, chloroform or dimethylformamide, depending on experiment) in a milling jar, containing two ball bearings. The samples were milled at 30 Hz for up to 90 min.

**Comparison of experimental and simulated PXRD patterns**

To start the PXRD comparisons, the experimental PXRD patterns in .raw format were converted into .xy format using the open source software PowDLL,[113] whereas all predicted structures were supplied in .res format. Comparisons were performed on background-subtracted PXRD patterns, such that each experimental PXRD pattern was compared individually with each predicted structure (9626 in total). Finally, the predicted CSP structures were ranked by ascending DIFF scores, where lower DIFF structures were accessed further to identify likely experimental matching structures.

**Conclusions**

We have presented the crystal structure prediction (CSP) study aimed at uncovering the crystal energy landscape of a highly polymorphic MOF material Zn**Im**$_2$. The major step forward in CSP methodology, presented herein, was the introduction of MLIPs for efficient geometry optimization and energy ranking of the trial structures, that allowed us to greatly extend the scope of the structural search, sampling millions of structures of highest complexity, reaching unit cells containing 16 ZIF Zn**Im**$_2$ formula units, whereas our previous searches, where we utilized periodic DFT calculations for energy ranking, were limited to four formula units.

The large search space enabled by the use of MLIP manifested itself in an unprecedented topological diversity of the predicted structures with 1484 unique topologies. Between full and partial matches, we have located all but one experimentally-reported polymorphs of Zn**Im**$_2$ falling within the boundaries of the defined search space. Moreover, the analysis of energy-density map and exploration of calculated void volume within the predicted structures allowed us to suggest some likely candidates that may lead to future new polymorphs of Zn**Im**$_2$, including a 2D form with **sql** topology.

We have then demonstrated the protocol of using CSP-generated structures for the assignment of mechanochemically-synthesized materials, by comparing the experimental and simulated powder diffraction patterns. Given the propensity of mechanochemistry to reveal new MOF solid forms, such an assignment approach is particularly important for the interpretation of experimental results.

Finally, releasing the entire CSP dataset will allow the readers to navigate the predicted structures, analyze their structural, topological and porosity characteristics. We hope that this will prove useful in guiding future experimental discovery of new ZIF materials.

To conclude, this work marks a major step in the development of CSP for MOFs, bringing it to the forefront of high-throughput computational discovery of new MOF structures with diverse packing arrangements, topological connectivities and functional properties.

## ASSOCIATED CONTENT

**Supporting Information**. Electronic Supplementary Information. Detailed description of computational and experimental procedures; CSP plots showing distributions of individual topologies; assignment of experimental PXRD patterns.

In addition, an accompanying dataset can be can be accessed via https://doi.org/10.5281/zenodo.20157677. This dataset includes all the predicted crystal structures in RES file format, as well as the associated data (energies, unit cell parameters, calculated porosity characteristics and network topologies) in a separate CSV file. In addition, a Chemiscope file is provided, for convenient data visualization via the file can be uploaded to www.chemiscope.org platform.

Finally, output files for the periodic DFT optimizations of structures deemed likely-synthesizable are also provided in the Zenodo dataset.

## AUTHOR INFORMATION

### Corresponding Authors

Andrew J. Morris: a.j.morris@bham.ac.uk;
Mihails Arhangelskis: m.arhangelskis@uw.edu.pl.

### Author Contributions

The manuscript was written through contributions of all authors. All authors have given approval to the final version of the manuscript.

### Funding Sources

YX and MA acknowledge the support of National Science Center (NCN) via grants 2018/31/D/ST5/03619, MA further acknowledges the grant 2023/51/B/ST5/01555.

This work has been supported by the “Developing Research Support” Program of the Croatian Ministry of Science and the Croatian Science Foundation, funded by the European Union from the NextGenerationEU program through grant NPOO.C3.2.R2-I1.06.0049.

AJM gratefully acknowledges networking support from CCP-NC (UKRI grant EP/T026642/1), CCP9 (EP/T026375/1), and UKCP (EP/P022561/1).

We gratefully acknowledge Poland's high-performance Infrastructure PLGrid ACC Cyfronet AGH for providing computer facilities and support within computational grant PLG/2025/018422.

This work was performed using resources provided by the Cambridge Service for Data Driven Discovery (CSD3) operated by the University of Cambridge Research Computing Service (www.csd3.cam.ac.uk), provided by Dell EMC and Intel using Tier-2 funding from the Engineering and Physical Sciences Research Council (capital grant EP/ P020259/1).

The authors acknowledge computational support from the UK national high performance computing service, ARCHER2, for which access was obtained via the UKCP consortium and funded by EPSRC grant ref EP/X035891/1.

## ACKNOWLEDGMENT

We would like to thank Professor Tomislav Friščić for the support and helpful discussions.

# Supporting Information

## Hierarchical Crystal Structure Prediction of Zeolitic Imidazolate Frameworks Using DFT and Machine-Learned Interatomic Potentials

Yizhi Xu,[a,b] Jordan Dorrell,[c] Katarina Lisac,[b] Ivana Brekalo,[b] James P. Darby,[d] Andrew J. Morris[e]* and Mihails Arhangelskis [a]*

[a]Faculty of Chemistry, University of Warsaw; 1 Pasteura Street, Warsaw 02-093, Poland.

[b]Division of Physical Chemistry, Ruđer Bošković Institute, Zagreb, Croatia

[c]School of Metallurgy and Materials, University of Birmingham, Edgbaston, Birmingham B15 2TT, U.K.

[d]Department of Engineering, University of Cambridge; Trumpington Street, Cambridge CB2 1PZ, UK.

[e]School of Metallurgy and Materials, University of Birmingham, Edgbaston, Birmingham B15 2TT, UK.

Corresponding authors:

*Prof. Andrew J. Morris, E-mail: a.j.morris@bham.ac.uk

*Dr. Mihails Arhangelskis, E-mail: m.arhangelskis@uw.edu.pl

## S1. Computational methods

### S1.1 Structure generation by Wyckoff Alignment of Molecules (WAM) method

Crystal structures were generated using AIRSS[1]+WAM[2], separately for 1, 2, 3, 4, 6, 8, 10, 12, 14 and 16 formula units of Zn(**Im**)$_2$ per primitive unit cell, with each formula unit containing one Zn atom and two imidazolate fragments. In each search, space groups with. up to 8 symmetry operations were uniformly sampled. Zn atoms and imidazolate linker fragments were placed at random positions within the trial cell, subject to space group symmetry constraints. In addition, minimal separation constraints were set to prevent overlap of different atomic fragments in the trial configurations. The total number of structures generated for each number of formula units is shown in Table S1.

Table S1. The number of structures generated for each number of Zn(**Im**)$_2$ formula units.

| Number of formula units per primitive cell | Number of generated structures |
|---|---|
| 1 | 500 |
| 2 | 1000 |
| 3 | 1500 |
| 4 | 3000 |
| 6 | 72047 |
| 8 | 295714 |
| 10 | 694635 |
| 12 | 743386 |
| 14 | 703491 |
| 16 | 1222749 |

### S1.2 Periodic DFT calculations

All putative randomly generated $Zn(\mathbf{Im})_2$ structures by WAM, containing 1 to 4 formula units per primitive cell were geometry optimized with the plane-wave periodic density-functional theory (DFT) code CASETP19.[3] The calculations were performed using LDA functional and the plane-wave cut-off was set to 400 eV. The ultrasoft pseudopotentials from the CASTEP internal library were used, while the first Brillion zone was sampled with a $2\pi$ x 0.07 $Å^{-1}$ Monkhorst Pack k-point grid.[4] Structures were optimized with respect to both lattice parameters and atomic positions, while enforcing the symmetry constraints defined by the WAM-assigned space group. The convergence criteria for the geometry optimizations were set to be maximum energy change of $2x10^{-5}$ eV $atom^{-1}$, maximum force on atom of 0.05 eV $Å^{-1}$, maximum atom displacement of 0.001 Å and residual stress of 0.1 GPa.

**S1.3 Training of machine-learned potentials**

Herein, the deep neural network-based software SchNetPack,[5] specifically with the polarizable interaction neural network (PaiNN) architecture[6] was employed to train the machine-learned interatomic potentials (MLIP). For the training, all 6000 aforementioned periodic DFT-optimized $Zn(\mathbf{Im})_2$ structures, containing 1 to 4 formula units per primitive cell, were utilized. The preparation of training data was performed via our internal code, ML-Tools. Specifically, geometry optimization snapshots, including unit cell parameters, atomic positions and forces, were extracted from individual .castep geometry optimization files, converted into an ASE[7] atoms object, and stored in a database format compatible with SchNetPack. The starting and final geometry configurations were always included, while the intermittent steps were sampled according to a two-part process. Initially, we discarded most steps, retaining only every $20^{th}$ step. Next, we randomly sampled the remaining steps, where the probability of a step being stored was weighted by the fractional distance through a geometry optimization walk according to Gaussian distribution, drawing more structures from the initial geometry steps, where the differences between successive energy steps are larger, ensuring greatest data diversity. We also ensured that no single geometry optimization trajectory is shared between training, validation and sets. In total, 19311 structural snapshots were extracted from the CASTEP geometry optimization output files, with the training, validation and test set split being 10972, 2739 and 5600. Separate MLIPs were constructed for energies and forces, the reasons for training separate potentials rather than a single one are discussed in section S2.

**S1.4 Geometry optimization and energy ranking using machine learned potentials**

The forces and energy MLIPs were used to perform geometry optimization and energy ranking of ZIF structures containing 6 to 16 formula units per primitive cell, which would be too computationally expensive to study by periodic DFT. All putative hundreds and thousands ZIF structures were generated by the aforementioned AIRSS+WAM method. Subsequently, each structure was optimized with the MLIP trained on atomic forces via ASE's geometry optimization function within the FrechetCellFilter class. Symmetry was constrained to a tolerance of 0.01 Å. The force tolerance for the initial geometry optimization was set to 0.05 eV $Å^{-1}$. Then, the energy

MLIP was applied for single point calculations for each optimized structure. All optimized structures were ranked by ascending energies, where the ones fall into an energy window of 45 kJ $mol^{-1}$ with respect to the global minimum structure were selected for further analysis. Furthermore, ZIF structures containing 1 to 4 formula units per primitive cell, which were used to train the two MLIPs, were also processed from the same procedure. In a CSP search, the presence of duplicate structures provides a major indication for the convergence of dataset, where sufficient amount of structures has been searched.

The software of Critic2[8,9] with the built-in function "Compare reduce 3e-2", in order to compare thousands of ML-optimized structures. The "reduce" option allow the algorithm to omit structures already shown to be equivalent to the others in the list. The tolerance for the comparison method to identify two duplicate structures was set to 3e-2. Consequently, a full set of ML-optimized unique structures containing 1 to 16 formula units per primitive were obtained. Followed by clustering, the final set of structures were re-optimized with a tighter force tolerance of 0.005 eV $Å^{-1}$, and the single-point energy calculations were subsequently performed for each structure.

Since the cutoff distance for pairwise interactions was set to 5 Å, it was shortly realized that some 1D and 2D structures inherited channels and layers separation distances significantly larger (i.e. 10 Å) to be chemically plausible. Therefore, 646 1D and 2D ML-optimized structures that exhibit minimal pore radius of 5 Å or above, were selected for additional calculations with external stress. These structures were optimized first with the force MLIP, under the stress of 1 GPa and force tolerance of 0.005 eV $Å^{-1}$. This resulted in the reduction of interlayer/interchain spacing, within the MLIP interaction cutoff. The structures were then reoptimized under zero pressure, to make them consistent with the remaining structures that did not require special treatment. Finally, the energy MLIP was used to obtain the single-point energy of each structure. Clustering using the simulated PXRD comparison, implemented in Critic2, was performed to remove any newly formed duplicate structures.

### S1.5 Post-processing of the predicted crystal structures

The final set of 9626 ML-optimized ZIF structures were obtained and ready for further analysis. The software PLATON[10] was used to convert all structures into the conventional crystallographic setting visa the command ADDSYM EXACT SHELX. The structural void volume and packing coefficient of each structure were extracted from the command CALC VOID in PLATON.

All pore-related properties such as total surface area per mass, network accessible surface area per mass, total helium volume, pore limiting diameter, max pore diameter and number of percolated dimensions were evaluated through the Pore Analyzer function from the CCDC packages. The CSD Python API[11] was employed for high throughput analysis for all CSP predicted structures.

The software ToposPro[12] was used to determine the network topologies and dimension for all the structures in the final CSP set, via the default settings of the AutoCN module. Subsequently, the resulting nets were simplified and analyzed with the default settings of the ADS module. Finally, the obtained topological descriptors were searched through the built-in TTD database, as well as http://topcryst.com/ online server.

CSD Editor software was used to construct a database out of the predicted structures, that was then analyzed in ConQuest[13] to determine bond length and bond angle distribution statistics.

**S1.6 Protocol for comparing experimental PXRD patterns with predicted CSP structures via Critic2**

As a prerequisite for the comparison protocol, the experimental PXRD patterns were converted to .xy format, and all CSP predicted structures were supplied in .res file format. The open source software PowDLL was used to convert experimental PXRD patterns from .raw, .xrdml or .brml files to .xy format. Subsequent analysis involved several steps in Critic2:

1) The background of the experimental PXRD pattern was calculated via the command "XRPD BACKGROUND experimental_pattern.xy background.xy".
2) The command "XRPD FIT background.xy" was used to obtain a list of background subtracted reflections and intensities.
3) The command COMPAREVC was used to compare the experimental PXRD pattern with each CSP predicted structure.

Overall, 9626 individual comparisons between all ML-optimized structures and the experimental PXRD pattern were performed, on the PLGrid high performance computer (HPC) HPC Helios. Finally, all CSP predicted structures were ranked by ascending DIFF scores, of which the structures with lower DIFF scores will be investigated further for possible experimental matching structures.

**S1.7 Periodic DFT optimization of the structures deemed likely experimental candidates**

The 982 predicted structures deemed to be synthesizable based on the energy-porosity criterion (Section S7) were geometry-optimized with periodic DFT. These calculations used PBE functional, combined with Grimme D3 dispersion correction.[14] The plane-wave basis set truncated at 800 eV cutoff, and CASTEP default ultrasoft pseudopotentials were used. The electronic k-point grid was sampled with a $2\pi \times 0.06$ Å$^{-1}$ Monkhorst-Pack k-point grid.[4] Convergence criteria were set as follows: maximum energy change $2 \times 10^{-5}$ eV atom$^{-1}$; maximum atomic force 0.05 eV Å$^{-1}$; maximum atom displacement $10^{-3}$ Å; maximum residual stress 0.05 GPa.

## S2. The use of Separate MLIPs for Energies and Forces

In this work two distinct MLIP's were trained to reproduce the DFT energies and forces respectively, rather than training a single model to reproduce both, as is common practice. This decision was made as in initial testing models targeting both energies and forces achieved very poor performance with approximately 10 times larger MAEs compared to separate models. After spending significant effort attempting to resolve this issue we proceeded using separate models. This decision was made pragmatically, as the MLIPs were used purely as a tool to accelerate the CSP, and the models trained separately to energies and forces where accurate enough to usefully rank the structures - MAEs of ~12 meV/atom and 66 meV $Å^{-1}$. For clarity, the MLIP trained on forces was used for structural relaxations whilst the MLIP trained on energies was then used for final energy rankings.

Subsequently this issue was revisited and, after updating the supplied neighbor list, a single MLIP was successfully fit to energies and forces simultaneously. As such, we hypothesize that the previous advantage observed for separate models was due to the energy and force training data appearing inconsistent and stress that in general there is no need for separate models.

In addition to the above, as fitting to energies and forces with separate models is unusual, we also fit a MACE model[15] to both energies and forces to check that substantially better accuracy could not be achieved with a commonly used modern architecture and standard setup. The resulting parity plots for energy and force predictions on the training, validation and testing sets are shown in Figure S1**.** The achieved accuracy is very comparable to the separate MLIPs fit in this work. Furthermore, given the recent success of foundation models, we fit a second MACE model by fine-tuning the MACE-MPA-0 foundation model[16] using the recommended multi-head-replay strategy to prevent catastrophic forgetting. The fine-tuned model achieved extremely similar validation MAE's of 9 meV/atom and 87 meV $Å^{-1}$ (compared to 7 meV/atom and 84 meV $Å^{-1}$ with vanilla fitting), indicating that fine-tuning offers little benefit here. We suggest this is likely due to the relatively large amount of training data (10972 configurations containing 444,907 atoms), for a single composition.

In summary, neither MACE model offered a substantial improvement in accuracy compared to the separate MLIP's fit to energies and forces.

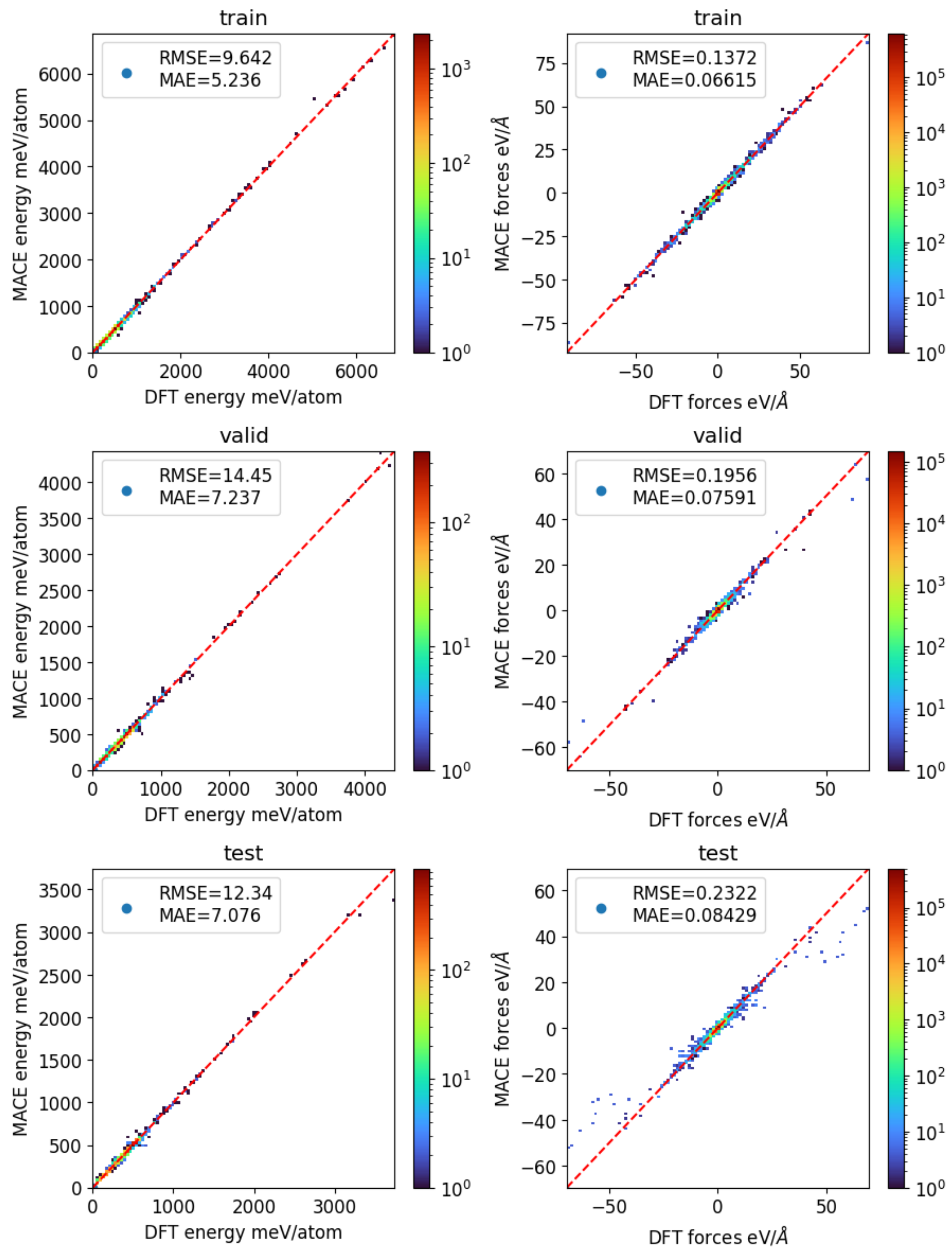

Figure S1. Parity plots of energies and forces comparing MACE predictions with reference data for the training, validation, and test sets. Errors are quantified using RMSE and MAE (shown inset). The energies were referenced to the lowest-energy structure, which was set to zero.

## S3. Mechanochemical synthesis

The experimental preparation of all materials used for the assignment of the PXRD data against the predicted structures was described in an earlier publication.[17] Ball milling reactions were conducted in a 14 mL jar with one 7 mm (1.4 g) and one 9 mm (3.5 g) diameter stainless steel ball bearing. In each liquid-assisted grinding (LAG) experiment, 100 μL of toluene, chloroform or dimethylformamide, depending on experiment (see Table S5), was added into a milling jar containing the ball bearings, zinc oxide (75.0 mg, 0.92 mmol) and imidazole (125.5 mg, 1.84 mmol). The samples were milled at 30 Hz for 90 minutes using a Retsch MM400 ball mill.

Guest-free porous material of unknown structure was prepared by the following procedure. Zinc oxide (75.0 mg, 0.922 mmol), imidazole (125.5 mg, 1.843 mmol) and 100 μL of acetophenone were placed in 14 mL Teflon jar along with one 7 mm (1.4 g) and one 9 mm (3.5 g) diameter stainless steel ball bearing. Reaction mixture was milled for 1 hours at frequency of 30 Hz on InSolido mixer mill. Product was washed with acetone and then dried for 3 hours at 150 °C in the vacuum oven.

## S4. Rietveld refinement

Rietveld refinement[18] was performed in Topas Academic 7.[19] Predicted crystal structure Znimid2_16_56_Pccn_YMQOEPiX (#2470 in the Chemiscope file) was refined against the PXRD pattern recorded for the material obtained via heating of **crbA**-Zn**Im**$_2$ (CSD: GAKXOX) at 150 °C (see section S3). Diffraction peak shapes were described with a pseudo-Voigt function, background was modelled with a 6$^{th}$ degree Chebyshev polynomial function. Unit cell parameters and zero shift were refined, while atom positions were kept fixed at the coordinates obtained from the theoretical structure.

## S5. Comparison of DFT and ML energies for the predicted structures

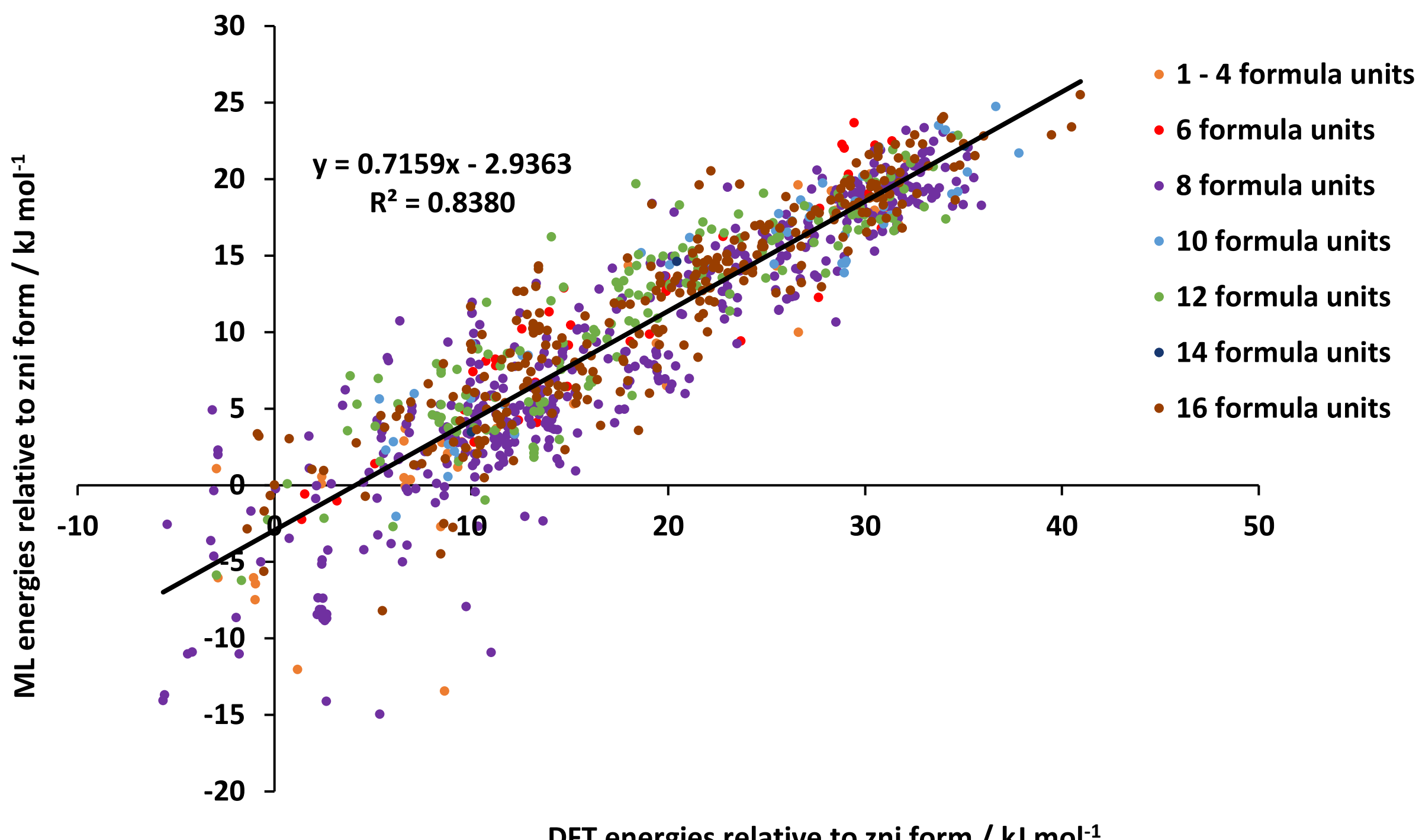

Figure S2. Plot of ML relative energies against the DFT relative energies. In both cases, the predicted structures matching with the experimental structure (CSD IMIDZB02) with **zni** topology was used as a reference. The regression line is drawn based on the energies of all structures, while different colors represent show datapoints separated by the number of formula units.

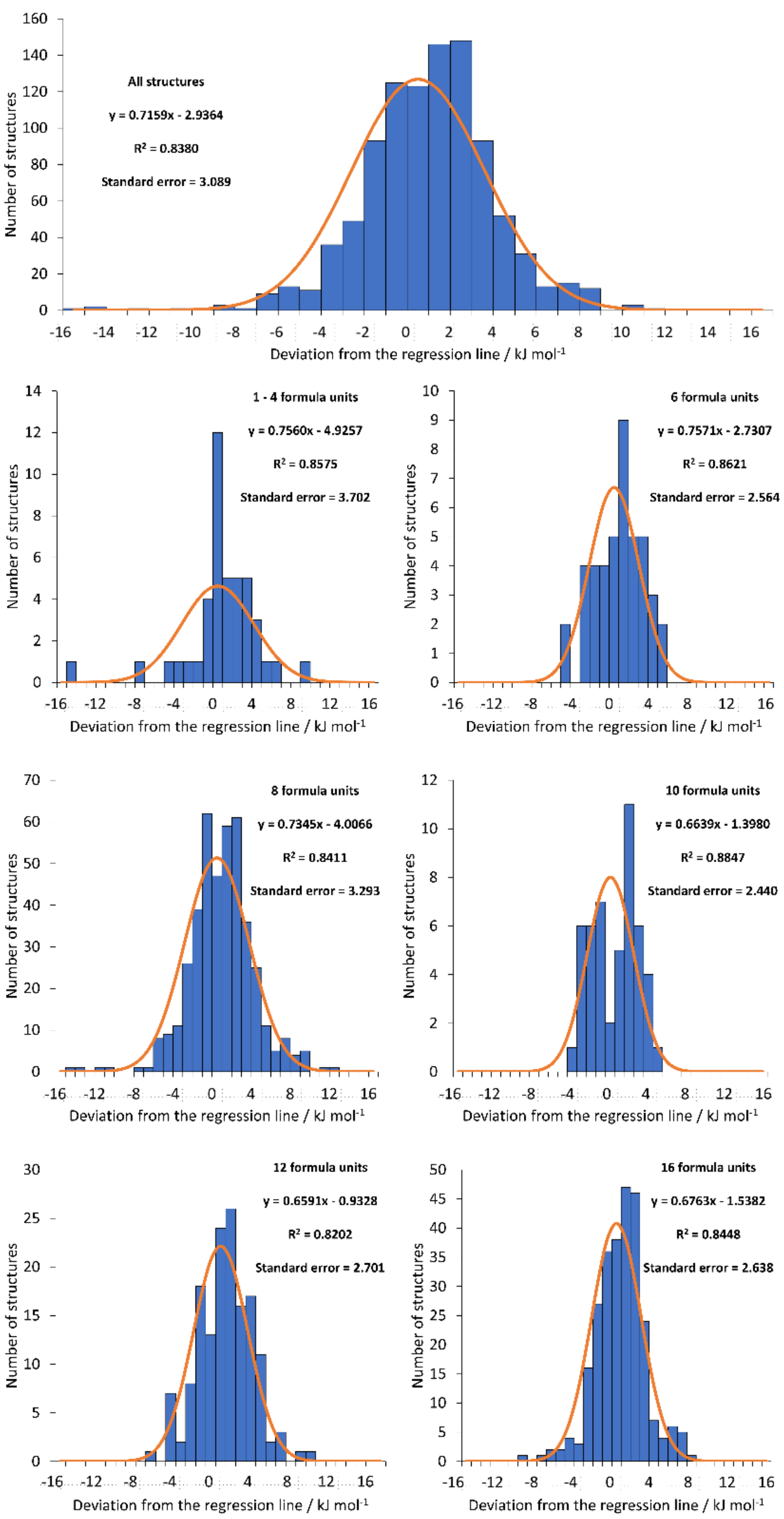

Figure S3. Deviations of relative ML energies from the regression line on Figure S1. At the top the overall distribution is shown, followed by the individual histograms grouped by the number of formula units. The orange curve in each histogram shows the normal distributions with the same standard deviation as the underlying data. It is evident that the pattern of energy deviations is similar across all groups, demonstrating that the MLIPs perform similarly for structures with all numbers of formula units per primitive cell.

## S6. Geometry analysis and distribution of framework topologies within the energy landscape

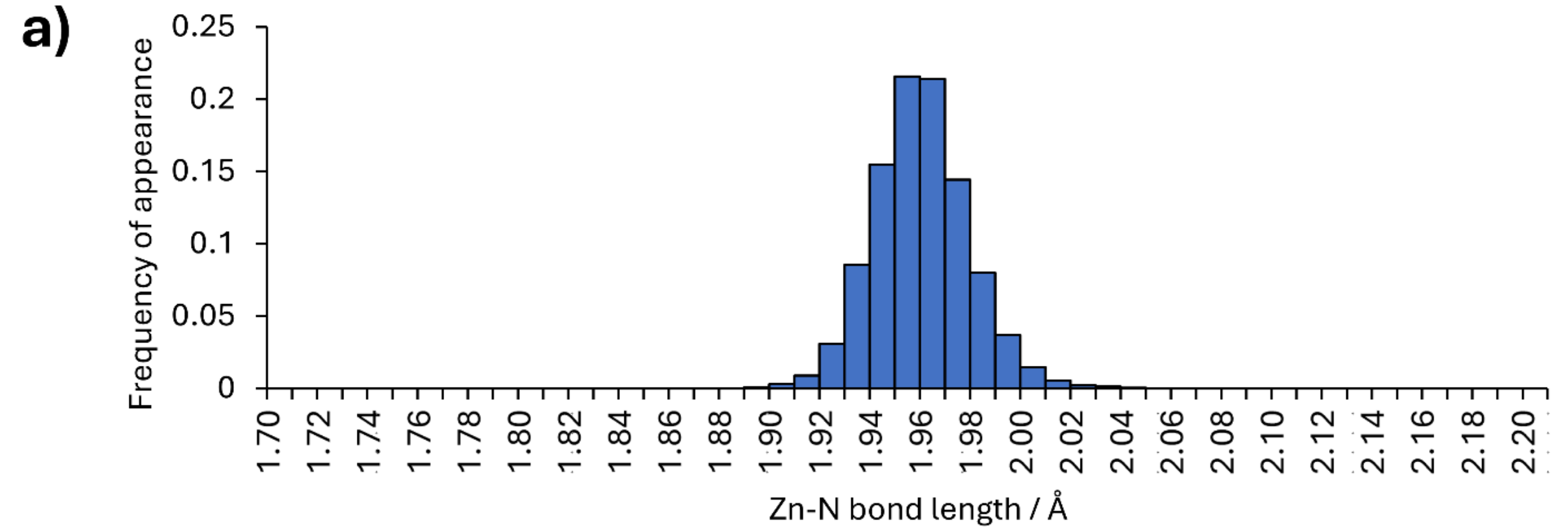

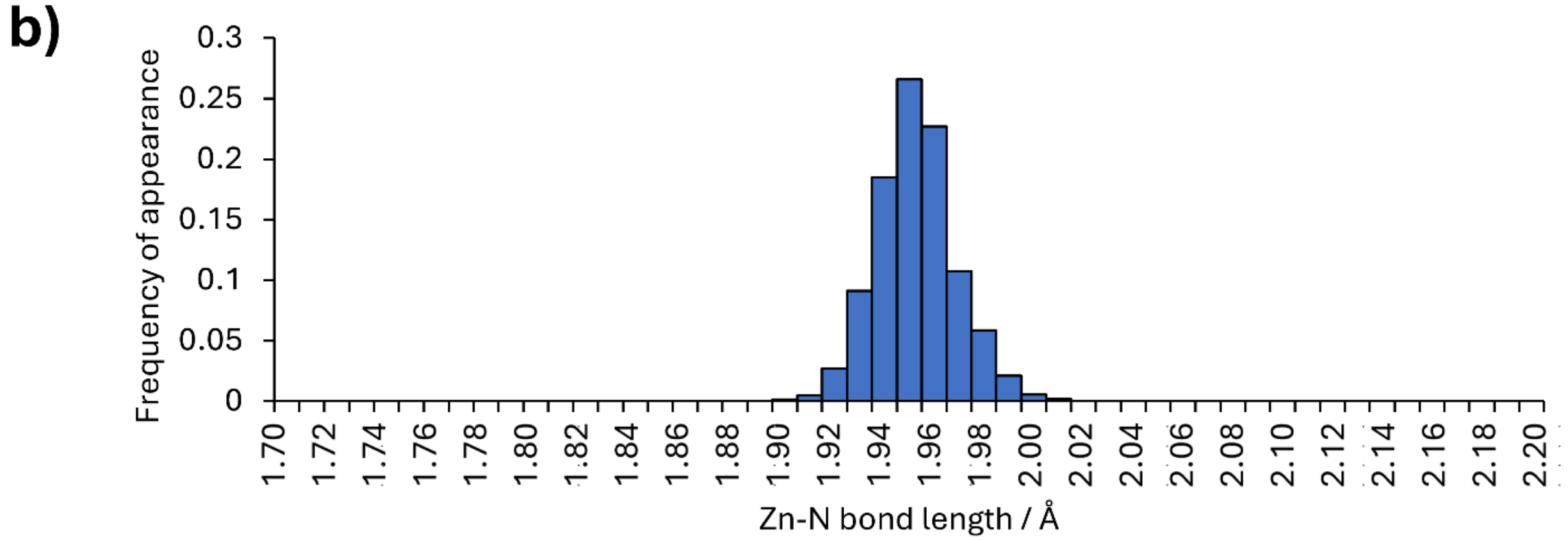

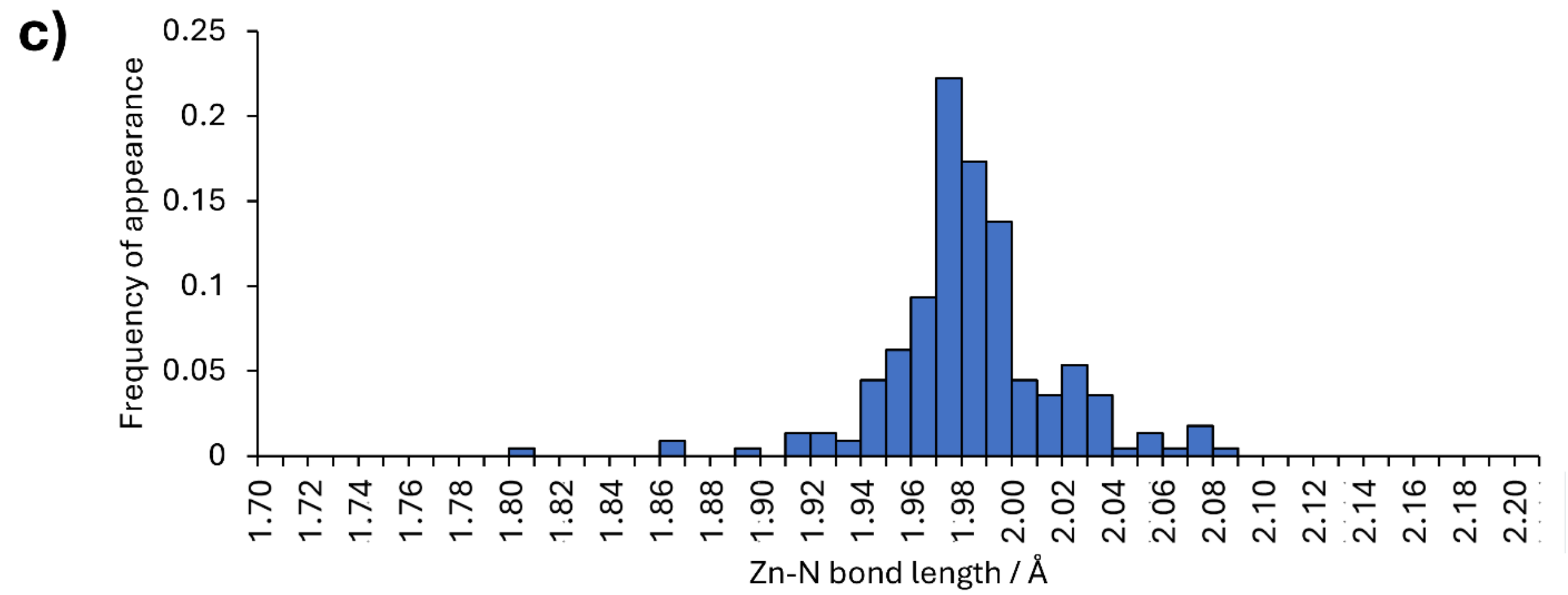

Figure S4. Comparison of the Zn-N bond length distributions: a) for all predicted structures; b) for the predicted structures from the likely-synthesizable group; c) for the experimental structures of Zn**Im**$_2$ found in CSD.

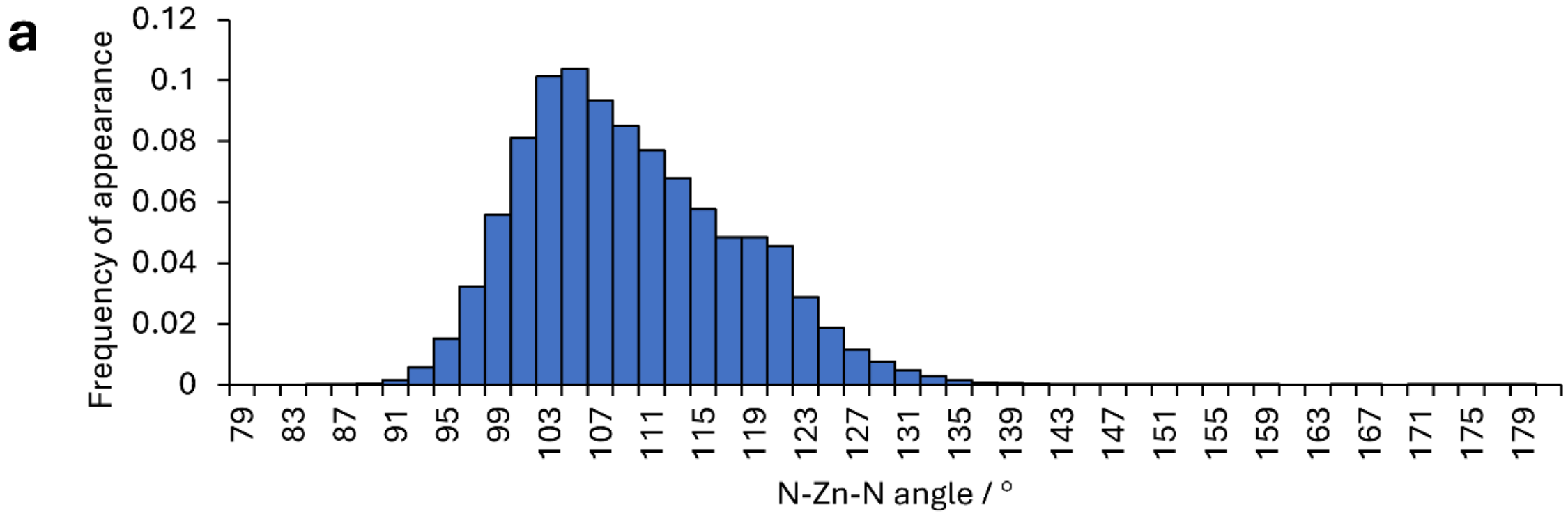

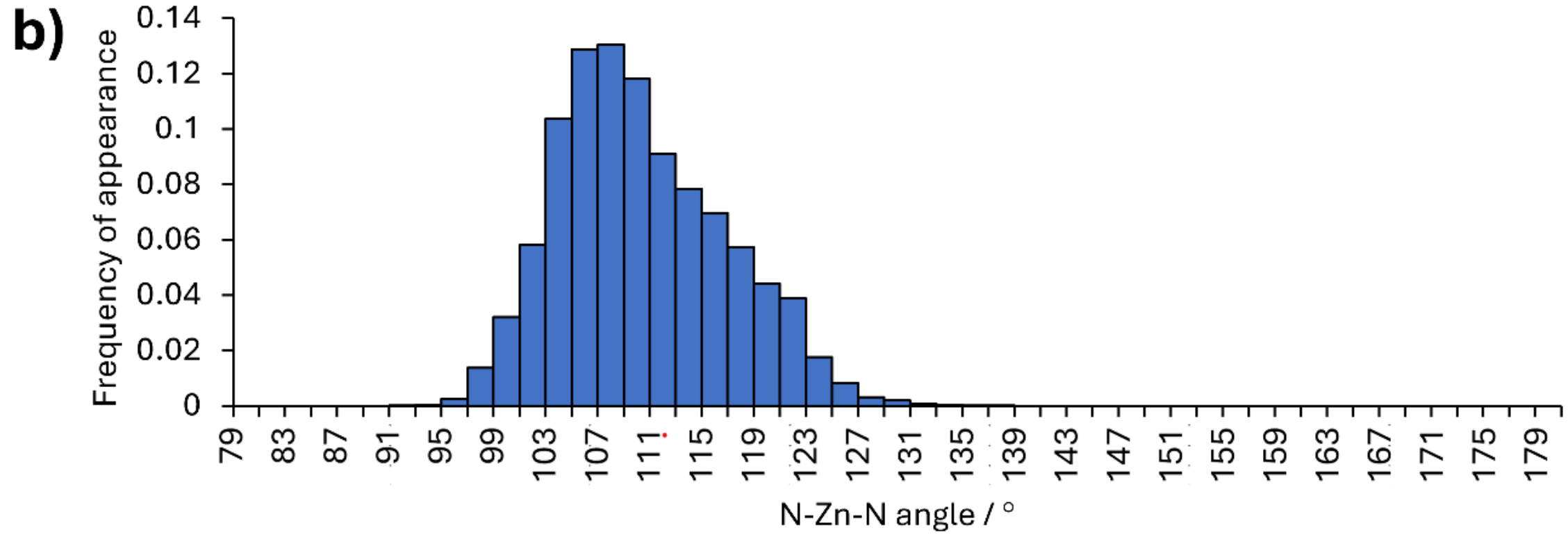

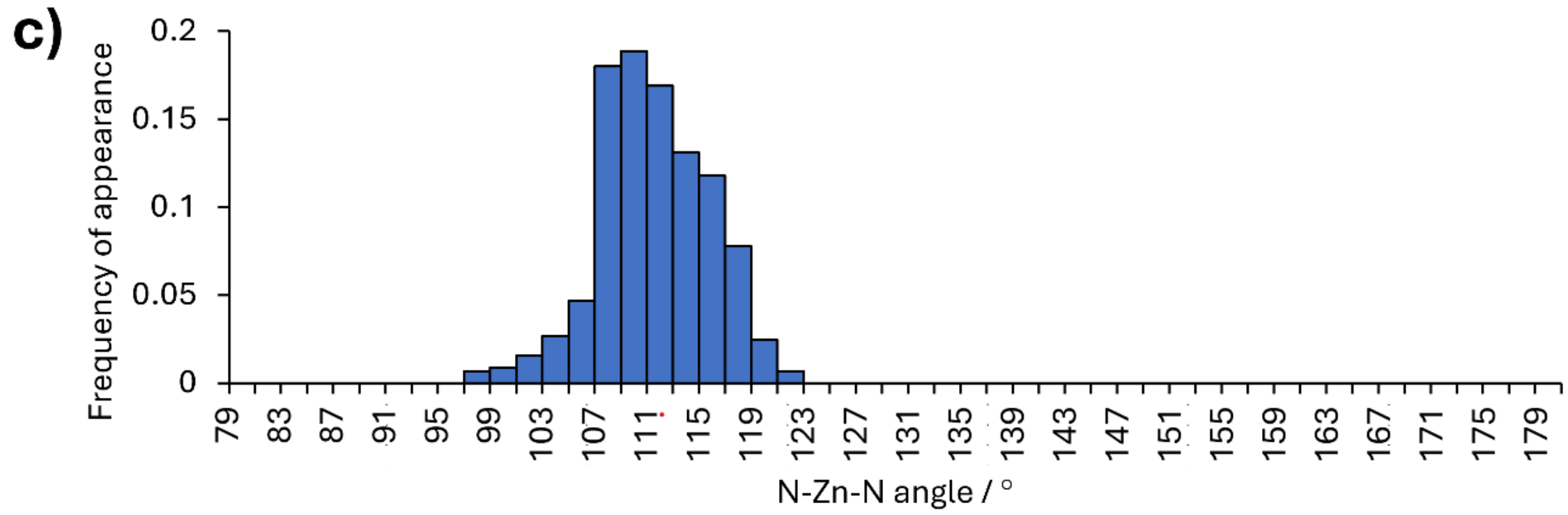

Figure S5. Comparison of the Zn-N bond length distributions: a) for all predicted structures; b) for the predicted structures from the likely-synthesizable group; c) for the experimental structures of Zn**Im**$_2$ found in CSD.

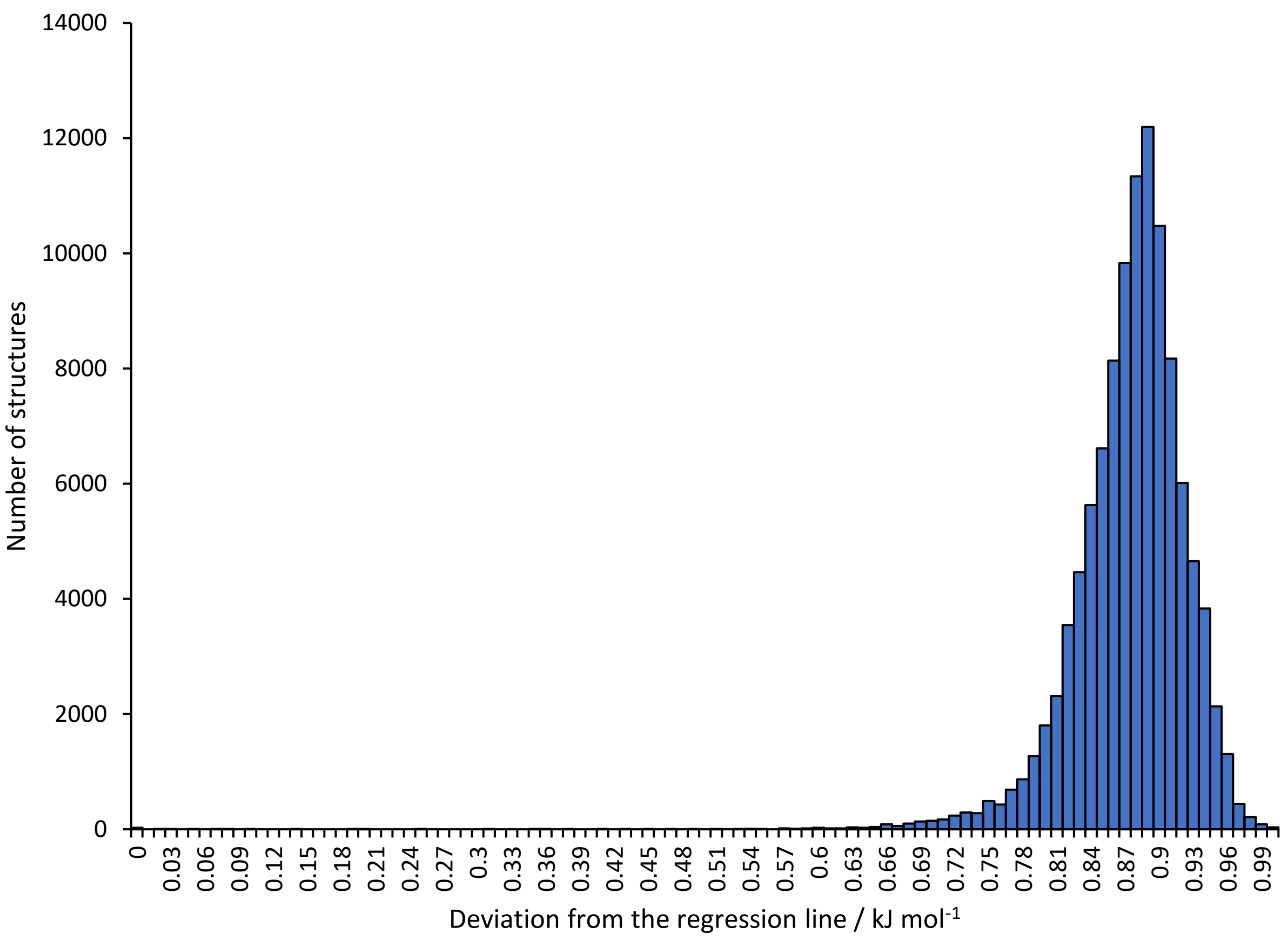

Figure S6. The τ geometry index calculated for all predicted structures. The characteristic values of τ are 0 for square planar geometry, 0.43 for seesaw geometry and 1 for tetrahedral geometry of the metal coordination center. The distribution maximum is located at 0.9, indicating that the coordination geometry in the predicted structures is predominantly tetrahedral, as would be expected for ZIF structures containing Zn nodes.

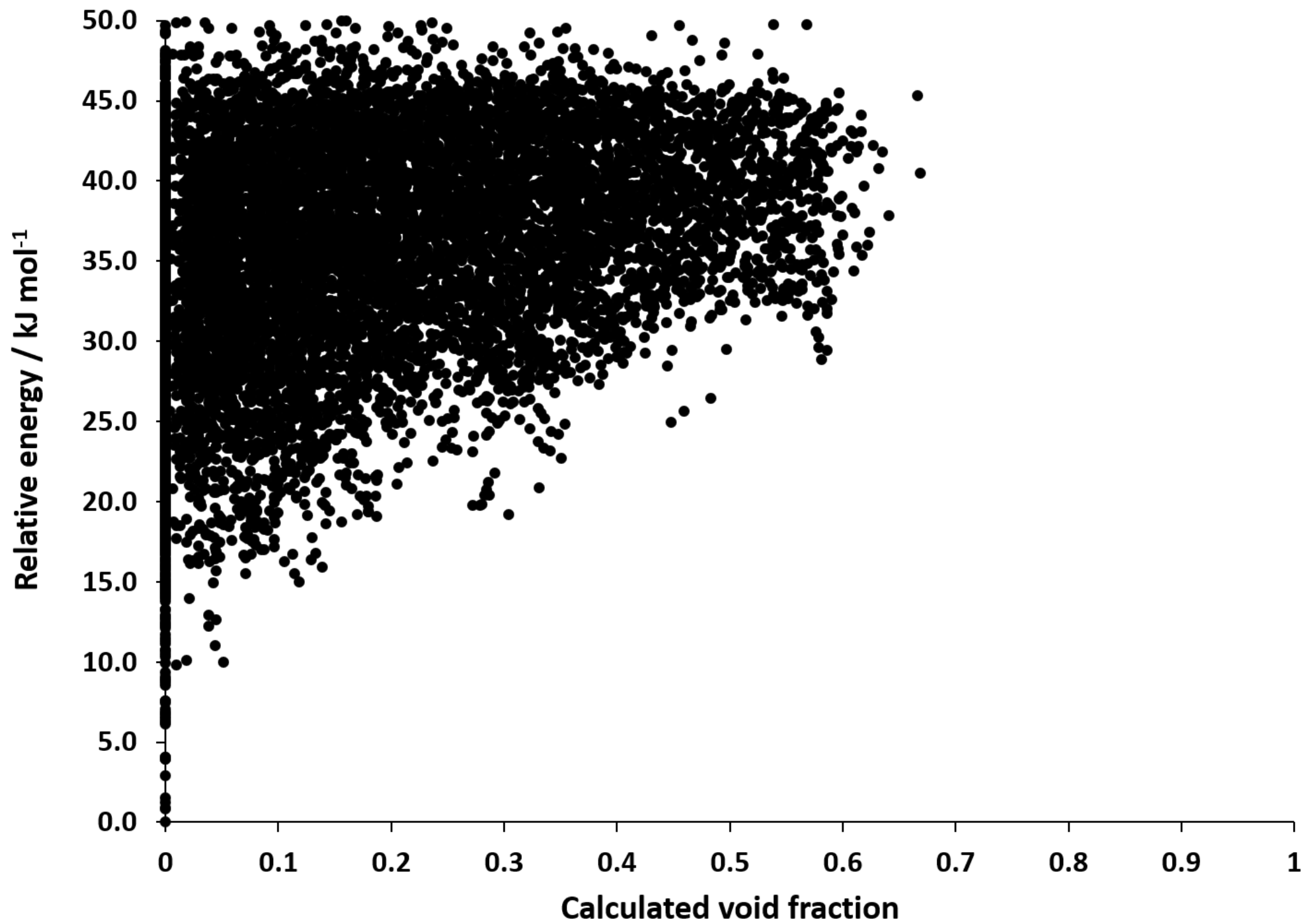

Figure S7. CSP energy landscape showing the relative energy of the predicted structures against the calculated void fraction.

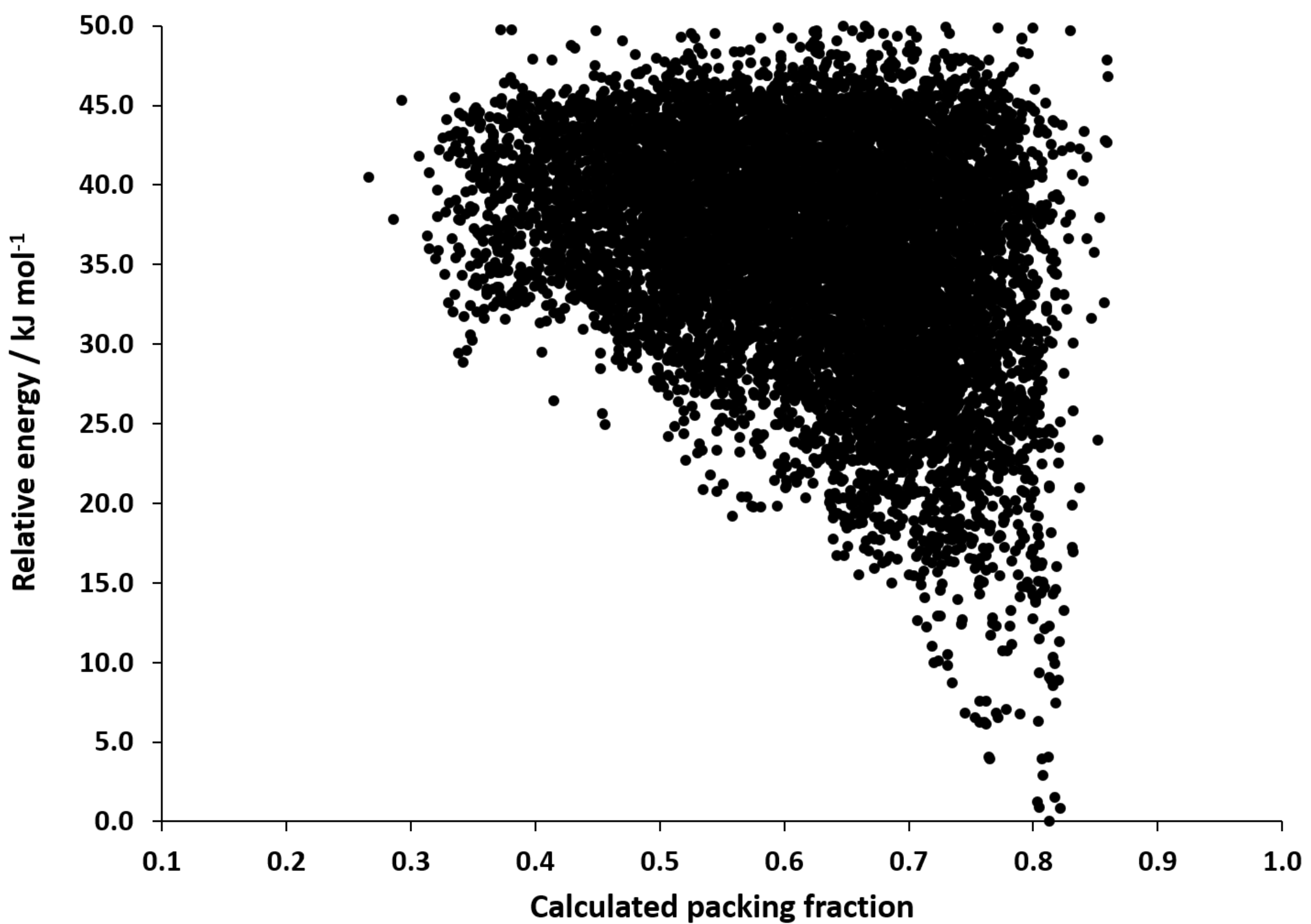

Figure S8. CSP energy landscape showing the relative energy of the predicted structures against the calculated packing fraction.

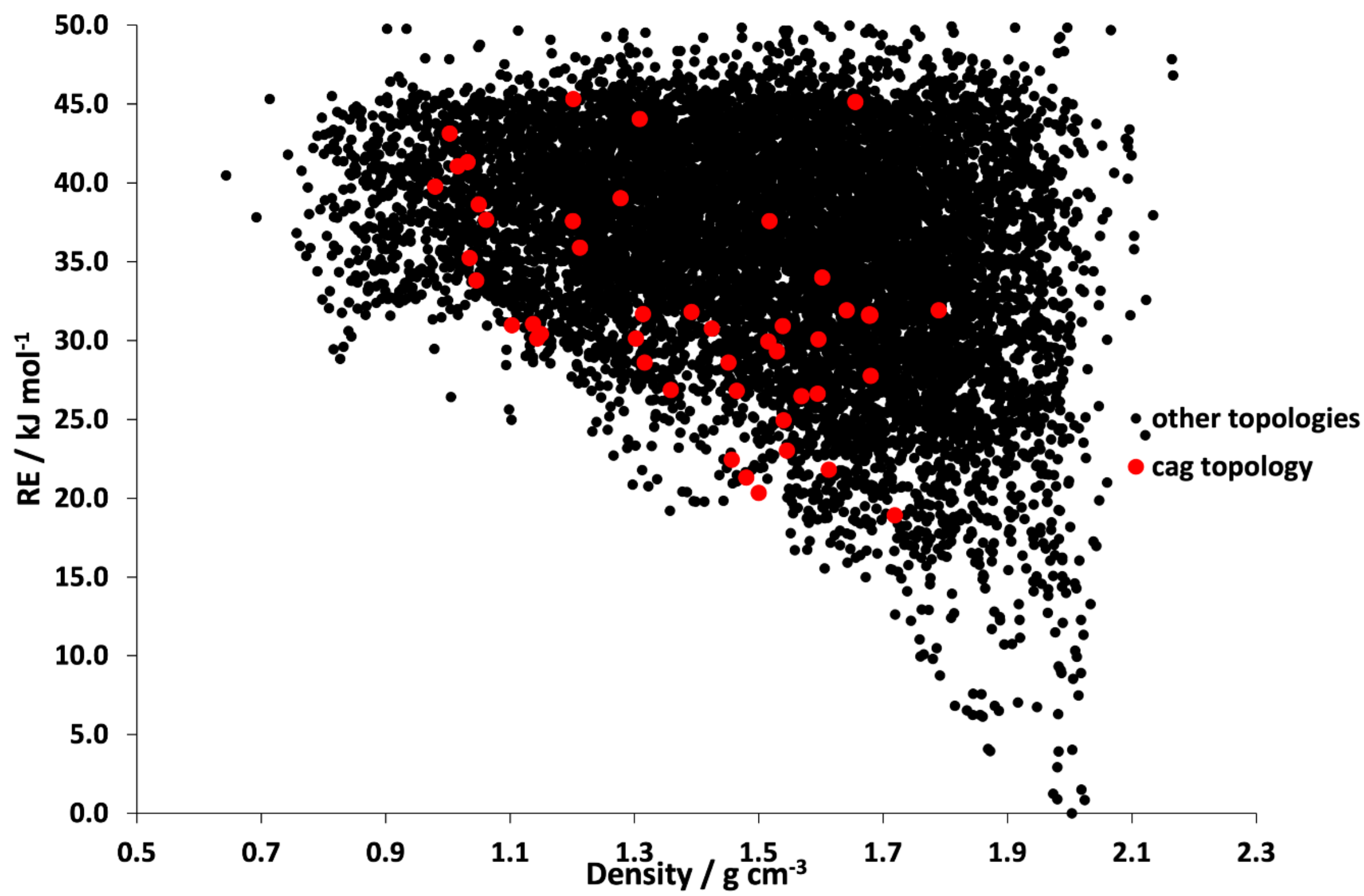

Figure S9. CSP energy landscape highlighting the structures with **cag** topology.

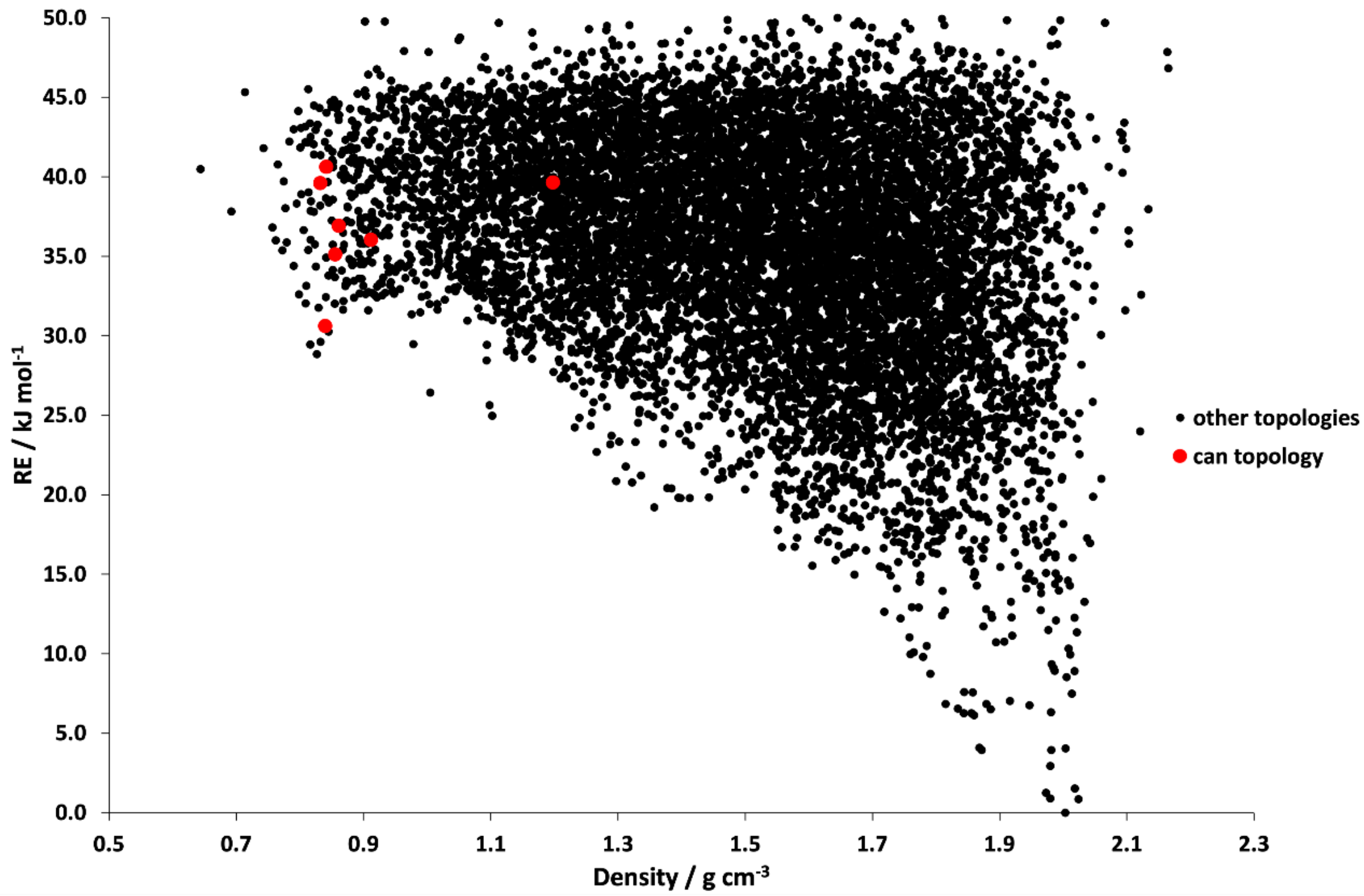

Figure S10. CSP energy landscape highlighting the structures with **can** topology.

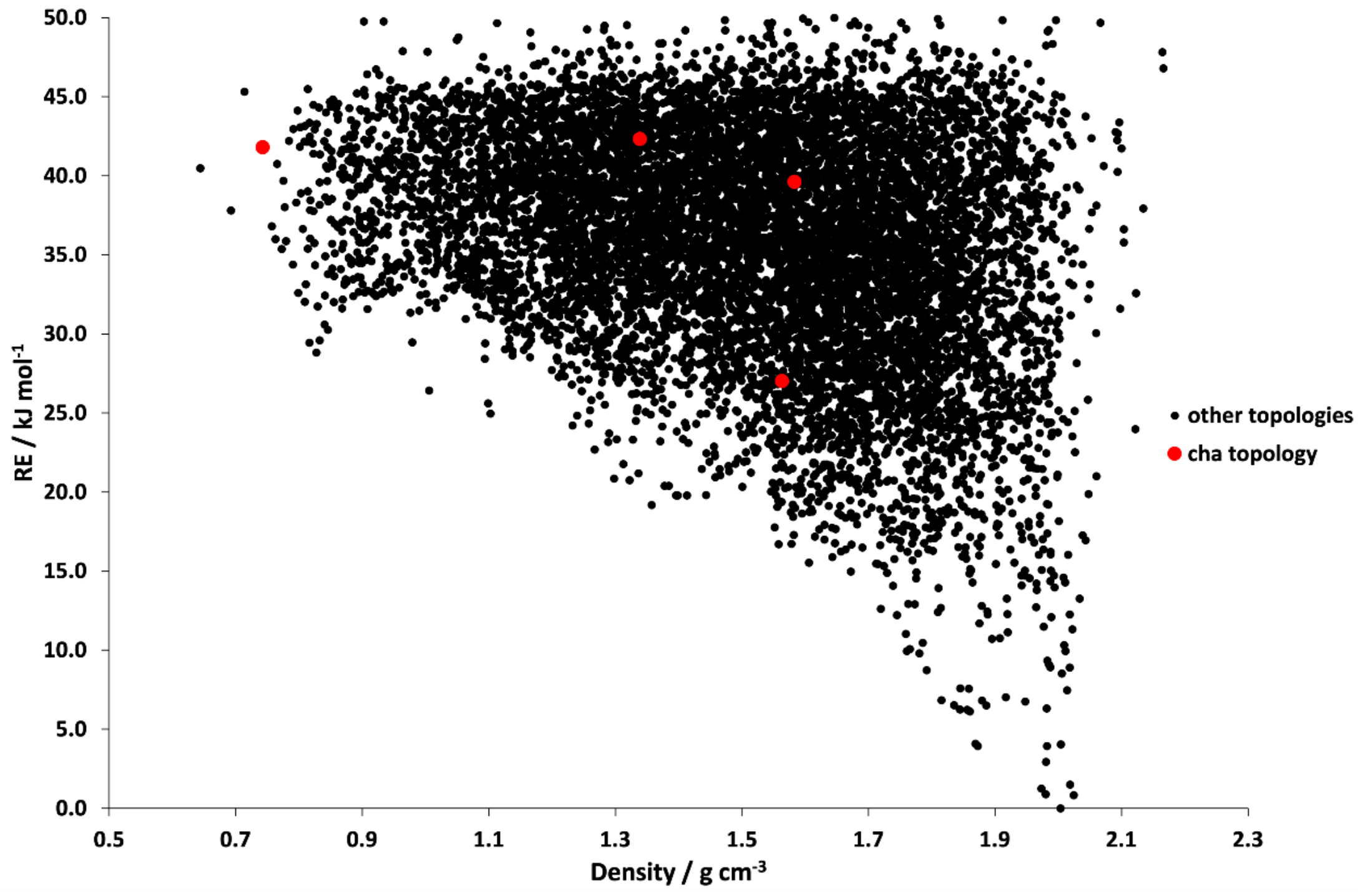

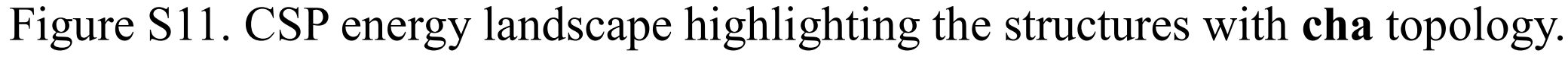

Figure S11. CSP energy landscape highlighting the structures with **cha** topology.

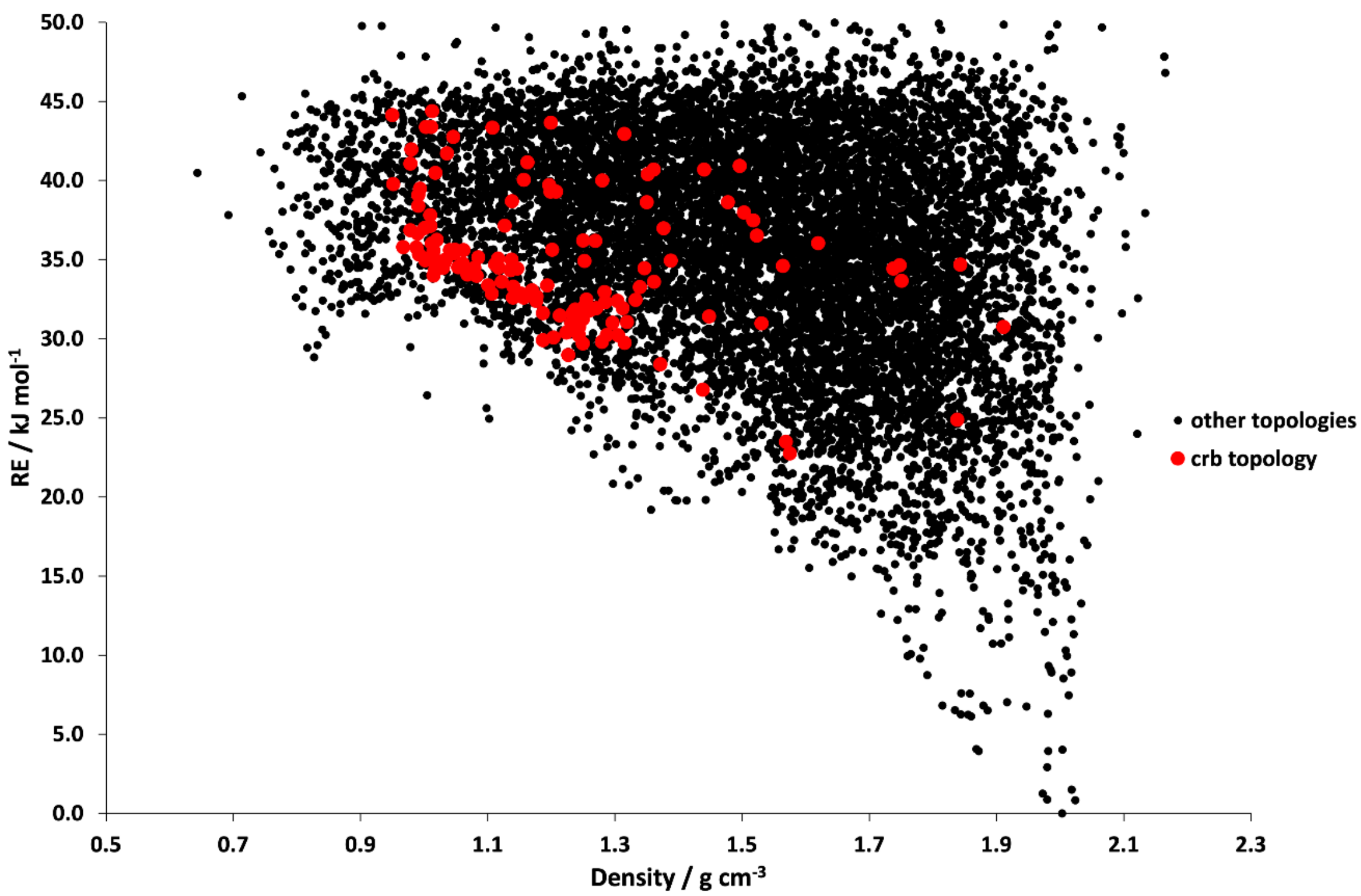

Figure S12. CSP energy landscape highlighting the structures with **crb** topology.

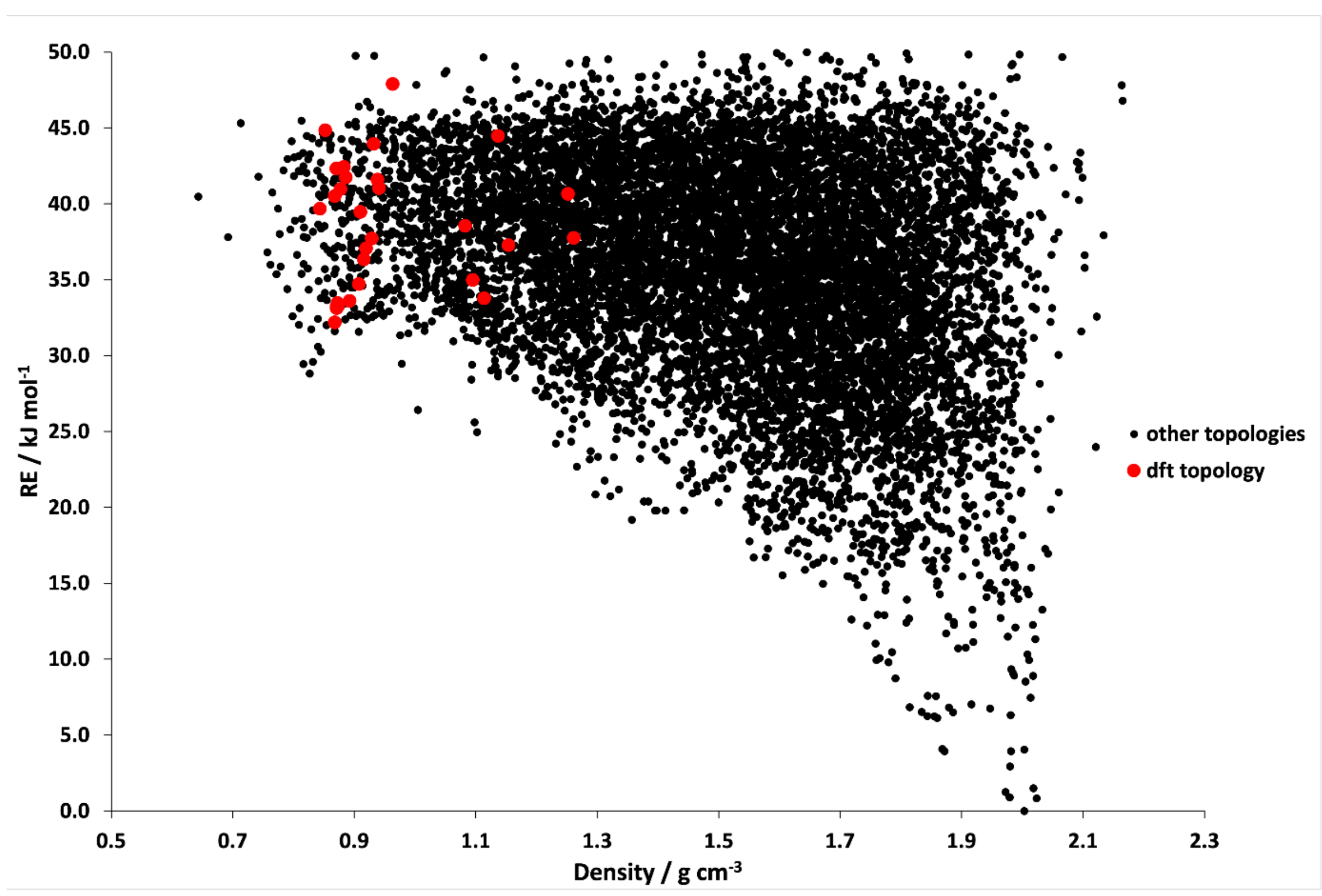

Figure S13. CSP energy landscape highlighting the structures with **dft** topology.

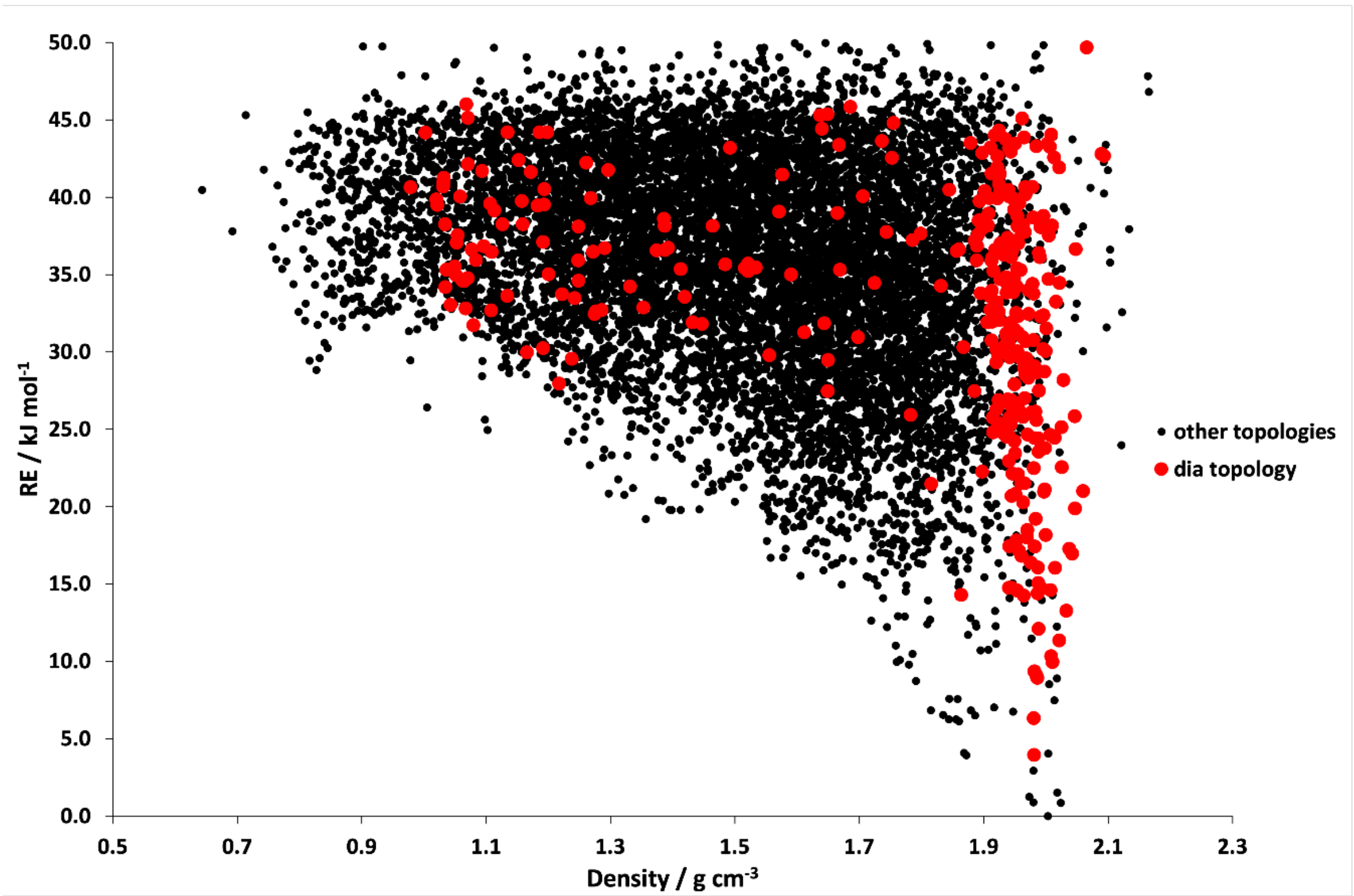

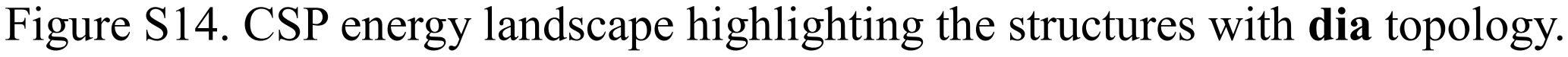

Figure S14. CSP energy landscape highlighting the structures with **dia** topology.

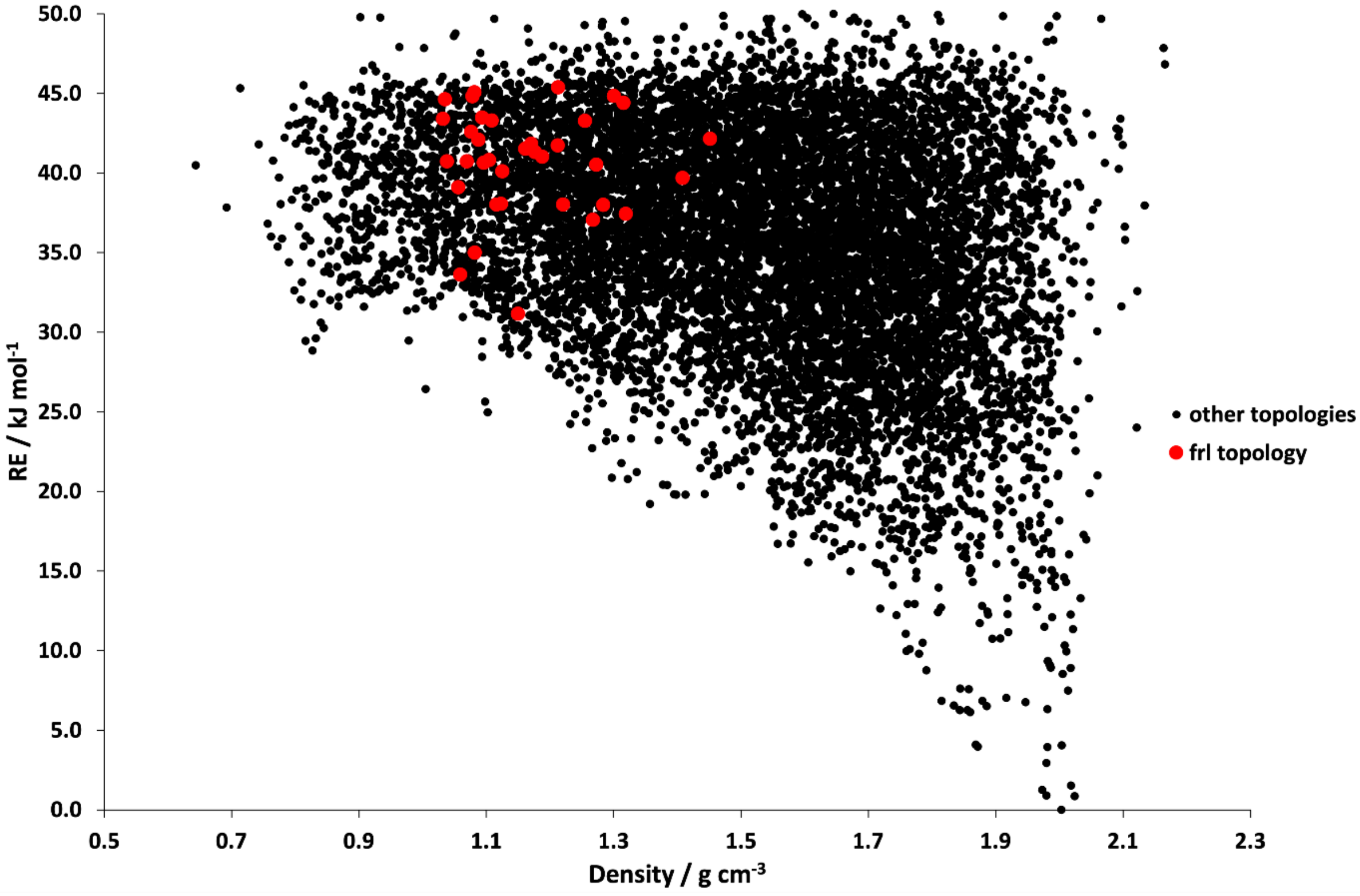

Figure S15. CSP energy landscape highlighting the structures with **frl** topology.

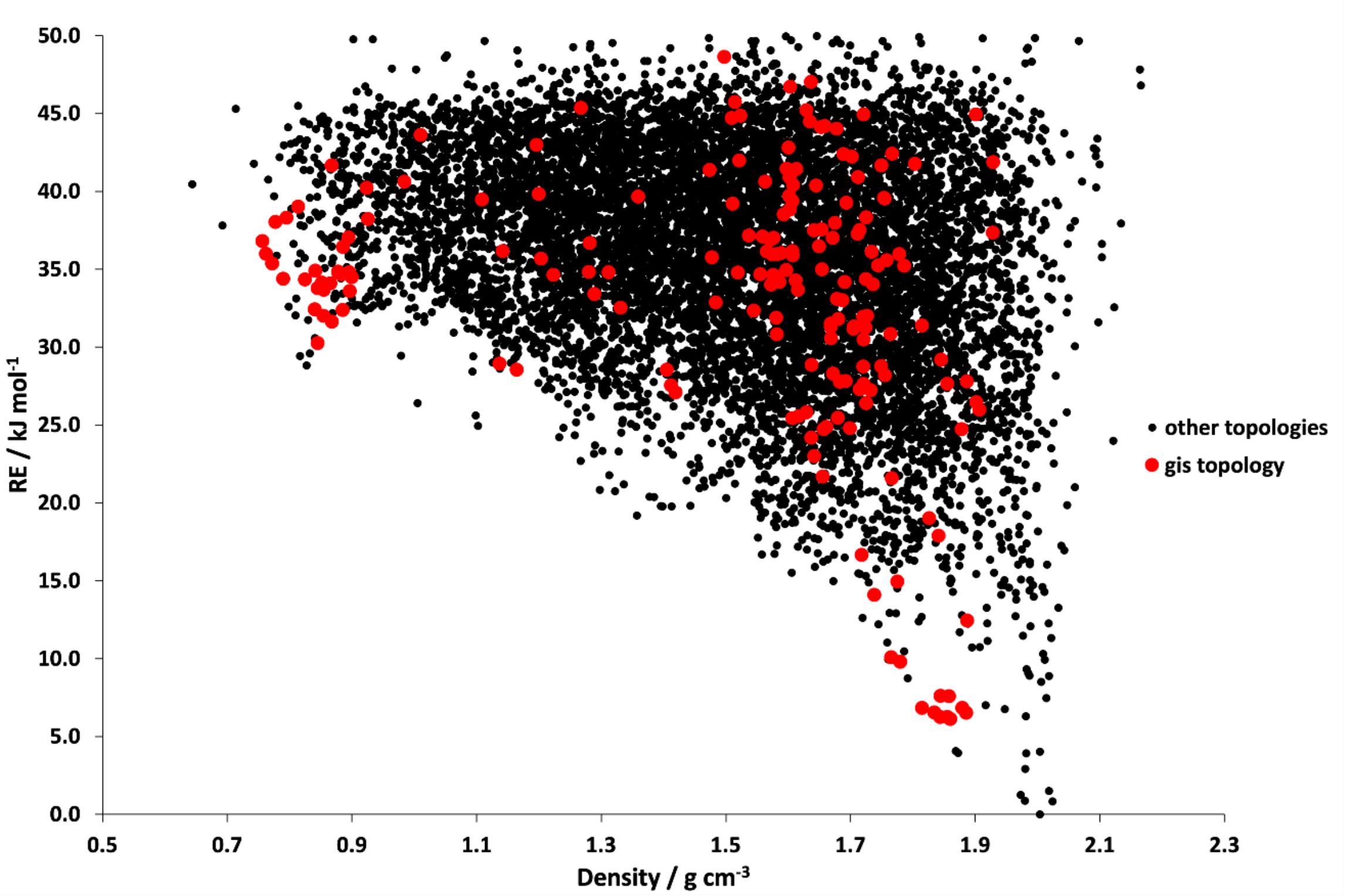

Figure S16. CSP energy landscape highlighting the structures with **gis** topology.

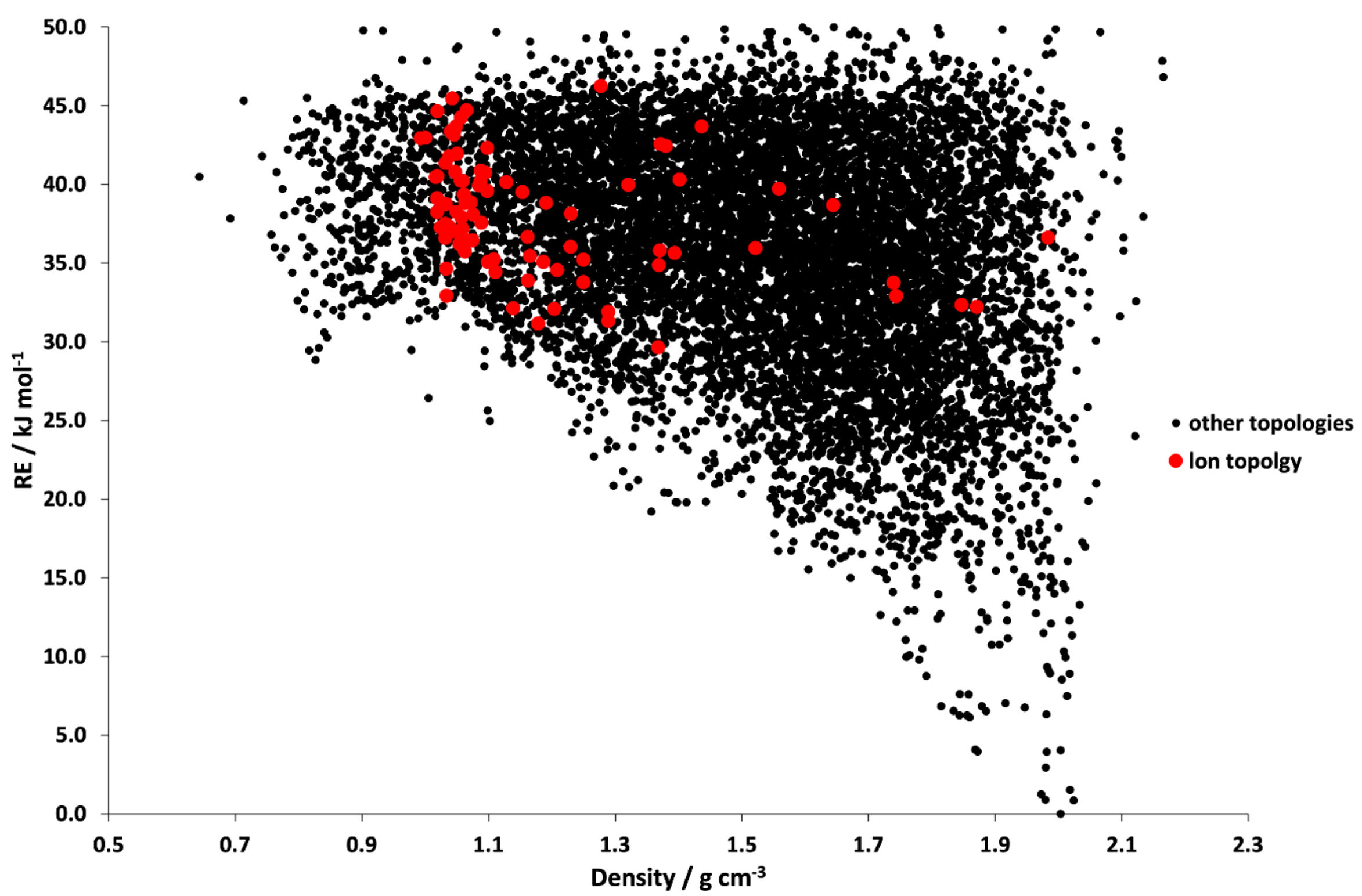

Figure S17. CSP energy landscape highlighting the structures with **lon** topology.

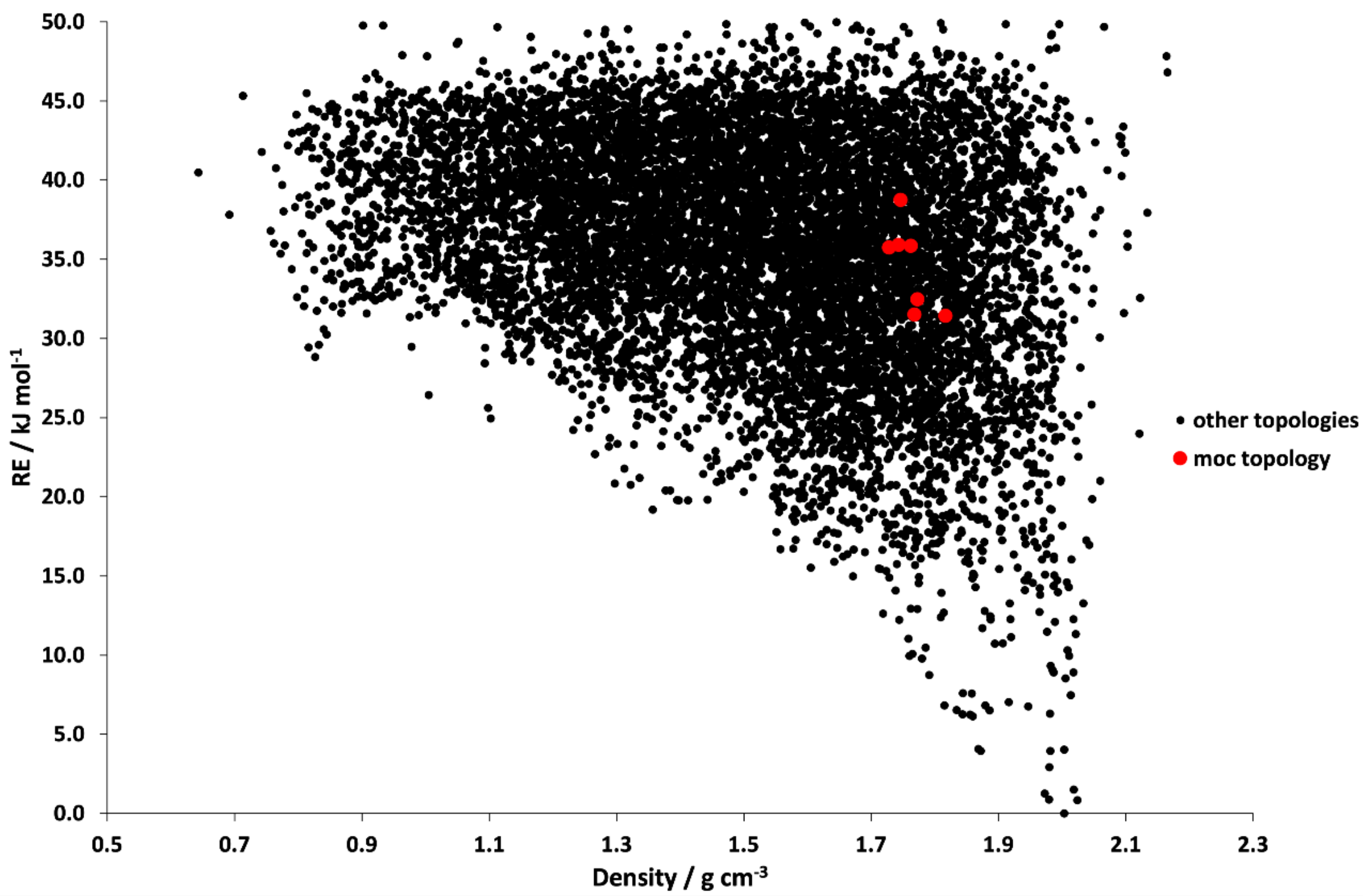

Figure S18. CSP energy landscape highlighting the structures with **moc** topology.

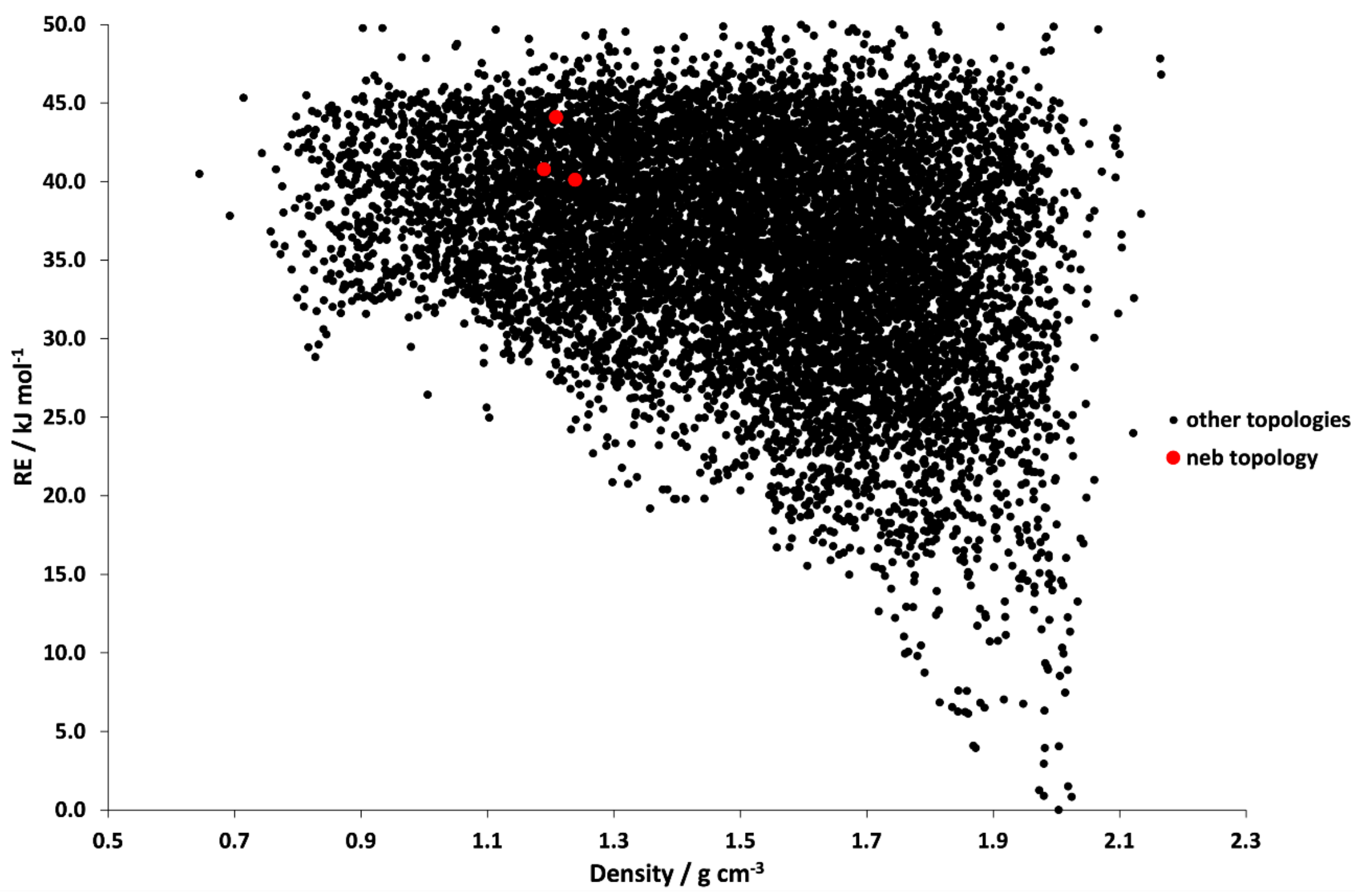

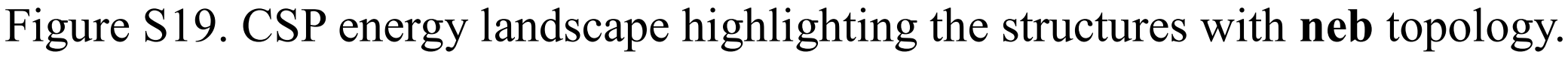

Figure S19. CSP energy landscape highlighting the structures with **neb** topology.

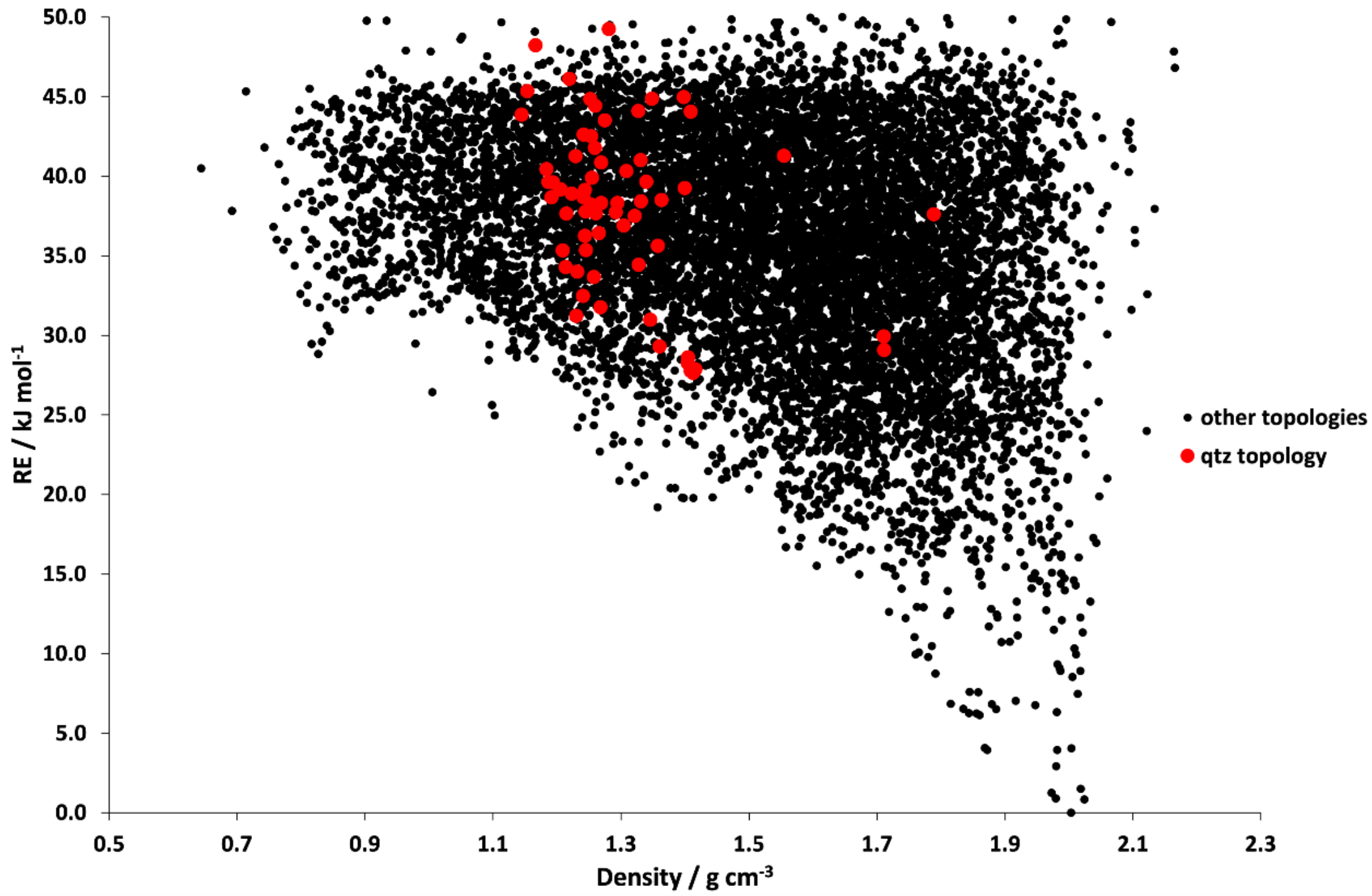

Figure S20. CSP energy landscape highlighting the structures with **qtz** topology

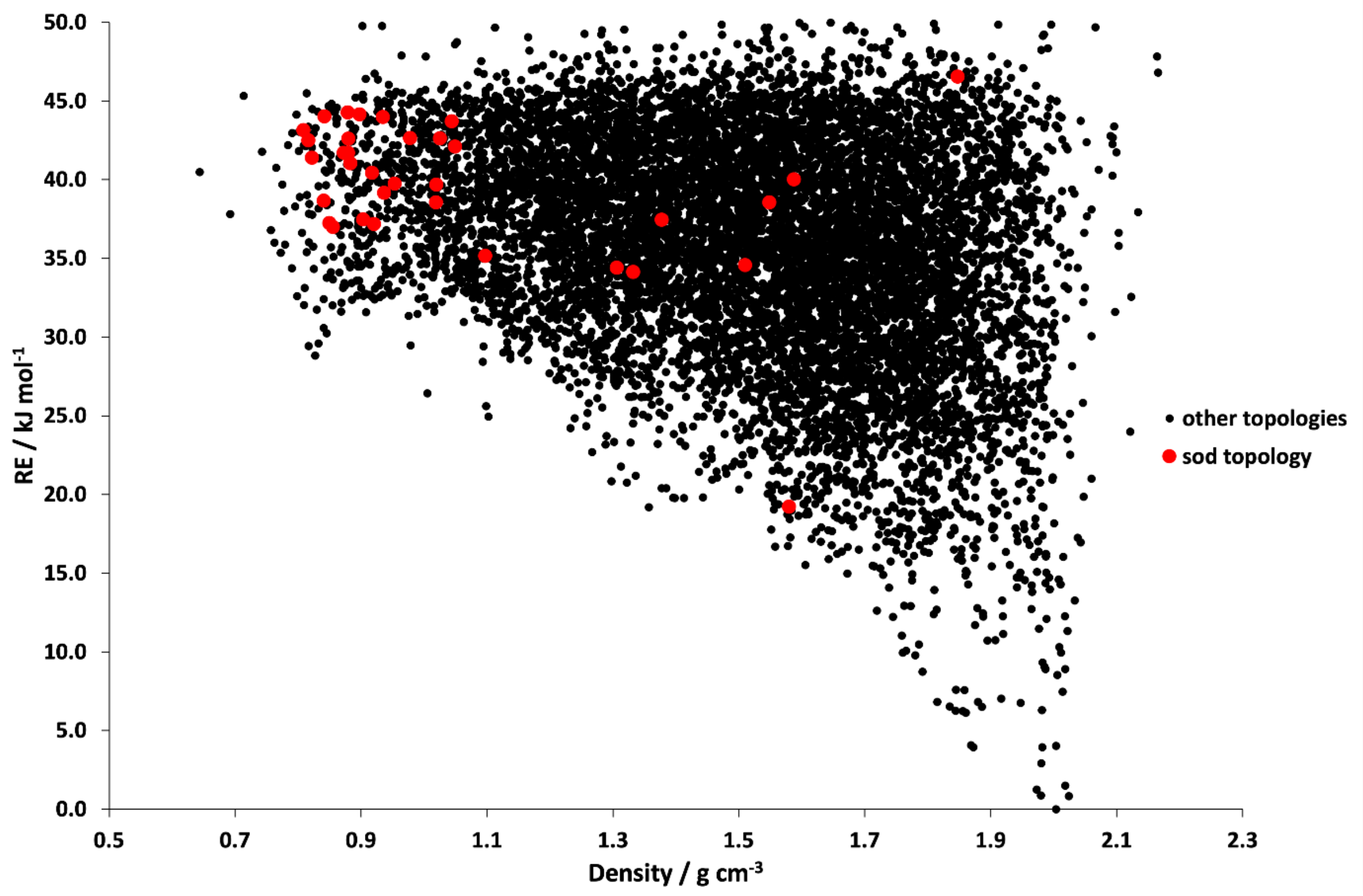

Figure S21. CSP energy landscape highlighting the structures with **sod** topology.

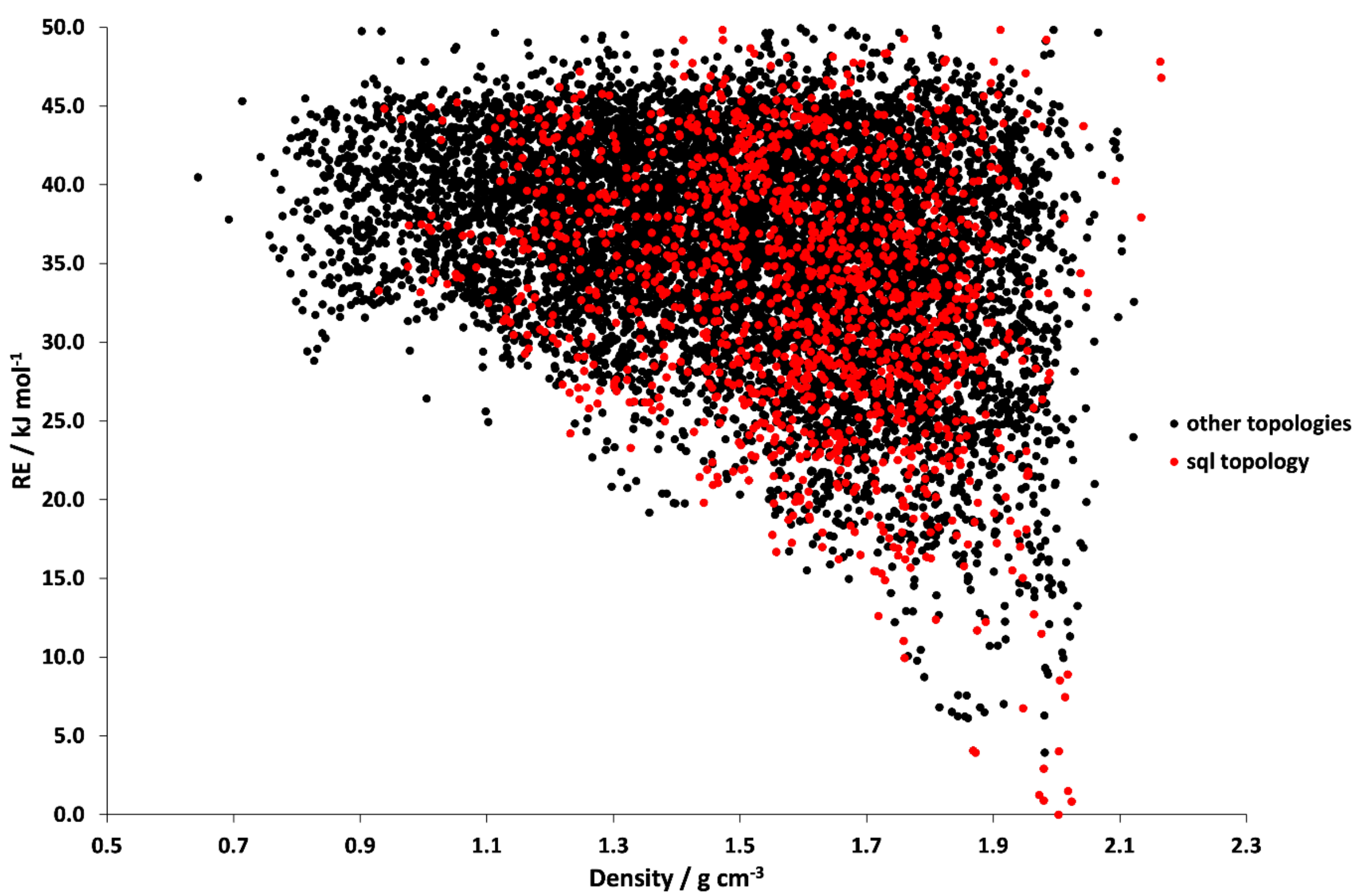

Figure S22. CSP energy landscape highlighting the structures with **sql** topology.

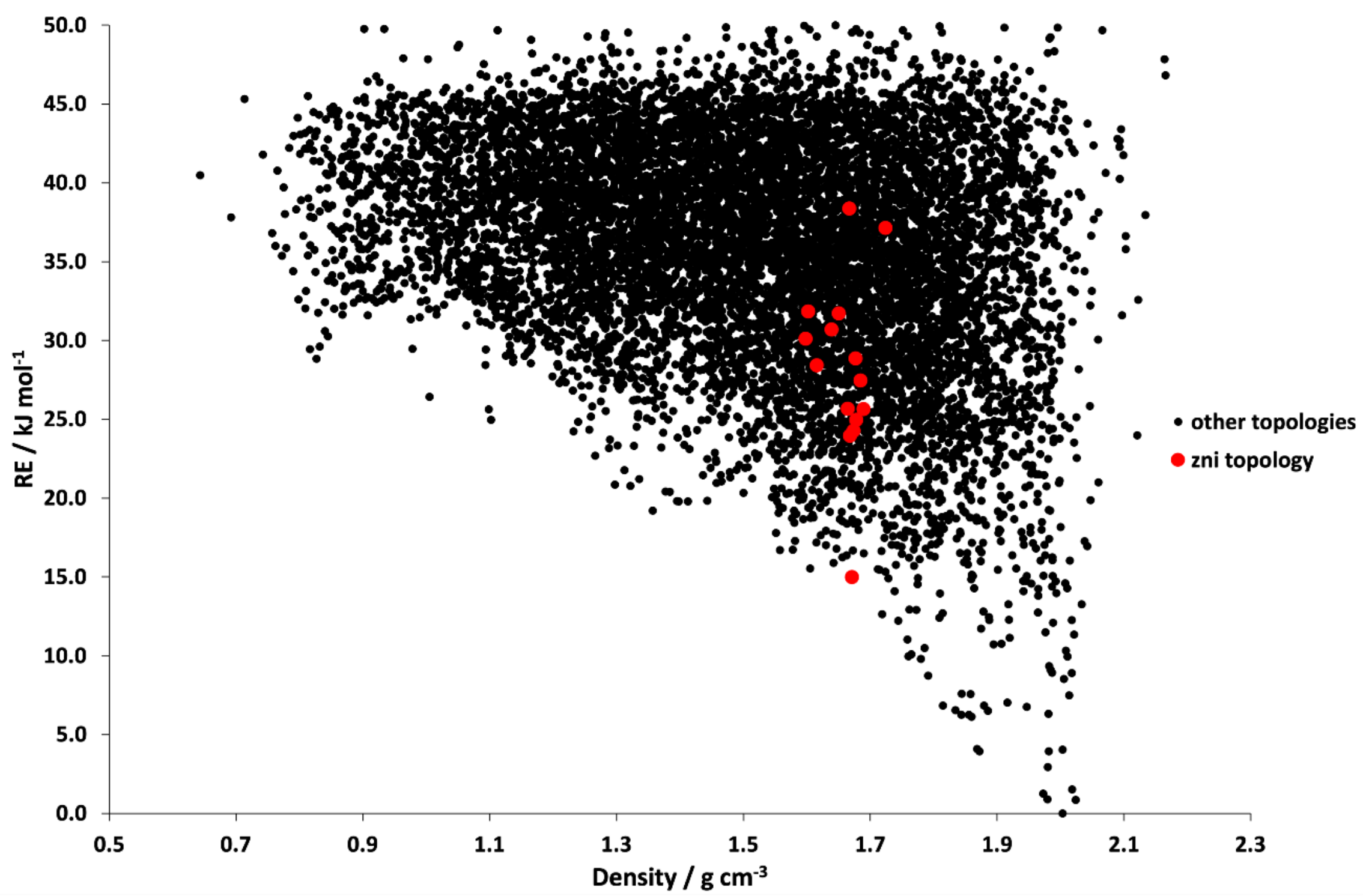

Figure S23. CSP energy landscape highlighting the structures with **zni** topology.

Table S2. Predicted structures identified as full matches to experimentally-observed polymorphs.

| Structure name | Energy relative to the global minimum / kJ $mol^{-1}$ | Calculated density / g $cm^{-3}$ | Network topology | CSD match |
|---|---|---|---|---|
| Znimid2_16_61_Pbca_FRXS3Ki7 | 31.04 | 1.137 | cag | GIZJOP |
| Znimid2_8_58_Pnnm_TNVhNp9e | 35.00 | 1.035 | crb | GAKXAW |
| Znimid2_16_94_P42212_3mPlivbU | 33.25 | 0.872 | dft | VEJYOZ |
| Znimid2_8_80_I41_sBzN9sN4 | 36.00 | 0.762 | gis | EQOCOC01 |
| Znimid2_12_114_P-421c_vYVwKfoC | 37.24 | 0.850 | sod | HIFVUO01 |
| Znimid2_16_110_I41cd_JC0of5zc | 14.98 | 1.672 | zni | IMIDZB02 |
| Znimid2_8_61_Pbca_3O6xYsHw | 17.26 | 2.038 | dia | IMIDZB14 |

Table S3. Predicted structures identified as partial matches to experimentally-observed polymorphs.

| Structure name | Energy relative to the global minimum / kJ $mol^{-1}$ | Calculated density / g $cm^{-3}$ | Network topology | CSD match |
|---|---|---|---|---|
| Znimid2_12_147_P-3_qO60Y1DX | 35.10 | 0.856 | can | PAJRUQ |
| Znimid2_8_14_P21_c_dDLJx0w7 | 29.73 | 1.316 | crb | GITTEJ |
| Znimid2_16_85_P4_n_HgEy04dz | 30.20 | 1.305 | crb | VEJYEP |
| Znimid2_16_56_Pccn_x64k7Fgx | 33.97 | 1.015 | crb | VEYJIT |
| Znimid2_8_23_I222_EAkeZmRJ | 35.39 | 0.813 | pcb | ZAVBAD |
| Znimid2_16_105_P42mc_twl6Oma1 | 36.81 | 0.897 | atn | USEKIP |

## S7. Scaling of relative energies with respect to calculated void fraction.

The experimentally-reported structures of Zn(**Im**)$_2$ found in Cambridge structural database (CSD) were geometry-optimized using and energy-ranked using the same MLIPs as in the CSP calculation. Analysis of these structures revealed a strong correlation between the calculated energy and void fraction (Figure S19, Table S2). The higher energy polymorphs contain larger solvent-accessible voids, while low energy structures contain smaller voids or are entirely close-packed.

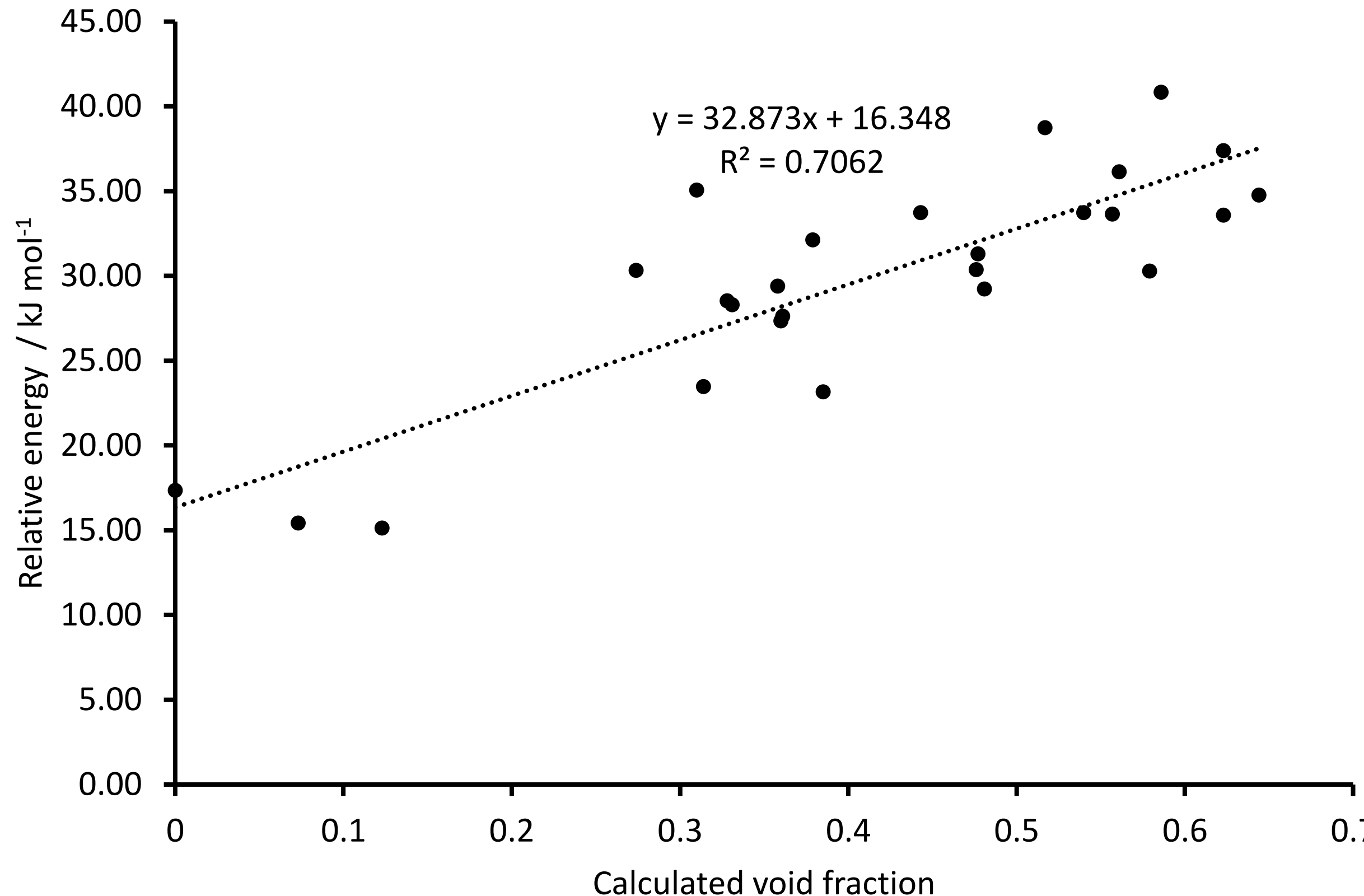

Figure S24. Correlation between the energy and void fraction for the experimentally-reported structures of ZnIm$_2$

Table S4. Calculated energies and void fractions for the experimental forms of $Zn(\mathbf{Im})_2$.

| CSD refcode | Topology | Relative energy, $E_{rel}$ / kJ mol$^{-1}$ | Void fraction, $f_{void}$ | Void-adjusted energy, $E'$ / $kJ\ mol^{-1}$ |
|---|---|---|---|---|
| IMIDZB02 | zni | 15.12 | 0.123 | 11.08 |
| IMIDZB07 | coi | 15.41 | 0.073 | 13.01 |
| IMIDZB14 | dia | 17.35 | 0.000 | 17.35 |
| HIFWAV | nog | 23.15 | 0.385 | 10.49 |
| GITTEJ | crb | 23.47 | 0.314 | 13.15 |
| VEJYUF01 | cag | 27.35 | 0.36 | 15.51 |
| GIZJOP | cag | 27.62 | 0.361 | 15.76 |
| VEJYUF07 | cag | 28.30 | 0.331 | 17.42 |
| GAKXOK | crb | 28.53 | 0.328 | 17.75 |
| USEKIP | atn | 29.22 | 0.481 | 13.41 |
| KUDJOK | neb | 29.39 | 0.358 | 17.62 |
| PAJRUQ | can | 30.28 | 0.579 | 11.25 |
| VEJYEP | crb | 30.32 | 0.274 | 21.31 |
| HICGEG | zec | 30.36 | 0.476 | 14.72 |
| GOQSIQ | 10mr | 31.30 | 0.477 | 15.62 |
| ZAVBUX | hlw1 | 32.12 | 0.379 | 19.66 |
| VEJZIU | mer | 33.57 | 0.623 | 13.09 |
| VEJYOZ | dft | 33.64 | 0.557 | 15.33 |
| ZAVBAD | pcb/aco | 33.72 | 0.54 | 15.97 |
| GAKXAW | crb | 33.73 | 0.443 | 19.17 |
| DOTCIC | gme | 34.77 | 0.644 | 13.60 |
| KEVLEE | neb | 35.07 | 0.31 | 24.88 |
| QOSXUS | aco | 36.14 | 0.561 | 17.69 |
| EQOCOC01 | gis | 37.37 | 0.623 | 16.89 |
| VEJYIT | crb | 38.75 | 0.517 | 21.75 |
| HIFVUO | gis | 40.83 | 0.586 | 21.57 |
| | | | | $E'_{average}$ 16.35±3.66 |

The energy-porosity relationship found in the experimental polymorphs can be used to assess the synthetic feasibility of the structures generated via CSP. The linear correlation between the lattice energy and void fraction allows us to introduce a void-adjusted energy descriptor under the equation:

$$E' = E_{rel} - k_V \times f_{void} \quad (1)$$

where $E_{rel}$ is the calculated energy of the structure relative to the global minimum; $k_V$ = 32.873 $kJ\ mol^{-1}$ - linear regression coefficient based on the energies and void fractions for the experimental structures; $f_{void}$ is the calculated void fraction.

The least squares analysis of the energies and void fractions of experimental polymorphs of Zn(**Im**)$_2$ gives the values of $E' = (16.348 \pm 3.664)\ kJ\ mol^{-1}$. Assuming the predicted structures follow the same trend as the experimental forms of Zn(**Im**)$_2$ and setting the upper limit of acceptable $E'$ one standard error above the mean value, we obtain a threshold value of

$$E'_{max} = 20.012\ kJ\ mol^{-1}$$

as an upper boundary for the structures that fulfil the synthesizability criterion, leaving 982 structures out of the total 9626 as likely candidates for experimental synthesis.

## S8. Comparison of the predicted structures against experimental powder diffraction patterns.

Table S5. Four experimental PXRD patterns from mechanochemical LAG method were selected to verify, how the software Critic2 can match experimental PXRD patterns against the predicted CSP structures. The variable-cell similarity score (DIFF) for each pattern is shown, along with the ranking of the matching predicted structure to the experimental pattern.

| Structure name and Chemiscope number | Topology | LAG additive | DIFF | Ranking |
|---|---|---|---|---|
| Znimid2_8_58_Pnnm_TNVhNp9e **3918** | crbT | toluene | 0.31 | 199 |
| Znimid2_16_110_I41cd_JC0of5zc **75** | zni | methanol | 0.02 | 6 |
| Znimid2_16_61_Pbca_FRXS3Ki7 **2263** | cag | dimethylformamide | 0.21 | 473 |
| Znimid2_16_61_Pbca_FRXS3Ki7 **2263** | cag | choloform | 0.23 | 763 |

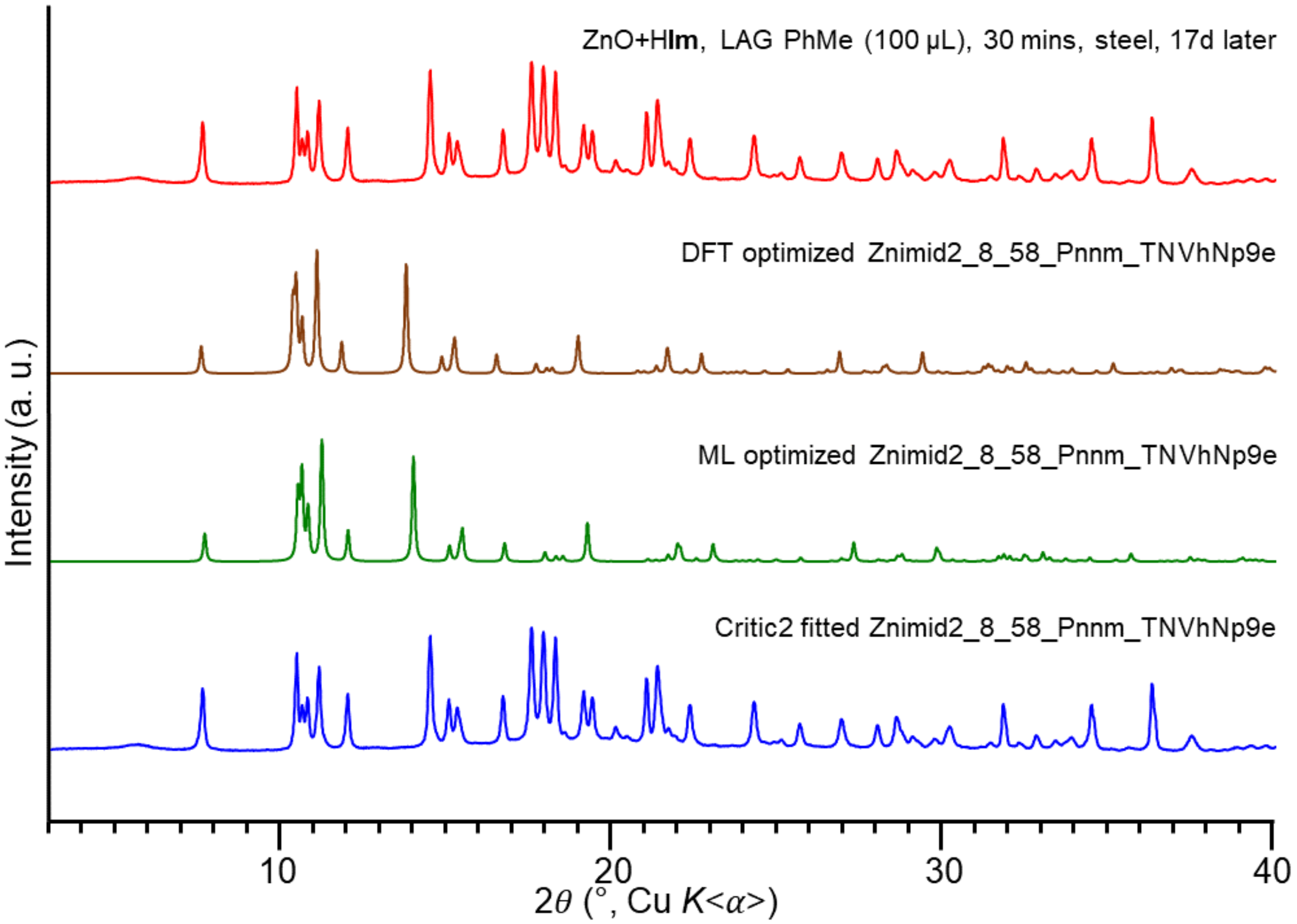

Figure S25. PXRD patterns overall from top to bottom for: experimentally measured pattern for Zn(**Im**)$_2$ with **crbT** (CSD GAKXAW) topology (red); simulated DFT optimized predicting structure pattern (brown); simulated pattern from ML optimized structure (green); Critic2 fitted PXRD pattern from the ML-optimized structure (blue).

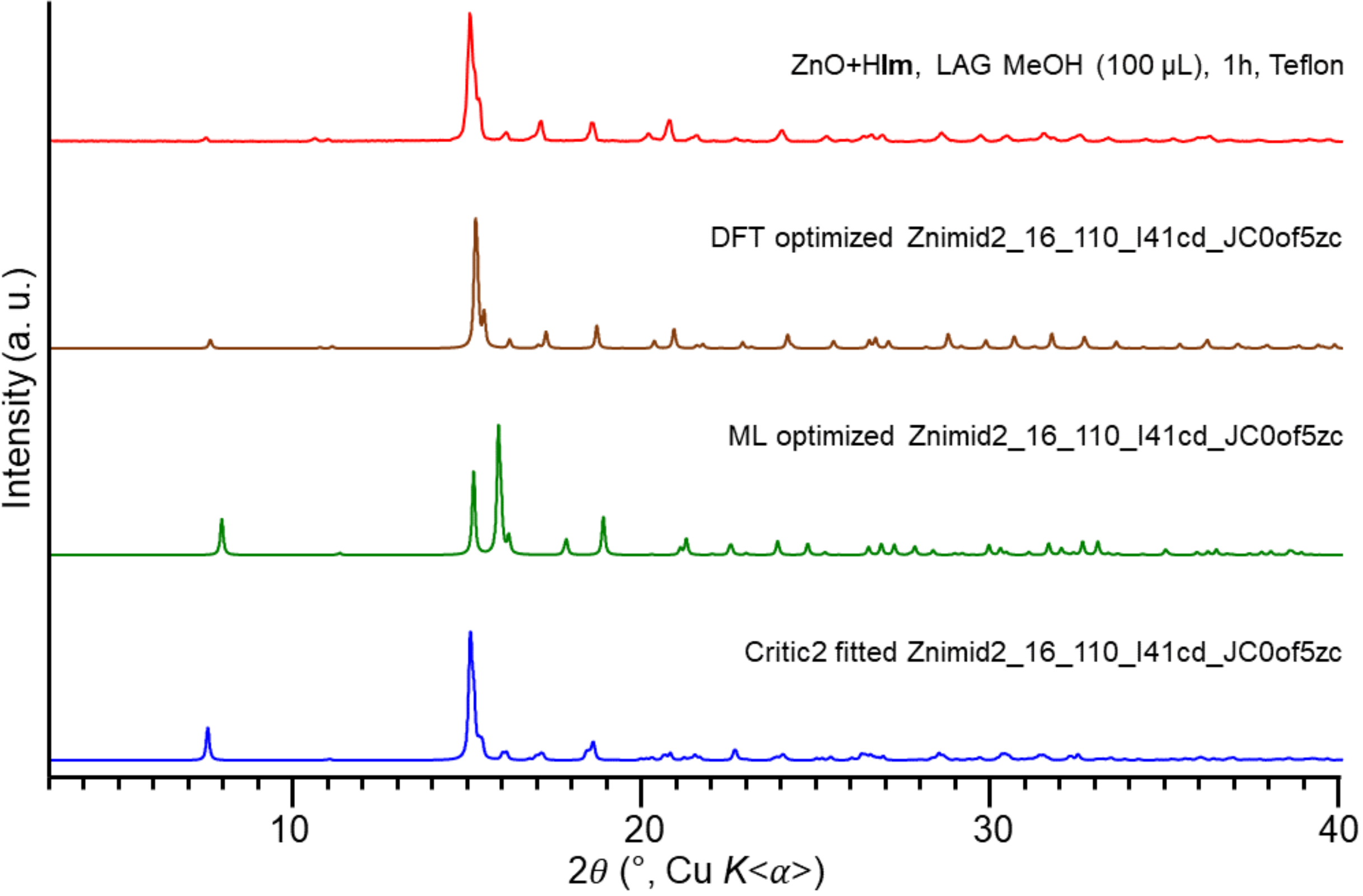

Figure S26. PXRD patterns overall from top to bottom for: experimentally measured pattern for Zn(**Im**)$_2$ with **zni** (CSD IMIDZB02) topology (red); simulated DFT optimized predicting structure pattern (brown); simulated pattern from ML optimized structure (green); Critic2 fitted PXRD pattern from the ML-optimized structure (blue).

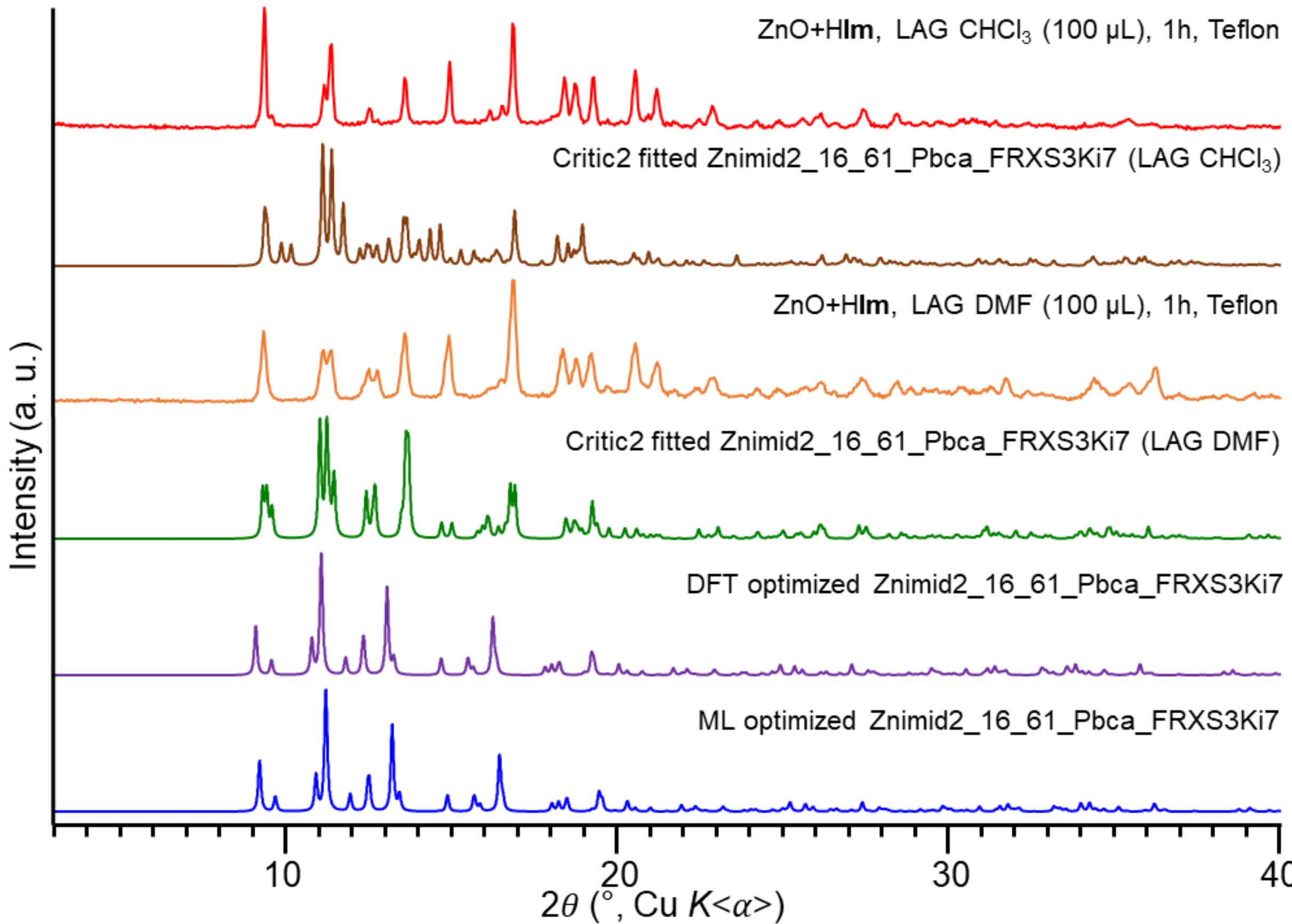

Figure S27. PXRD patterns overall from top to bottom for: experimentally measured pattern, using chloroform as LAG, for Zn(**Im**)$_2$ with **cag** (CSD GAKXEA) topology (red); ); Critic2 fitted PXRD pattern with chloroform as LAG (brown); experimentally measured PXRD pattern using DMF as LAG (orange); Critic2 fitted PXRD pattern with chloroform as LAG (green); simulated DFT optimized predicting structure pattern (purple); simulated pattern from ML optimized structure (blue).

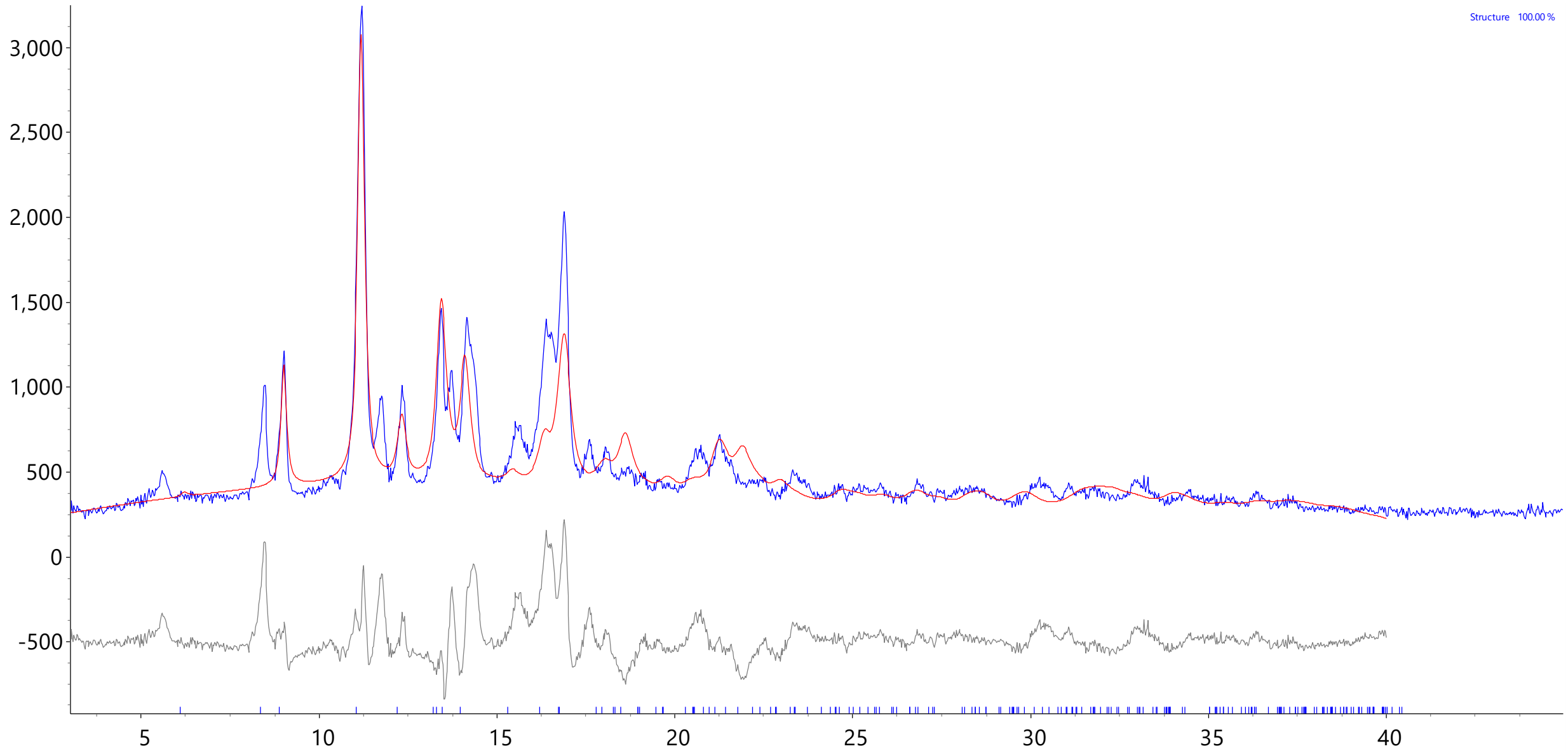

Figure S28. Rietveld refinement plot for the best matching predicted structure for the material obtained by heating **crb**-$ZnIm_2$ at 150 °C for 3 hours. The structure identifier is Znimid2_16_56_Pccn_YMQOEPiX, located in the Chemiscope file under the number 9563. The experimental profile is shown in blue, the calculated profile is shown in red, and the difference curve is shown in grey. Evidently there are significant discrepancies between the experimental and calculated diffraction profile, indicating that the current structural model requires more work before it can be counted as a fully-accurate structure determination.